\documentclass[letterpaper,11pt]{article}

\usepackage{jheppub}
\allowdisplaybreaks
\usepackage{booktabs}
\usepackage{subcaption}
\usepackage{caption}
\usepackage{graphicx}
\usepackage{blkarray}
\usepackage{xspace}
\usepackage{stackengine}[2013-09-11]
\usepackage{adjustbox}

\usepackage{tikz}
\usetikzlibrary{decorations.pathreplacing}
\usepackage{slashed}
\usepackage{feynmp}
\usepackage{color}

\DeclareRobustCommand{\Sec}[1]{Sec.~\ref{sec:#1}}
\DeclareRobustCommand{\Secs}[2]{Secs.~\ref{sec:#1} and \ref{sec:#2}}

\DeclareRobustCommand{\Fig}[1]{Fig.~\ref{fig:#1}}

\DeclareRobustCommand{\Eq}[1]{Eq.~(\ref{eq:#1})}
\DeclareRobustCommand{\Eqs}[2]{Eqs.~(\ref{eq:#1}) and (\ref{eq:#2})}

\newcommand{\eqn}[1]{\begin{align}#1\end{align}}
\newcommand{\eqna}[1]{\begin{align}\begin{aligned}#1\end{aligned}\end{align}}

\newcommand{\Pythia}{{\sc Pythia}\xspace}
\newcommand{\FastJet}{{\sc FastJet}\xspace}
\newcommand{\JUNIPR}{{\sc Junipr}\xspace}
\newcommand{\JUNIPRtitle}{JUNIPR\xspace}
\newcommand\slashzero{{\stackinset{c}{}{c}{}{\tiny /}{0}}}
\newcommand{\JUNIPRzero}{{\sc Junipr$_{\scriptsize \slashzero}$}\xspace}

\title{\JUNIPRtitle: a Framework for Unsupervised Machine Learning in Particle Physics}

\author[a]{Anders Andreassen,}
\author[b]{Ilya Feige,}
\author[a]{Christopher Frye,}
\author[a]{Matthew D.~Schwartz}

\affiliation[a]{Department of Physics, Harvard University, Cambridge, MA 02138}
\affiliation[b]{ASI Data Science, 54 Welbeck Street, London, W1G 9XS}

\emailAdd{anders@physics.harvard.edu}
\emailAdd{ilya@asidatascience.com}
\emailAdd{frye@physics.harvard.edu}
\emailAdd{schwartz@physics.harvard.edu}

\abstract{
In applications of machine learning to particle physics, a persistent challenge is how to go beyond discrimination to learn about the underlying physics.
To this end, a powerful tool would be a framework for unsupervised learning, 
where the machine learns the intricate high-dimensional contours of the data upon which it is trained, without reference to pre-established labels. 
In order to approach such a complex task, an unsupervised network must be structured intelligently, based on a qualitative understanding of the data.
In this paper, we scaffold the neural network's architecture around a leading-order model of the physics underlying the data. 
In addition to making unsupervised learning tractable, this design actually alleviates existing tensions between performance and interpretability.
We call the framework \JUNIPR: ``Jets from UNsupervised Interpretable PRobabilistic models''.
In this approach, the set of particle momenta composing a jet are clustered into a binary tree that the neural network examines sequentially.
Training is unsupervised and unrestricted: the network could decide that the data bears little correspondence to the chosen tree structure.
However, when there is a correspondence, the network's output along the tree has a direct physical interpretation.
\JUNIPR models can perform discrimination tasks, through the statistically optimal likelihood-ratio test, 
and they permit visualizations of discrimination power at each branching in a jet's tree.
Additionally, \JUNIPR models provide a probability distribution from which events can be drawn, providing a data-driven Monte Carlo generator.
As a third application, \JUNIPR models can reweight events from one (e.g.~simulated) data set to agree with distributions from another (e.g.~experimental) data set.
}

\begin{document} 
\maketitle
\flushbottom
\newpage


\section{Introduction}
\label{sec:Introduction}

Machine learning models based on deep neural networks have revolutionized information processing over the last decade. 
Such models can recognize objects in images \cite{krizhevsky2012imagenet, ResNet, DenseNet}, perform language translation \cite{NMT, GoogleNMT}, transcribe spoken language \cite{Transcription}, and even speak written text \cite{Wavenet} at approaching human level. 
The truly revolutionary aspect of this progress is the generality of deep neural networks:
a broad diversity of network architectures can be created from basic building blocks that allow for efficient calculation of gradients via back propagation, 
and thus efficient optimization through stochastic gradient descent \cite{BackProp}. 
These methods are arbitrarily expressive and can model extremely high dimensional data. 

The architecture of a neural network should be designed to  process information efficiently, 
from the input data all the way through to the network's final output.
Indeed, it empirically seems to be the case that networks that process information evenly layer-by-layer perform very well.
One example of this empirical result is that deep convolutional networks for image processing seem to perform sequentially more abstract operations as a function of depth \cite{krizhevsky2012imagenet}. 
Similarly, recurrent networks perform well on time series data, 
as their recurrent layers naturally describe step-by-step evolution in time \cite{mikolov2010recurrent}. 

The power and generality of deep neural networks has been leveraged across the sciences, and in particular in particle physics.
The simplest architecture explored has been the fully-connected network, which has successfully been applied in a wide variety of contexts,
such as in identifying and splitting clusters from multiple particles in the pixel detector \cite{Aad:2014yva}, 
in $b$-tagging \cite{Aad:2015ydr}, 
and in $\tau$-identification \cite{Chatrchyan:2012zz}.
In these basic applications, the neural network optimizes its use of some finite number of relevant physical observables
for the task at hand.\footnote{
For recent work on constructing a basis for neural network inputs, see \cite{Datta:2017rhs,Datta:2017lxt,Luo:2017ncs}, 
and see \cite{Komiske:2017aww} for a linear approach that does not require neural network methods.}
One drawback of such an approach is that the neural network is limited by the observables it is given. 
In fact, for these applications, other multivariate methods such as boosted decision trees often have comparable performance using the same
inputs, but train faster and can be less sensitive to noise~\cite{Gallicchio:2010dq,ATL-PHYS-PUB-2017-004}.

As an alternative to feeding a neural network a set of motivated observables, one can feed it raw information. By doing so, one allows
the network to take advantage of useful features that physicists have yet to discover.
One way of preprocessing the raw data in a fairly unbiased way is through the use of jet images, which contain as pixel intensities the energy deposited by jet constituents in calorimeter cells \cite{Cogan:2014oua}.
Jet images invite the use of techniques from image recognition to discriminate jets of different origins.
In \cite{Cogan:2014oua}, the pixel intensities in the two-dimensional jet image were combined into a vector, 
and a Fisher linear discriminant was then used to find a plane in the high-dimensional space that maximally separates two different jet classes. 
Treating a 2-dimensional jet image as an unstructured collection of pixel intensities, however, ignores the spatial locality of the problem, 
i.e.~that neighboring pixels should have related intensities.
Convolutional neural networks (CNNs), which boast reduced complexity by leveraging this spatially local structure,
have since been adopted instead, and they generally outperform fully-connected networks due to their efficient feature detection.
In the first applications of CNNs to jet images, on boosted $W$ detection \cite{deOliveira:2015xxd} and quark/gluon discrimination \cite{Komiske:2016rsd}, 
it was indeed found that simple CNNs could generally outperform previous techniques.
Since then, a number of studies have aimed to optimize various discrimination tasks using CNNs \cite{Komiske:2017ubm,Kasieczka:2017nvn, Bhimji:2017qvb, ATL-PHYS-PUB-2017-017, Macaluso:2018tck, Chien:2018dfn}. 

While the two-dimensional detector image acts as a natural representation of a jet, especially from an experimental standpoint, 
the 4-momenta of individual jet constituents provide a more fundamental representation for the input to a neural network. 
One complication in transitioning from the jet image to its list of momenta is that, 
while the image is a fixed-size representation, the list of momenta will have different sizes for different jets.
To avoid this problem, one could truncate the list of momenta in the jet to a fixed size, and zero-pad jets smaller than this size \cite{Pearkes:2017hku}.
Alternatively, there are network architectures, namely recursive (RecNNs) and recurrent neural networks (RNNs), that handle variable length inputs naturally. 
With such methods, one also has the freedom to choose the order in which constituent momenta are fed into the network.
In \cite{Louppe:2017ipp}, a RecNN was used to build a fixed-size representation of the jet, 
and the authors explored various ways of ordering the momenta as input to the network: by jet clustering algorithms, by transverse momentum, and randomly. 
The resulting representation of the jet was then fed to a fully-connected neural network for boosted $W$ tagging. 
RecNNs and RNNs have also been used in similar ways for quark/gluon discrimination \cite{Cheng:2017rdo}, top tagging \cite{Egan:2017ojy}, and jet charge \cite{Fraser:2018ieu}. 
See also \cite{Guest:2016iqz, ATL-PHYS-PUB-2017-003} for jet flavor classification using tracks.

To date, the majority of applications of machine learning to particle physics employ supervised machine learning techniques. 
Supervised learning is the optimization of a model to map input to output based on labeled input-output pairs in the training data.
These training examples are typically simulated by Monte Carlo generators, in which case the labels come from the underlying physical processes being generated.
Most of the classification studies mentioned above employ this style of supervised learning, and similar techniques have also been utilized for regression tasks such as pileup subtraction \cite{Komiske:2017ubm}.
Alternatively, training data can be organized in mixed samples, each containing different proportions of the different underlying processes. In this case, labels correspond to the mixed samples, and learning is referred to as weakly supervised. 
While full and weak supervision are very similar as computational techniques, the distinction is exceptionally important in particle physics, where the underlying physical processes are unobservable in real collider data.
Early studies of weakly supervised learning in particle physics show very promising results: performance comparable to fully supervised methods was found both with low-dimensional inputs~\cite{Metodiev:2017vrx, Cohen:2017exh} (a few physical observables) 
and with very high-dimensional  inputs~\cite{Komiske:2018oaa} (jet images).

With supervised learning, there is a notion of absolute accuracy: 
since every training example is labeled with the desired output, the network predicts this output either correctly or incorrectly.
This is in contrast to {\bf unsupervised learning}, where the machine learns underlying structure that is unlabeled in the training data. 
Without output-labeled training examples, there is no notion of absolute accuracy. 
Several recent studies have employed unsupervised learning techniques in particle physics.
In \cite{Metodiev:2018ftz}, borrowing concepts from topic modelling in text documents,
the authors extract observable distributions of underlying quark and gluon jets from two mixed samples.
In \cite{deOliveira:2017pjk,Paganini:2017hrr,Paganini:2017dwg}, generative adversarial networks (GANs) are used to efficiently generate realistic jet images and calorimeter showers.

In this work, we explore another approach to unsupervised machine learning in particle physics, 
in which a deep neural network learns to compute the relative differential cross section of each data point under consideration, 
or equivalently, the probability distribution generating the data.
The power of having access to the probability distribution underlying the data should not be underestimated. 
For example, likelihood ratios would provide optimal discriminants \cite{Neyman289}, 
and sampling from the probability distribution would provide completely data-driven simulations.

In this paper, we introduce a framework named \JUNIPR: ``Jets from UNsupervised Interpretable PRobabilistic models''.
We also present a basic implementation of this framework using a deep neural network.
This network directly computes the general probability distribution underlying particle collider data using unsupervised learning.

The task of learning the probability distribution underlying collider data comes with challenges due to the complexity of the data.
Some past studies have aimed to process collider information efficiently by using neural network architectures inspired by physics techniques already in use \cite{Louppe:2017ipp,Cheng:2017rdo,Egan:2017ojy,Fraser:2018ieu,Guest:2016iqz,Butter:2017cot}.
In this paper, we take this idea one step further. 
We scaffold the neural network architecture around a leading-order description of the physics underlying the data, 
from first input all the way to final output.
Specifically, we base the \JUNIPR framework on algorithmic jet clustering trees.
The tree structure is used, both in processing input information, and in decomposing the network's output.
In particular, \JUNIPR's output is organized into meaningful probabilities attached to individual nodes in a jet's clustering tree.
In addition to reducing the complexity and increasing the efficiency of the corresponding neural network,
this approach also forces the machine to speak a language familiar to physicists,
thus enabling its users to interpret the underlying physics it has learned.
Indeed, one common downside associated with machine learning techniques in physics is that, 
though they provide powerful methods to accomplish the tasks learned in training, 
they do little to clarify the underlying physics that underpins their success. 
Our approach minimizes this downside.

Let us elaborate on the tree-based architecture used for \JUNIPR's implementation.
In particle physics, events at colliders are dominated by the production of collimated collections of particles known as jets. 
The origin of jets and many of their properties can be understood through the fundamental theory of strong interactions, quantum chromodynamics (QCD).
One insight from QCD is that jets have an inherently fractal structure, inherited from the approximate scale invariance of the fundamental theory. 
The fractal structure is made precise through the notion of factorization, 
which states that the dynamics in QCD stratify according to soft, collinear, and hard physics \cite{Coleman:1965xm, Collins:1985ue, Collins:1988ig, feige2014hard, feige2013shell}, with each sector being separately scale invariant. 
To capture this structure efficiently in \JUNIPR, we use a kind of factorized architecture, with a dense network to describe local branchings (well-suited for collinear factorization), and a global RNN superstructure general enough to encode soft coherence and any factorization-violating effects. 

One might naively expect this setup to require knowledge of the sequence of splittings that created the jet. 
Although there is a sequence of splittings in parton-shower simulations, 
the splittings are only a semi-classical approximation used to model the intensely complex and essentially incalculable distribution of final state particles.  
Real data is not labelled with any such sequence. 
In fact, there are many possible sequences which could produce the same event, 
and the cross section for the event is given by the square of the quantum mechanical sum of all such amplitudes, including effects of virtual particles. 
A proxy for this fictitious splitting history is a clustering history that can be constructed in a deterministic way using a jet-clustering algorithm, 
such as the $k_t$ algorithm \cite{Catani:1993hr,Ellis:1993tq} or the Cambridge/Aachen (C/A) algorithm \cite{Dokshitzer:1997in,Wobisch:1998wt}. 
There is no {\it correct} algorithm: each is just a different way to process the momenta in an event. 
Indeed, there seems to be useful information in the multiple different ways that the same event can be clustered~\cite{Ellis:2012sn,Kahawala:2013sba,Mackey:2015hwa}. 
Any of these algorithms, or any algorithm at all that encodes the momenta of an event into a binary tree, can be used to scaffold a neural network in the \JUNIPR approach.

For practical purposes, \JUNIPR is implemented with respect to a fixed jet clustering algorithm.
Without a fixed algorithm, the probability of the final-state particles constructed through $1\to2$ branchings would require marginalization over all possible clustering histories 
--- an extremely onerous computational task.
In principle, fixing the algorithm used to implement \JUNIPR should be inconsequential for its output, namely the probability distribution over final-state momenta,
as these momenta are independent of clustering algorithm.
To reiterate, the \JUNIPR approach does not require the chosen clustering algorithm to agree with the underlying data-generation process;
this is demonstrated in \Secs{DiscussionPrinterJets}{DiscussionAntikt} below.
On the other hand, the \emph{sequence} of probabilities assigned to each branching in a clustering tree certainly depends on the algorithm used to define the tree.
For example, the same final probability $P=10^{-22}$ could be reached with one clustering algorithm through the sequence $P = 10^{-5}\cdot 10^{-6}\cdot 10^{-8} \cdot 10^{-3}$, 
or with another algorithm through $P= 10^{-15}\cdot 10^{-2}\cdot 10^{-1} \cdot 10^{-4}$.
The key idea is that, if an algorithm is chosen which does correspond to a semi-classical parton shower, the resulting sequence of probabilities may be understandable.  
This provides avenues for users to interpret what physics the machine learns, and we expect that dissecting \JUNIPR will be useful in such cases.
We will demonstrate this throughout the paper. 

It is worth emphasizing one fundamental aspect of our approach for clarity. 
The \JUNIPR framework yields a {\bf probabilistic model}, not a generative model. 
The probabilistic model allows us to directly compute the probability density of an individual jet, as defined by its set of constituent particle momenta.
To be precise, this is the probability density for those particular momenta to arise in an event, conditioned on the event selection criteria used to select the training data.
As a complementary example of this, shower deconstruction \cite{Soper:2011cr,Soper:2014rya} provides a theory-driven approach to probabilistic modeling in particle physics,
in which probabilities are calculated using QCD rather than a neural network.
In contrast, a generative model would output an example jet, taking random noise as input to seed the generation process.
Given a distribution of input seeds, the jets output from a generative model should follow the same distribution as the training data.
While this means that the probability distribution underlying the data is internally encoded in a generative model, this underlying distribution is hidden from the user.
Examples of generative models in particle physics include Monte Carlo event generators and,
more recently, GANs used to generate jet images and detector simulations \cite{deOliveira:2017pjk,Paganini:2017hrr,Paganini:2017dwg}.

The direct access to the probability distribution that is enabled by a probabilistic model comes with several advantages.
If two different probabilistic models are trained on two different samples of jets,
they can be used to compute likelihood ratios that distinguish between the two samples. 
Likelihood ratios provide theoretically optimal discriminants \cite{Neyman289}, which is indeed a major motivation for \JUNIPR's probabilistic approach. 
One can also sample from a probabilistic model in order to generate events, 
though generative models are better-suited for this application \cite{deOliveira:2017pjk,Paganini:2017hrr,Paganini:2017dwg}. 
In addition, one can use a probabilistic model to reweight events generated by an imperfect simulator, 
so that the reweighted events properly agree with data.

In this paper, as a proof-of-concept, we use simulated $e^+e^-$ data to train a basic implementation of the \JUNIPR framework described above.
We have not yet attempted to optimize all of this implementation's hyperparameters; 
however, we do find that a very simple architecture with no fine tuning is adequate.
This is confirmed by its impressive discrimination power and its effective predictivity for a broad class of observables,
but more rigorous testing is needed to determine whether this approach can provide state-of-the-art results on the most pressing physics problems.

The general probabilistic model, its motivation, and a specific neural network implementation of it are discussed in \Sec{Model}. 
A comprehensive discussion of training the model, including the data used and potential subtleties in extending the model are covered in \Sec{Training}. 
Results on discrimination, generation, and reweighting are presented in \Sec{Results}. 
We provide robustness tests and some conceptually interesting results related to factorization in \Sec{Discussion}, including the counterintuitive anti-$k_t$ shower generator. 
There are many ways to generalize our approach, as well as many applications that we do not fully explore in this work. 
We leave a discussion of some of these possible extensions to \Sec{Conclusions}, where  we conclude.


\section{Unsupervised Learning in Jet Physics}
\label{sec:Model}

To establish the framework clearly and generally, \Sec{ProbModel} begins by describing \JUNIPR as a general probabilistic model, independent of the specific parametric form taken by the various functions it involves. From this perspective, such a probabilistic model could be implemented in many different ways. \Sec{NNModel} then describes the particular neural network implementation of \JUNIPR used in this paper, which has a simple but QCD-customized architecture and minimal hyperparameter tuning.

\subsection{General Probabilistic Model}
\label{sec:ProbModel}

\begin{figure}[t]
\centering
\includegraphics[width=0.8\linewidth]{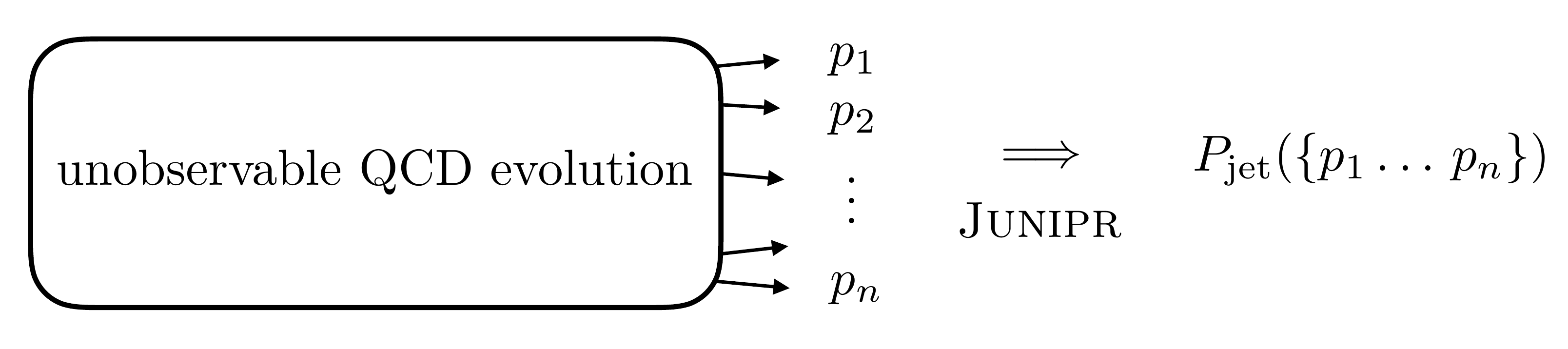}
\caption{\JUNIPR predicts the probability density $P_{\text{jet}}(\{p_1,\cdots,p_n\})$ of finding a given set of momenta $\{p_1,\ldots,p_n\}$ in a jet, conditioned on the jet selection criteria used to select the training data. No assumptions are made about the underlying quantum-mechanical processes that generated the jet.}
\label{fig:UnClusteredJet}
\end{figure}

Consider a set of final-state 4-momenta $p_1, \ldots, p_n$ that we hereafter refer to as ``the jet".
\JUNIPR computes the probability density $P_\text{jet}(\{p_1, \ldots, p_n\})$ of this set of momenta arising in an event,
assuming the event selection criteria used to select the training data.
This probability distribution is normalized so that, abstractly,
\begin{equation}
    \sum_{n=1}^\infty \, \int d^4p_1 \cdots d^4p_n \, P_\text{jet}(\{p_1, \ldots p_n\}) = 1\,,
\end{equation}
where the integral extends over the physical region of phase space. 
(In practice, in implementing \JUNIPR we discretized the phase space into cells and assigned a measure of unity to each discrete cell. 
This results in $P_\text{jet}$ being a discrete cell-size-dependent probability distribution, but this choice is conceptually unimportant here.)
A high-level schematic of \JUNIPR is shown in \Fig{UnClusteredJet}, which emphasizes that the model does not attempt to learn the quantum-mechanical evolution that created the jet, but only meaningfully predicts the likelihood of its final-state momenta.

\begin{figure}[t]
\centering
\includegraphics[width=\linewidth]{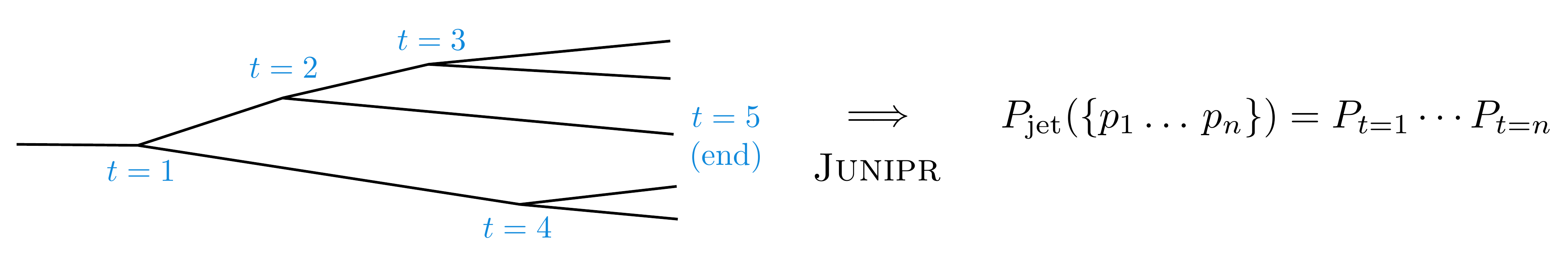}
\caption{With any fixed clustering algorithm, the probability distribution over final-state momenta can be decomposed into a product of distributions. Each factor in the product corresponds to a different step in the clustering tree. Subsequent probabilities are conditioned on the outcomes from previous steps, so this decomposition entails no loss of generality.}
\label{fig:ClusteredJet}
\end{figure}

An unstructured model of the above form would ignore the fact that we know jet evolution is well-described by a semi-classical sequence of $1\to2$ splittings, 
due to factorization theorems \cite{Coleman:1965xm, Collins:1985ue, Collins:1988ig, feige2014hard, feige2013shell}. 
A model that ignores factorization would be much more opaque to interpretation, 
and have many more parameters than needed due to its unnecessary neutrality. 
Thus, we propose a model that describes a given configuration of final-state momenta using sequential $1\to2$ splittings. 
Such a sequence is defined by a jet clustering algorithm, which assigns a clustering tree to any set of final-state momenta,
so that a sequential decomposition of the probability distribution can be performed without loss of generality.
We imagine fixing a specific algorithm to define the trees, 
so that there is no need to marginalize over all possible trees in computing a probability, a computation that would be intractable. 
While a deterministic clustering algorithm cannot directly describe the underlying quantum-mechanical parton evolution, that is not the goal for this model.
With the algorithm set, the model as shown in \Fig{UnClusteredJet} becomes that shown in \Fig{ClusteredJet}.

We will now formalize this discussion into explicit equations. 
For the rest of this section we assume that the clustering tree is determined by a fixed jet algorithm (e.g.~any of the generalized $k_t$ algorithms \cite{Cacciari:2008gp,Cacciari:2011ma}). 
The particular algorithm chosen is theoretically inconsequential to the model, as the same probability distribution over final states will be learned for any choice. 
Practically speaking, however, certain algorithms may have advantages over others. 
We will discuss the choice of clustering algorithm further in \Secs{DiscussionPrinterJets}{DiscussionAntikt}.

The application of a clustering algorithm on the jet constituents $p_1, \ldots, p_n$ defines a sequence of ``intermediate states'' $k^{(t)}_1, \ldots, k^{(t)}_{t}$. 
Here the superscript $t = 1, \ldots, n$ labels the intermediate state after the $(t-1)^\text{th}$ branching in the tree (where counting starts at 1) 
and the subscript $i = 1, \ldots, n$ enumerates momenta in that state. 
To be explicit,
\begin{itemize}
\item the ``initial state'' consists of a single momentum: $k^{(1)}_1 = p_1 + \cdots + p_n$;
\item at subsequent steps $\{k^{(t)}_1,\ldots, k^{(t)}_{t}\}$ is gotten from $\{k^{(t-1)}_1,\ldots, k^{(t-1)}_{t-1}\}$ by a single momentum-conserving $1 \to 2$ branching;
\item after the final branching, the state is the physical jet: $\{k^{(n)}_1,\ldots, k^{(n)}_{n}\} = \{p_1,\ldots, p_n\}$.
\end{itemize}
In this notation, the probability of the jet (as shown in \Fig{ClusteredJet}) can be written as 
\eqn{
\label{eq:BasicProductAssumption}
P_\text{jet}(\{p_1,\ldots p_n\}) &= \Bigg[\prod_{t=1}^{n-1} 
P_t\big(k^{(t+1)}_1,\ldots, k^{(t+1)}_{t+1} \big| k^{(t)}_1,\ldots, k^{(t)}_{t}\big)
\Bigg] \\
&\times P_{n}\big(\text{end} \big| k_1^{(n)}, \ldots, k_n^{(n)}\big).\nonumber}

\Eq{BasicProductAssumption} allows for a natural, sequential description of the jet. 
However, it obscures the factorization of QCD which predicts an approximately self-similar splitting evolution. 
Thus we decompose the model further, so that each $P_t$ in \Eq{BasicProductAssumption} is described by a $1\to2$ branching function that only indirectly receives information about the rest of the jet. 
The latter is achieved via an unobserved representation vector $h^{(t)}$ of the global state of the jet at step $t$. 
To be explicit, let $k_m^{(t)} \to k_{d_1}^{(t+1)} \, k_{d_2}^{(t+1)}$ denote the branching of a mother into daughters that achieves the transition from $k^{(t)}_1,\ldots, k^{(t)}_{t}$ to $k^{(t+1)}_1,\ldots, k^{(t+1)}_{t+1}$ in the clustering tree. 
Then we can write 
\begin{align}
\label{eq:IndividTimeStepModel}
    P_t\big(k^{(t+1)}_1,\ldots \big| k^{(t)}_1,\ldots\big) &= 
    P_\text{end}\big(0\big|h^{(t)}\big) \,
    P_\text{mother}\big(m^{(t)}\big|h^{(t)}\big) \,
    P_\text{branch}\big(k_{d_1}^{(t+1)}, k_{d_2}^{(t+1)} \big| k_m^{(t)}, h^{(t)}\big) \nonumber\\[5pt]
    P_{n}\big(\text{end} \big| k_1^{(n)}, \ldots\big) &= P_\text{end}\big(1\big|h^{(n)}\big)
\end{align}
where $m^{(t)}$ is the mother's discrete index in the $t^\text{th}$ intermediate state. 
We thus have a sequential model that at each $t$ step predicts
\begin{itemize}
\item $P_\text{end}\big(0\big|h^{(t)}\big)$: probability over binary values for whether or not the tree ends;
\item $P_\text{mother}\big(m^{(t)}\big|h^{(t)}\big)$: probability over $m\in\{1,\ldots, t\}$ indexing candidate mother momenta;
\item $P_\text{branch}\big(k_{d_1}^{(t+1)}, k_{d_2}^{(t+1)} \big| k_m^{(t)}, h^{(t)}\big)$: probability over possible $k_m \to k_{d_1}, k_{d_2}$ branchings.
\end{itemize}
Note that we have left the conditioning on $\text{end}=0$ implicit in $P_\text{mother}$ and $P_\text{branch}$, since we will never need to use these functions when $\text{end}=1$. 
In the product of \Eq{IndividTimeStepModel}, each subsequent factor is thus conditioned on the outcomes of previous factors, 
so that breaking up $P_\text{jet}$ in this way is without loss of generality.
In particular, no assumption has been made about the underlying physical processes that generate the data.

With these choices, we force the hidden representation $h^{(t)}$ to encode all global information about the tree, 
since it must predict whether the tree ends, which momentum branches next, and the branching pattern. 
In fact, providing $P_\text{branch}$ with the momenta that directly participate in the $1\to2$ branching means that $h^{(t)}$ only needs to encode global information. 
We show that the global structure stored in $h^{(t)}$ is crucial for the model to predict the correct branching patterns in \Sec{DiscussionGlobal}.

\subsection{Neural Network Implementation}
\label{sec:NNModel}

\begin{figure}[t]
\centering
\includegraphics[width=0.9\linewidth]{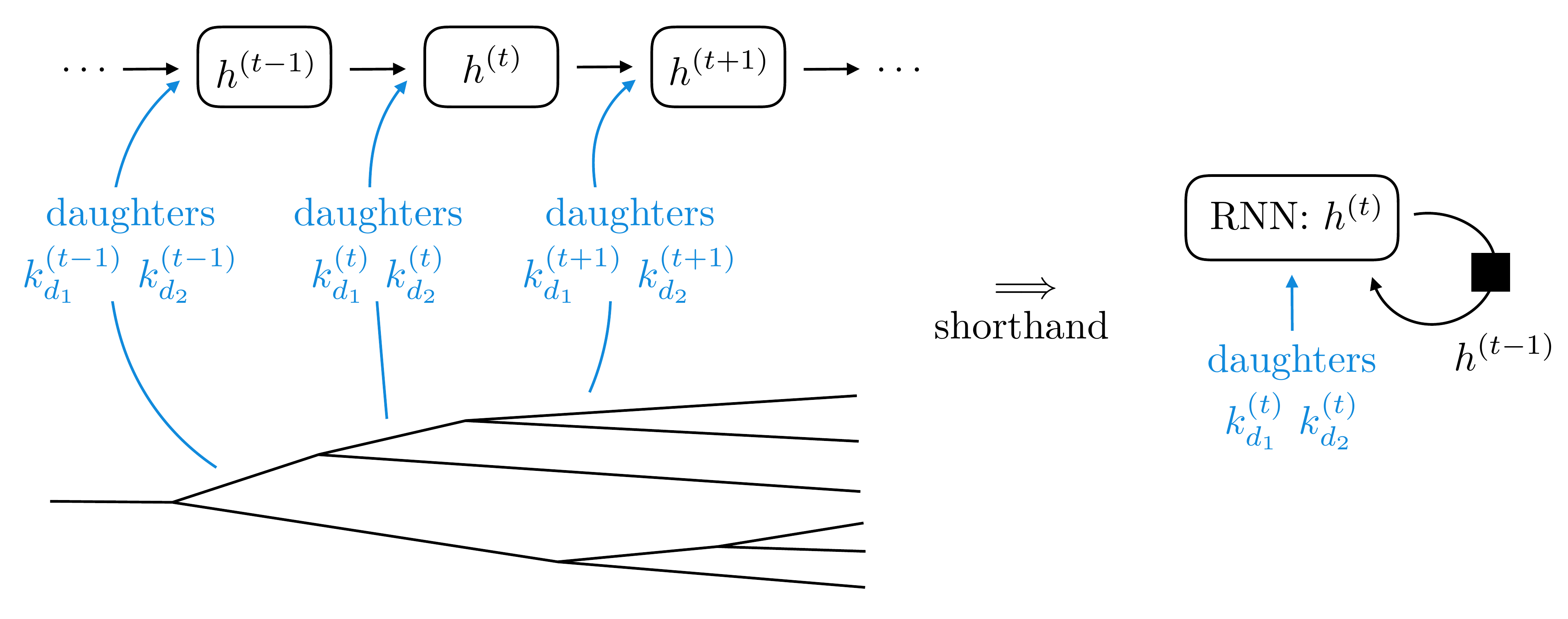}
\caption{Information about the clustering tree is embedded in the hidden state $h^{(t)}$ of the RNN. For brevity, this recurrent structure is simplified on the right using a shaded box to indicate stepping from $t-1$ to $t$. At each step, the next two daughter momenta emerging in the tree and the previous hidden state $h^{(t-1)}$ are inputs to the updated hidden state $h^{(t)}$.}
\label{fig:RNN}
\end{figure}

For a neural network based implementation of the model defined by \Eqs{BasicProductAssumption}{IndividTimeStepModel}, 
we use an RNN with hidden state $h^{(t)}$ augmented by dense neural networks for each of the three probability distributions in \Eq{IndividTimeStepModel}. 
The recurrent structure of this implementation is shown in \Fig{RNN}, 
which emphasizes how the RNN's hidden representation $h^{(t)}$ keeps track of the global state of the jet, 
by sequentially reading in the momenta that branched most recently. 

\begin{figure}[t]
\centering
\includegraphics[width=0.75\linewidth]{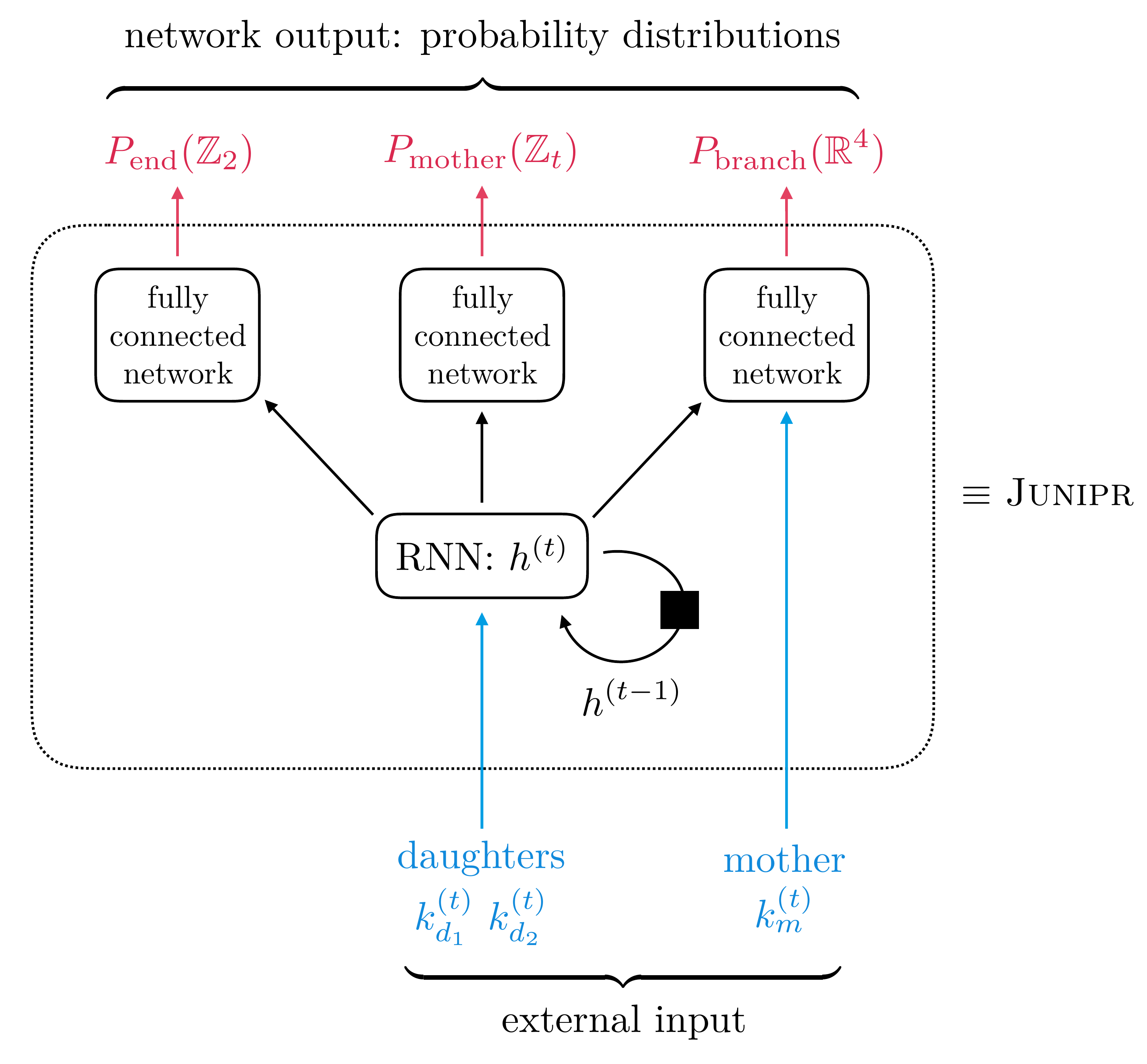}
\caption{Neural network implementation of the general probabilistic model proposed in \Eqs{BasicProductAssumption}{IndividTimeStepModel}. The network takes as external inputs two daughter momenta and one mother momentum. The global RNN then passes only its representation vector $h^{(t)}$ to each of the dense networks shown. The networks output three full probability distributions, which predict the end of the tree, the next mother to branch, and its daughter momenta.}
\label{fig:network_diagram}
\end{figure}

The fact that $h^{(t)}$ learns and remembers the full jet, despite only being shown the two new momenta at step $t$, 
is ensured by the tasks for which $h^{(t)}$ is responsible. 
These are shown in the detailed network diagram of \Fig{network_diagram}.
There one can see that $h^{(t)}$ is the only input into the components of the model that predict when the tree ends and which momentum is next to branch.
The domains of the three probability functions in \Eq{IndividTimeStepModel} are shown in \Fig{network_diagram} as well: 
$P_\text{end}$ is defined over the binary set $\mathbb{Z}_2$ corresponding to ``end'' or ``not'';
$P_\text{mother}$ is multinomial over the set $\mathbb{Z}_t$ of candidate mothers; and
$P_\text{branch}$ is defined on the space of possible $1 \to 2$ branchings, which is (a subset of) $\mathbb{R}^4$ by momentum conservation.
At each step, the model outputs the full probability distributions, which in mathematical notation are $P_\text{end}(\mathbb Z_2|h^{(t)})$, $P_\text{mother}(\mathbb Z_t|h^{(t)})$, and $P_\text{branch}(\mathbb R^4|k_m^{(t)}, h^{(t)})$.

\begin{figure}[t]
\centering
\includegraphics[width=0.8\linewidth]{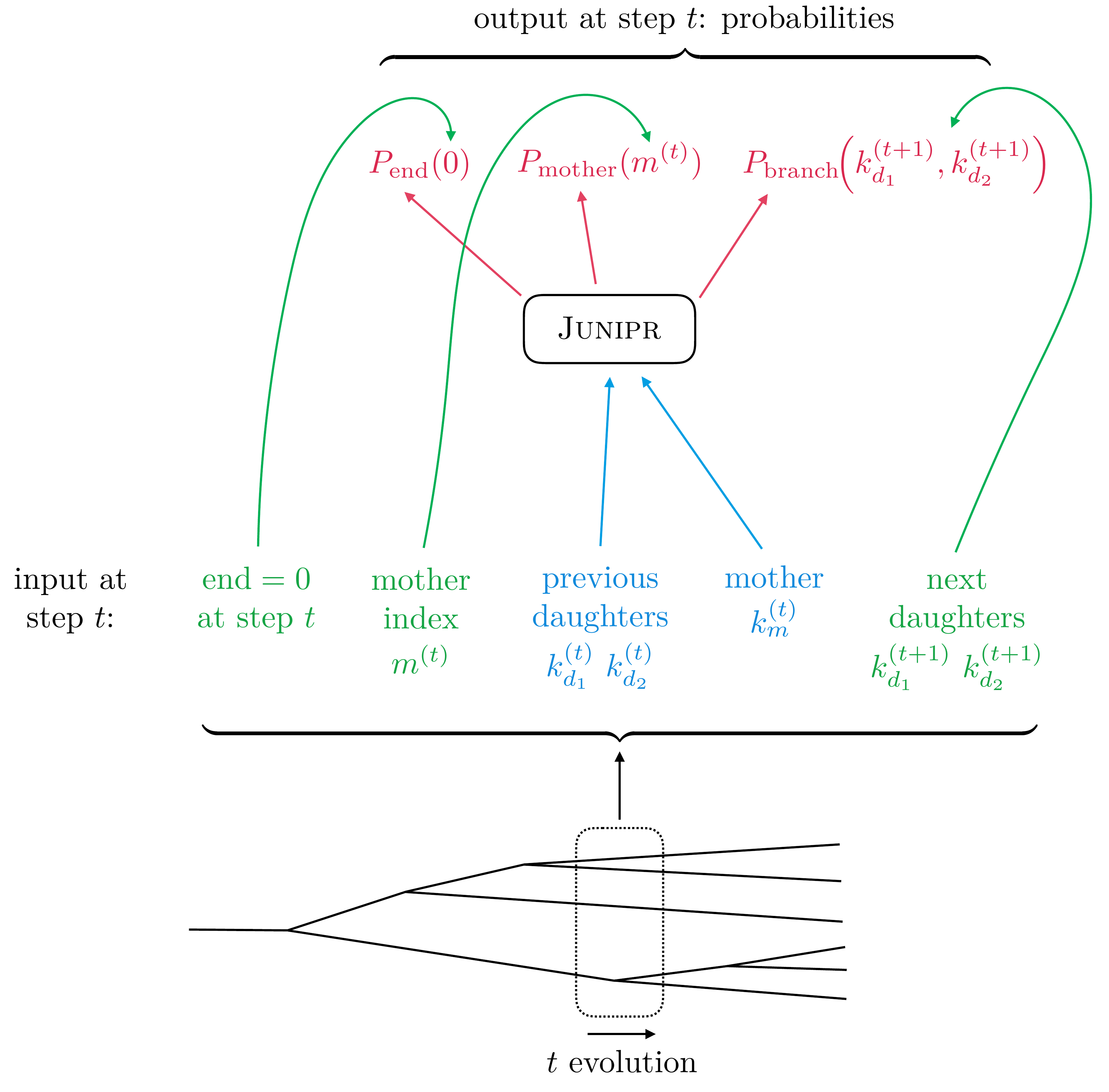}
\caption{Using \JUNIPR to evaluate the probability density over final-state momenta in a jet. 
For a given jet and its particular clustering tree, the values associated with the tree ending, which momenta branch, and the emerging daughters are all known 
and plugged into the probability distributions directly. The probability density of the jet is then the product over the three distributions, over all splitting steps $t$.}
\label{fig:Pjet_evaluation}
\end{figure}

\Fig{RNN} and \Fig{network_diagram} show how \JUNIPR provides a probability distribution at each step $t$ given the momenta emerging from the preceding branching. For clarity, \Fig{Pjet_evaluation} separately shows how \JUNIPR is used to evaluate the full probability density $P_\text{jet}(\{p_1,\ldots, p_n\})$ over final-state momenta in a jet. At each step $t$, the point in $\mathbb Z_2$ representing whether the tree ends, the point in $\mathbb Z_t$ representing which mother momentum branches, and the point in $\mathbb R^4$ representing its daughters are plugged into the probability distributions to obtain the probabilities that should be assigned to the jet under consideration. The product of these three probabilities, taken over all $t$ steps, leads to $P_\text{jet}(\{p_1,\ldots, p_n\})$.

Let us now go into detail about the neural network architecture used.
We use basic RNN cells \cite{DeepLearningBook} with $\tanh$ activation,
\begin{equation}
\label{eq:ht_recur}
    h^{(t)} = \tanh \big( W \cdot (k_{d_1}^{(t)}, k_{d_2}^{(t)}) + V \cdot h^{(t-1)} + b\big),
\end{equation}
and found that a hidden representation vector $h^{(t)}$ of generic size 100 was sufficient for our needs.
We found GRU \cite{GRUs} and LSTM \cite{LSTMs} cells to be unnecessarily complex and high-capacity for the tasks carried out in this paper. 
This is in contrast to language modelling, for which basic RNN cells are underpowered.
To see why this might heuristically be expected, note that a sentence containing 20 words is much more complex than a jet containing 20 momenta,
because the words in the sentence are ordered, whereas the momenta in the jet are not.
This introduces an additional factor of $20! \sim 10^{18}$ to the complexity of language modelling.
It is thus reasonable to expect that jet physics will not require all the high-powered tools designed for natural language processing.

For $P_\text{end}$ we use a fully-connected network with $h^{(t)}$ as input, a single hidden layer of size 100 with ReLU activation, and a sigmoid output layer. We use the same setup for $P_\text{mother}$, the only difference being that the output layer is a softmax over the $t$ candidate mother momenta, ordered by energy. These choices are generic and not highly tuned. We found that \JUNIPR works well for a very general set of architectures and sizes, so we stick with this simple setup.

\begin{figure}
\centering
\includegraphics[scale=0.35]{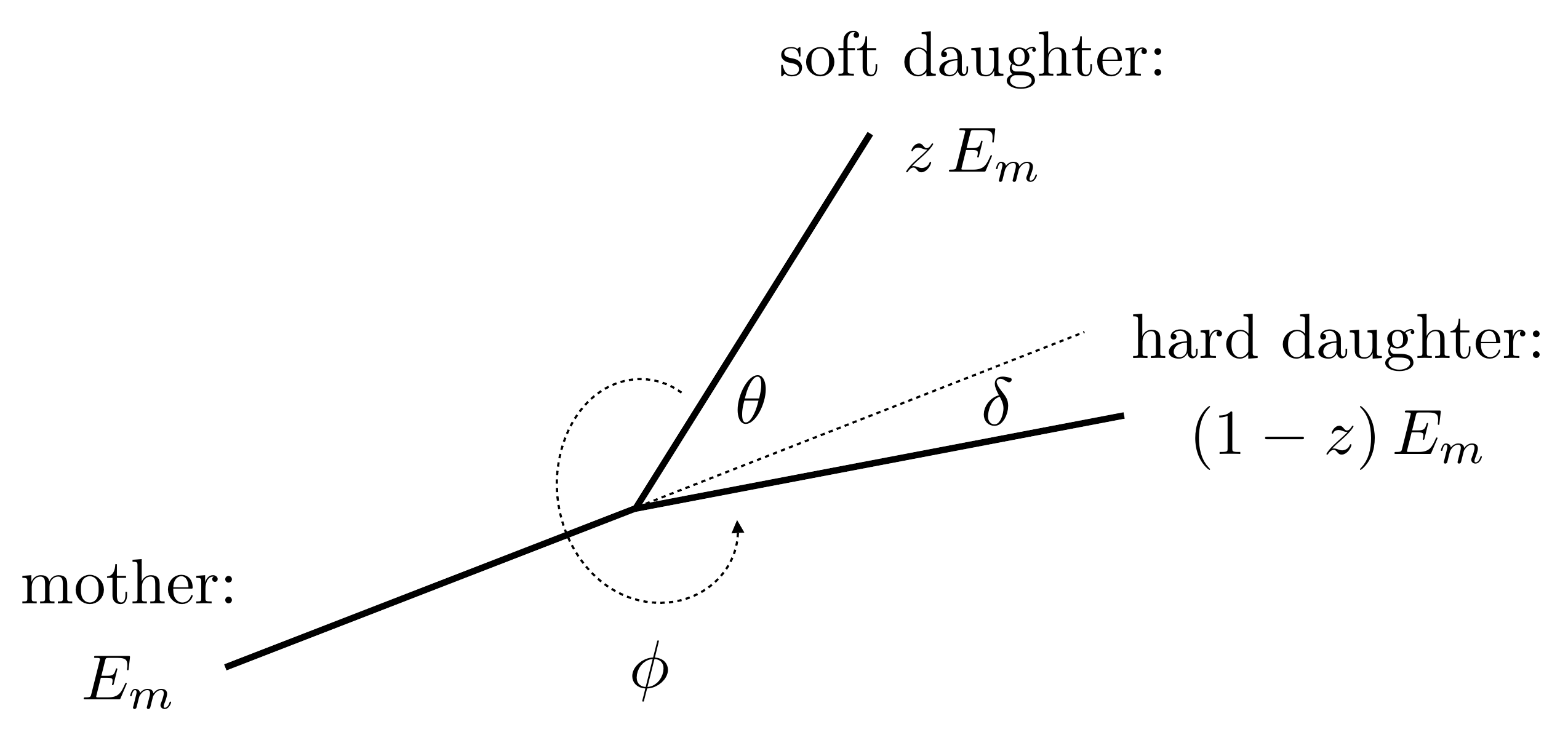}
\caption{Local coordinates $x=(z,\theta,\phi,\delta)$ that parameterize the momentum-conserving $1\to2$ branching at each step in the clustering tree of a jet.}
\label{fig:Coordinates}
\end{figure}

For the branching function $P_\text{branch}$ we must describe the probability distribution over all possible configurations of daughter momenta $k_{d_1}^{(t+1)}, k_{d_2}^{(t+1)}$ consistent with the mother momentum $k_m^{(t)}$. For this system, we use coordinates $x=(z,\theta,\phi,\delta)$ centered around the mother, where $z$ is the energy fraction of the softer daughter, $\theta$ ($\delta$) is the opening angle of the softer (harder) daughter, and $\phi$ specifies the plane in which the branching occurs. See \Fig{Coordinates} for a visualization of these coordinates.  

There are two separate approaches one could take to model the branching function $P_\text{branch}$. Firstly, the variables $x$ could be treated as discrete, with $P_\text{branch}$ outputting a softmax probability over discrete cells representing different $x$ values. Secondly, one could treat $x$ as a continuous variable and use an ``energy model'' of the form
$
    P_\text{branch} \sim {e^{E(x)} / Z}\,,
$
where $Z$ is a normalizing partition function. In this work we predominantly adopt the former approach, as it is much faster, and most distributions are insensitive to the discretization of $x$. However, we do train an energy model to show that models with continuous $x$ are possible, which we discuss in \Sec{TrainingResolution}.

In the discrete case, we bin the possible values of $x$ into a 4-dimensional grid with 10 bins per dimension, so that the entire grid has $10^4$ cells. For a given value of $x$, we place a 1 in the bin corresponding to that value, and we place 0's everywhere else. This 1-hot encoding of the possible values of $x$ allows us to use a softmax function at the top layer of the neural network describing $P_\text{branch}$ (see \Fig{network_diagram}). Furthermore, we use a dense network with a single hidden layer of size 100 and ReLU activation for $P_\text{branch}$, just as we did for $P_\text{end}$ and $P_\text{mother}$. The hidden units in this network receive $h^{(t)}$ as input, as well as the mother momentum $k_m^{(t)}$.

Thus we have a neural network implementation of \Eqs{BasicProductAssumption}{IndividTimeStepModel}, with a representation of the evolving global jet state stored in $h^{(t)}$, and with fully-connected networks describing $P_\text{end}$, $P_\text{mother}$, and $P_\text{branch}$. 
As defined above, the model has a single $10^6$ parameter matrix, mapping the branching function's 100 dimensional hidden layer to its $10^4$ dimensional output layer, and has $6\times 10^4$ parameters elsewhere.
One might refer to this implementation as \JUNIPRzero, as one can imagine many alternative implementations within the \JUNIPR framework that may prove useful in future applications.
We will continue to use the term \JUNIPR for brevity, to refer both to the framework and to the basic implementation described here.


\section{Training and Validation}
\label{sec:Training}

We now describe how to train the model outlined in \Sec{NNModel}. We begin by discussing the training data used, followed by our general approach to training and validation.
Finally we discuss an alternative model choice that allows higher resolution on the particle momenta.

\subsection{Training Data}
\label{sec:TrainingData}

To enable proof-of-concept demonstrations of \JUNIPR's various applications, we train the implementation described in \Sec{NNModel} using jets simulated in \Pythia v8.226 \cite{Sjostrand:2006za,Sjostrand:2014zea} and clustered using \FastJet v3.2.2 \cite{Cacciari:2011ma}. 
We simulated 600k hemisphere jets in \Pythia using the process $e^+e^- \to q\bar q$ at a center-of-mass energy of 1 TeV, with hemispheres defined in \FastJet using the exclusive $k_t$ algorithm \cite{Catani:1993hr,Ellis:1993tq}, and with an energy window of 450--550 GeV imposed on the jets. To create the deterministic trees that \JUNIPR requires, we reclustered the jets using the C/A clustering algorithm \cite{Dokshitzer:1997in,Wobisch:1998wt}, with $E_\text{sub}=1$ GeV and $R_\text{sub}=0.1$. The nonzero values of $E_\text{sub}$ and $R_\text{sub}$ make the input to \JUNIPR formally infrared-and-collinear safe, but this is by no means necessary. Furthermore, our approach is formally independent of the reclustering algorithm chosen. We demonstrate this by showing results using an absurd reclustering algorithm inspired by a 2D printer in \Sec{DiscussionPrinterJets}, as well as for anti-$k_t$ \cite{Cacciari:2008gp} reclustering in \Sec{DiscussionAntikt}.

Thus we have 600k quark jets with $E_\text{jet}\sim 500$ GeV and $R_\text{jet}\sim\pi/2$. We use 500k of these jets for training, with 10k set aside as a test set to monitor overfitting, and we use the remaining validation set of 100k jets to make the plots in this paper. 

In the applications of \Sec{Results}, we also make use of several other data sets produced according to the above specifications, with small but important changes. We list these modifications here for completeness. In one case, quark jets from $e^+e^- \to q\bar q$ were required to lie in a very tight mass window of 90.7--91.7 GeV. A sample of boosted $Z$ jets from $e^+e^- \to ZZ$ events was also produced with the same mass cut. And finally, another sample of quark jets was produced, as detailed above, but with the value of $\alpha_s(m_Z)$ in the final state shower changed from \Pythia's default value of 0.1365 to 0.11.

Before being fed to \JUNIPR, jets in these data sets must be clustered, so that each jet becomes a tree of $1\to2$ branchings ending in the $n$ final-state momenta of the jet:\vspace{-2mm}
\eqn{
\adjustbox{valign=c}{\includegraphics[width=0.9\linewidth]{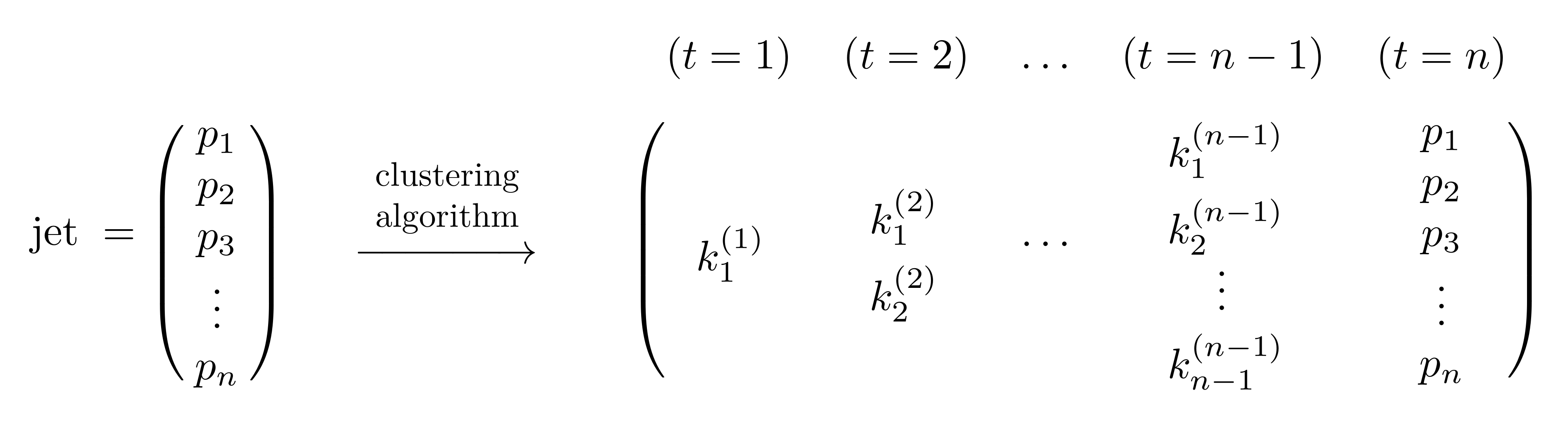}}
}
where the momenta in one column are equal to those of the next column except for a single $1\to2$ branching. 
At each step $t$, only the momenta associated with this $1\to2$ branching are fed into \JUNIPR, as detailed in \Sec{Model}.
With this setup, \JUNIPR requires minimal parameters; it learns to update $h^{(t)}$ as the tree evolves by focusing only on the step-by-step changes to the jet. Note also that jets of arbitrary length can be considered.

Note that in implementing \JUNIPR, we do not directly evaluate the branching function $P_\text{branch}\big(k_{d_1}^{(t+1)},k_{d_2}^{(t+1)}\big|k_m^{(t)},h^{(t)}\big)$ on the momenta $k_{d_1}^{(t+1)}, k_{d_2}^{(t+1)}$ but instead use the parameterization $x=(z, \theta, \phi, \delta)$ shown in \Fig{Coordinates}. In fact, we use a nonlinear transformation of this parameterization:
\eqna{
\tilde z &= \frac{\log z - \log \frac{E_\text{sub}}{E_\text{jet}}}{\log\frac12 - \log \frac{E_\text{sub}}{E_\text{jet}}}
\hspace{2cm}
&\tilde \theta &= \frac{\log \theta - \log \frac{R_\text{sub}}{2}}{\log R_\text{jet} - \log \frac{R_\text{sub}}{2}}
\vspace{4mm}\\[5pt]
\tilde \phi &= \frac{\phi}{2\pi}
\hspace{2cm}
&\tilde \delta &= \frac{\log \delta - \log \frac{E_\text{sub}R_\text{sub}}{E_\text{jet}}}{\log \frac{R_\text{jet}}2 - \log \frac{E_\text{sub}R_\text{sub}}{E_\text{jet}}}
\label{eq:TransformedCoordinates}
}
This invertible transformation simply maps the range of each coordinate onto $[0,1]$, which reduces the amount of global parametric shift required in optimization. Similarly, we perform a transformation on the components of $k_{d_1}^{(t)}, k_{d_2}^{(t)}$ before feeding them into the update rule for $h^{(t)}$ in \Eq{ht_recur}; we do the same for $k_m^{(t)}$, the input to the branching function $P_\text{branch}$.
This is a technical point that is not conceptually important.

\subsection{Approach to Training}
\label{sec:TrainingApproach}

To train \JUNIPR, we maximize the log likelihood over the full set of training data:
\eqn{
\label{eq:LogLike}
\text{log likelihood} ~= \sum_{\text{jet $i$ in data}} \log P_\text{jet}(\{p^{(i)}_1,\ldots,p^{(i)}_n\})\,.
}
For a particular jet with final-state momenta $p_1,\ldots,p_n$ we use \Eqs{BasicProductAssumption}{IndividTimeStepModel} to compute
\eqn{
\label{eq:LossFunction}
\log P_\text{jet}(\{p_1,\ldots,p_n\}) = \sum_{t=1}^{n-1} \bigg[& \log P_\text{end} \big(0 \big| h^{(t)} \big) \\
+& \log P_\text{mother} \big(m^{(t)} \big| h^{(t)} \big) \nonumber \\
+& \log P_\text{branch} \big(k^{(t+1)}_{d_1}, k^{(t+1)}_{d_2} \big| k_m^{(t)}, h^{(t)} \big) \bigg] + \log P_\text{end} \big(1 \big| h^{(n)} \big)
\nonumber
}
where $m^{(t)}$ is the index of the mother momentum at step $t$ in the training example and $k^{(t+1)}_{d_1}, k^{(t+1)}_{d_2}$ are its daughters. 
Maximizing the log likelihood in this way allows the model to learn each $t$ step in parallel, providing computational efficiency and stability. 

For all models presented in this paper, we use basic stochastic gradient descent with the following learning rate and batch size schedule, where training proceeds from left to right: 
\begin{center}
 \begin{tabular}{c || c c c c c c} 
 Schedule & 5 epochs & 5 epochs & 5 epochs & 5 epochs & 5 epochs & 5 epochs \\ 
 \hline
 learning rate & $10^{-2}$ & $10^{-3}$ & $10^{-4}$ & $10^{-3}$ & $10^{-4}$ & $10^{-5}$ \\ 
 batch size & 10 & 10 & 10 & 100 & 100 & 100
\end{tabular}
\end{center}
We follow such a schedule to slowly increase the resolution and decrease the stochasticity of gradient descent throughout training. Decreasing the learning rate reduces the step size, thereby allowing finer details of the cost surface to be resolved. Increasing the batch size reduces the stochasticity by improving the sample estimates of the true gradients. 

We wrote \JUNIPR in Theano \cite{Theano} and trained it on 16-core CPU servers using the SherlockML technical data science platform. Training \JUNIPR on 500k jets according to the above schedule took an average of 4 days.

\subsection{Validation of Model Components}
\label{sec:ResultsBasic}

\JUNIPR is constructed as a probabilistic model for jet physics by expanding $P_\text{jet}$ as a product over steps $t$ in the jet's clustering tree, as shown in \Eq{BasicProductAssumption}. Each step involves three components: the probability $P_\text{end}$ that the tree will end, the probability $P_\text{mother}$ that a given momentum will be the next mother to branch, and the probability $P_\text{branch}$ over the daughter momenta of the branching, as shown in \Eq{IndividTimeStepModel}. We now validate each of \JUNIPR's components using our validation set of 100k previously unseen \Pythia jets. In this section, we present histograms of actual outcomes in the \Pythia validation set (i.e.~frequency distributions) as well as \JUNIPR's probabilistic output when evaluated on the jets in this data set (i.e.~marginalized probability distributions) to check for agreement.

\begin{figure}[t]
\centering
\includegraphics[width=0.5\linewidth]{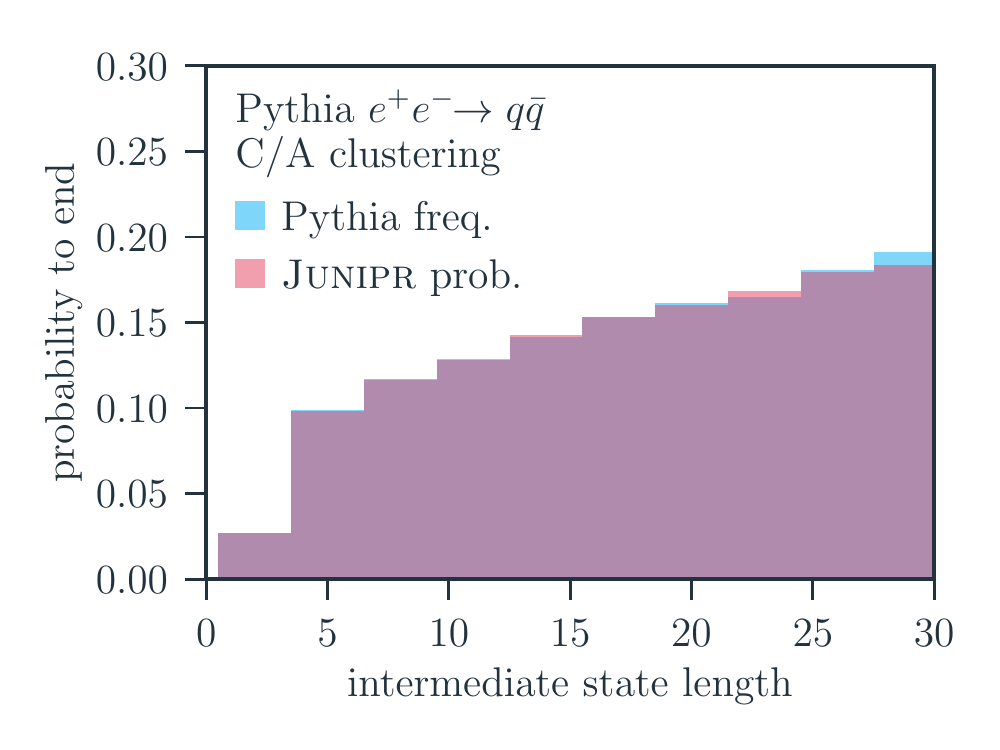}~
\includegraphics[width=0.5\linewidth]{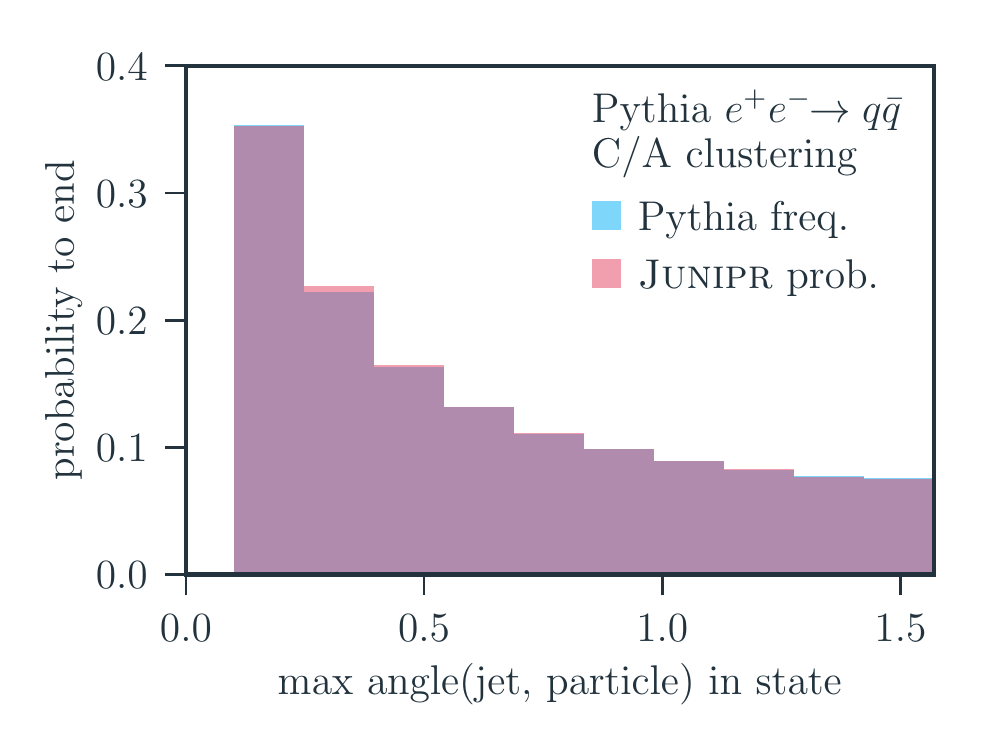}
\caption{Validation of $P_\text{end}$, the probability that the tree should end. 
Comparison is made between actual outcomes in the validation set of \Pythia jets and \JUNIPR's probabilistic predictions for these jets. (Left) $P_\text{end}$ as a function of intermediate state length. (Right) $P_\text{end}$ as a function of the maximum angle between the jet axis and momenta in the intermediate state.}
\label{fig:validate_end_shower}
\end{figure}

In \Fig{validate_end_shower} we show the probability $P_\text{end}$ that the tree should end, as a function of both intermediate state length and maximum particle-to-jet-axis angle. In both cases we see excellent agreement with the validation data, demonstrating a good model fit with low underfitting and no overfitting. Note that \Fig{validate_end_shower} (left) is in one-to-one correspondence with the jet constituent multiplicity, and that the shape of \Fig{validate_end_shower} (right) is a direct consequence of C/A clustering with $R_\text{sub}=0.1$\,. Indeed, if an opening angle near $R_\text{sub}$ already exists in an angular-ordered tree, then there are likely no remaining branchings in the clustering tree.

\begin{figure}[t]
\centering
\includegraphics[width=0.5\linewidth]{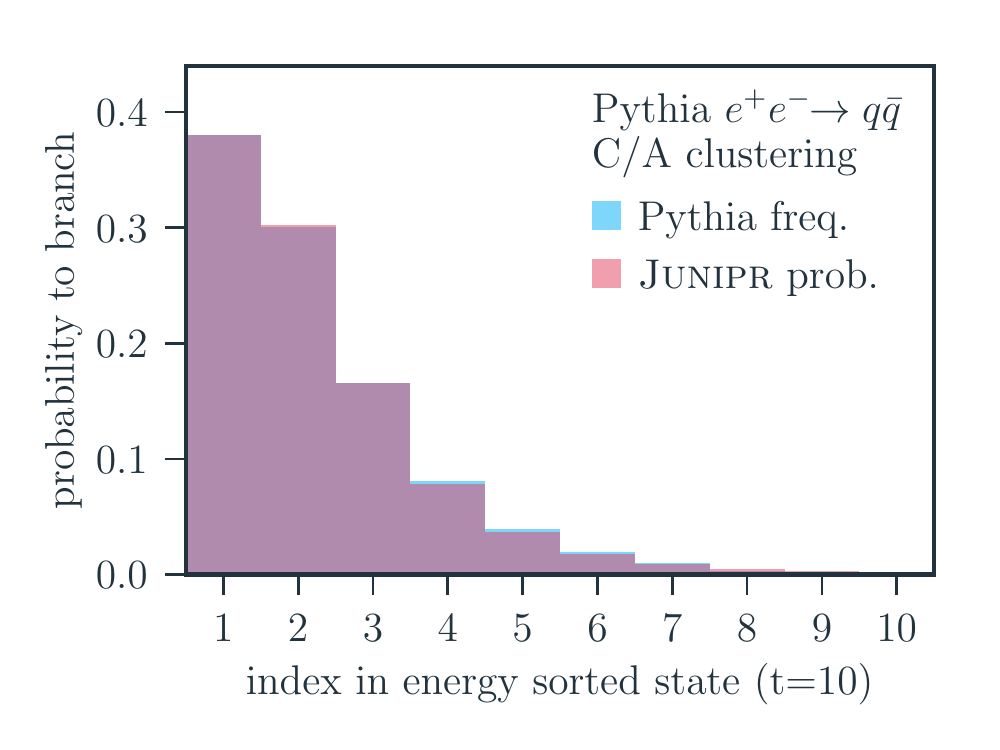}~
\includegraphics[width=0.5\linewidth]{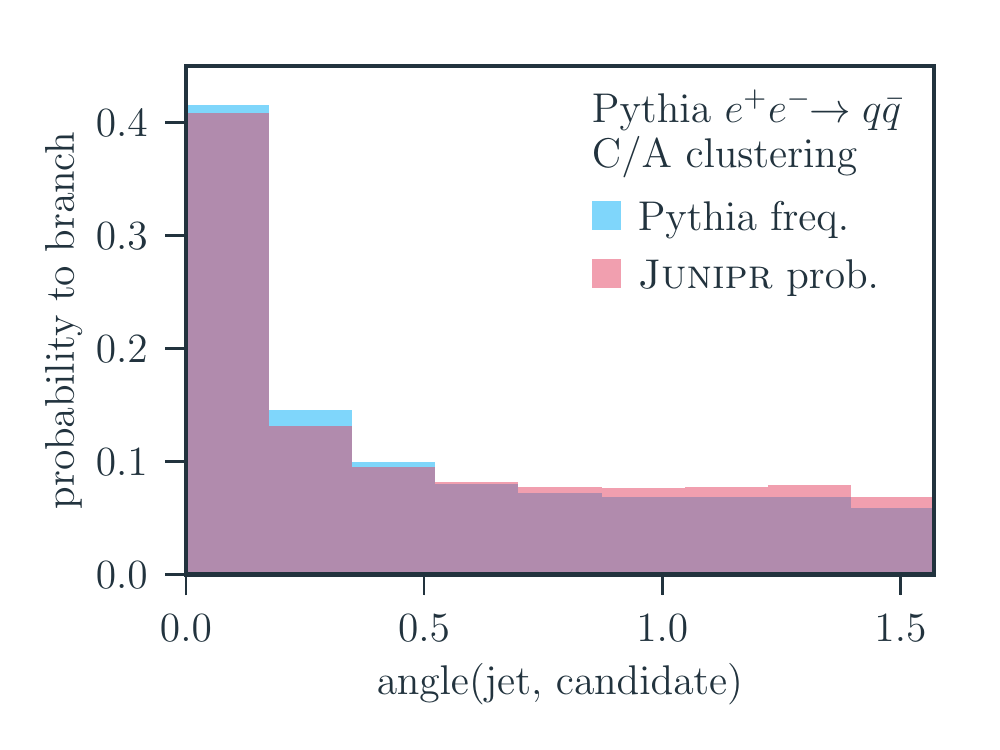}\\
\caption{Validation of $P_\text{mother}$, the probability that a given candidate will branch next in the clustering tree. Comparison is made between actual outcomes in the validation set of \Pythia jets and \JUNIPR's probabilistic predictions for these jets. (Left) $P_\text{mother}$ at $t=10$, as a function of a candidate's index in the energy ordered intermediate state. (Right) $P_\text{mother}$ averaged over all $t$'s, as a function of a candidate's angle relative to the jet axis.}
\label{fig:validate_choose_parent}
\end{figure}

In \Fig{validate_choose_parent} we show the probability $P_\text{mother}$ that a given candidate will be the next mother to branch in the clustering tree, as a function of both the candidate's index (which is sorted to be decreasing in energy) and the candidate's angle from the jet axis. The first of these results is shown in particular for the $t=10^{\text{th}}$ step in the clustering trees. We observe again that the model fits the validation data well. Note from \Fig{validate_choose_parent} (left) that the highest energy branches of the clustering tree are most likely to undergo subsequent branchings, in line with the expectation at leading logarithmic accuracy. \Fig{validate_choose_parent} (right) shows consistent predictions, since the highest energy branches also lie at the narrowest angles to the jet axis.

\begin{figure}[t]
\centering
\includegraphics[width=0.5\linewidth]{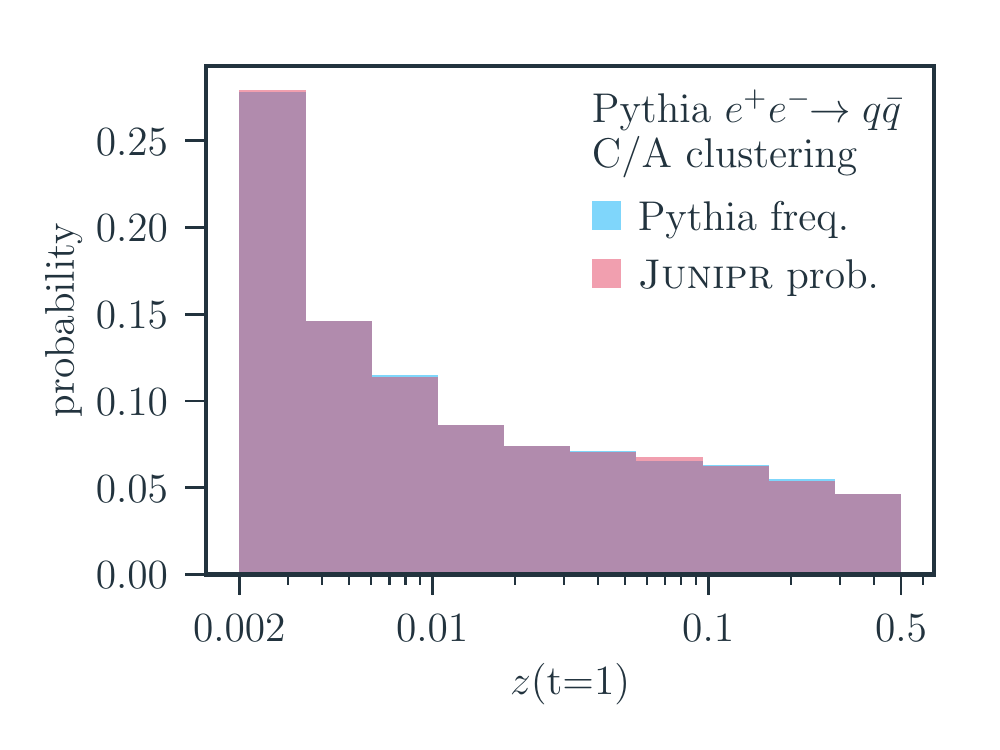}~
\includegraphics[width=0.5\linewidth]{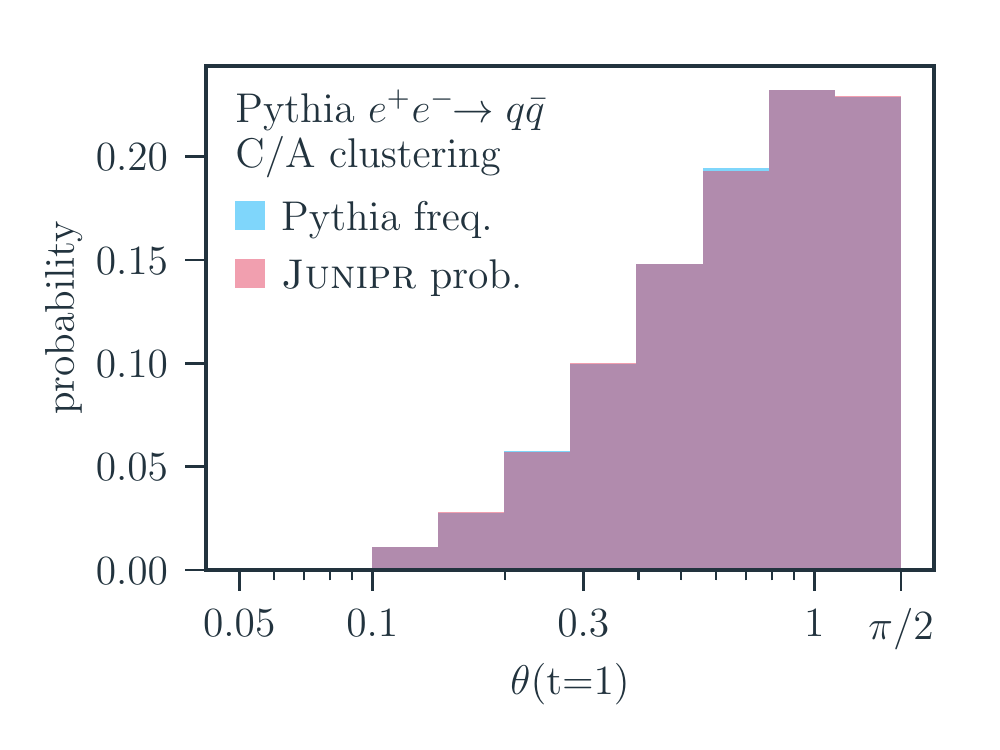}\\
\includegraphics[width=0.5\linewidth]{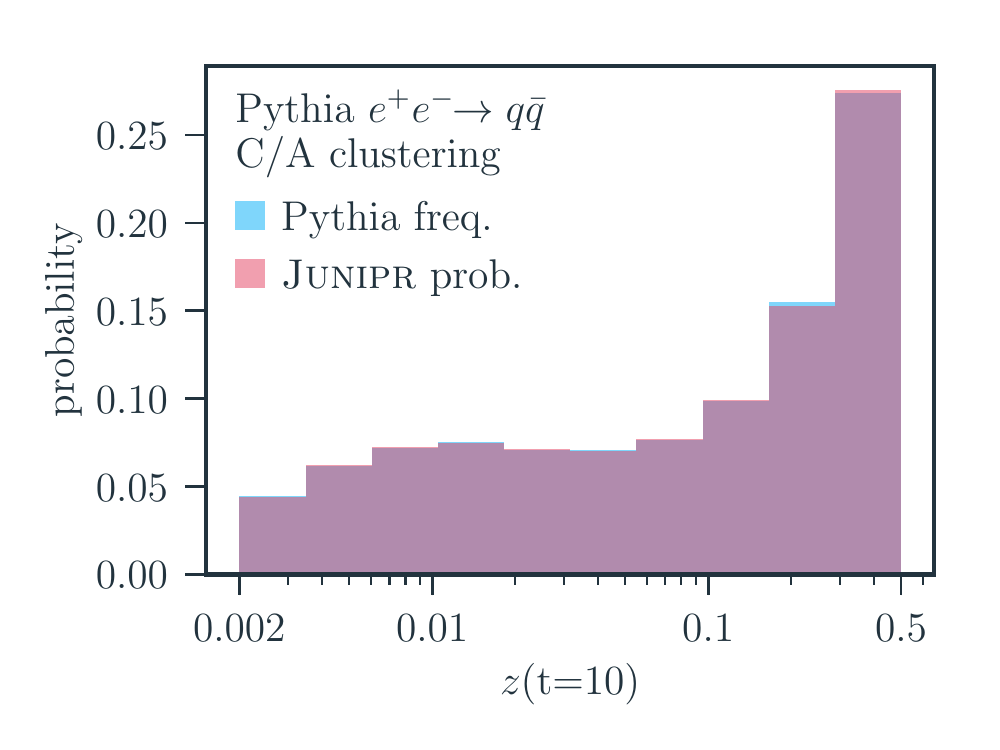}~
\includegraphics[width=0.5\linewidth]{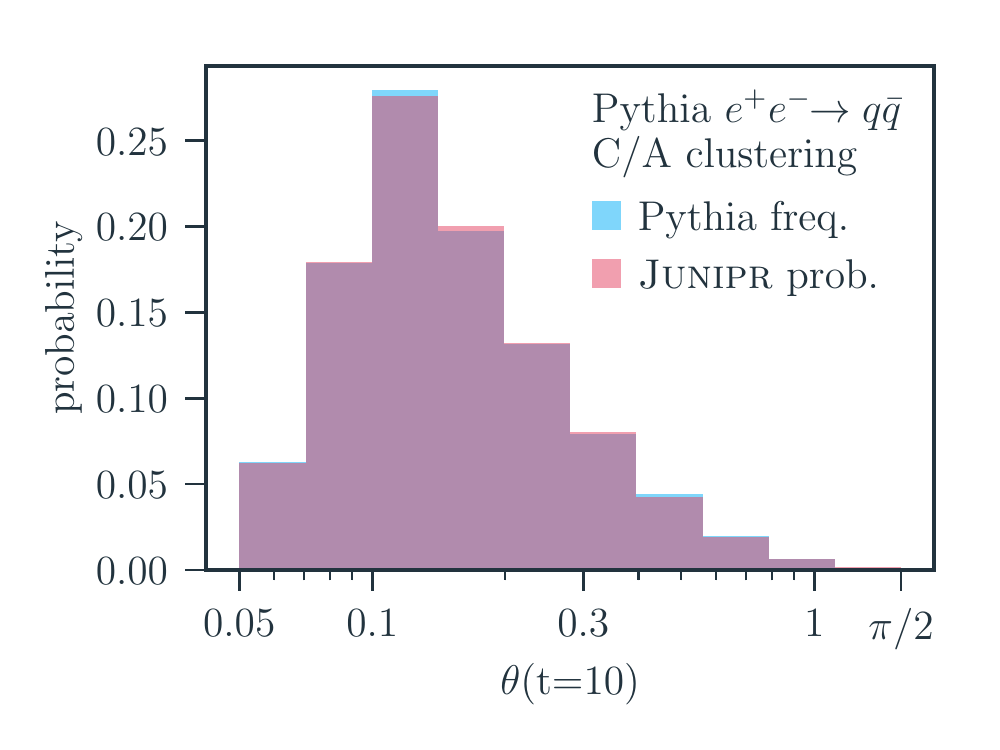}
\caption{Validation of $P_\text{branch}$, the 4-dimensional probability distribution over $1 \to 2$ branchings. Comparison is made between actual outcomes in the validation set of \Pythia jets and \JUNIPR's probabilistic predictions for these jets. Results are shown for energy fraction $z$ (left) and branching angle $\theta$ (right) as defined in \Fig{Coordinates}. Evolution step $t=1$ is shown (top) where soft wide-angle emissions are the norm, as expected in the C/A tree. Evolution step $t=10$ (bottom) gives rise to harder more-collinear branchings.}
\label{fig:validate_splitting_fns}
\end{figure}

In \Fig{validate_splitting_fns} we show the branching function $P_\text{branch}$, the component of the model that predicts how a mother momentum should split into a pair of daughter momenta. We show the branching function results for $z$ and $\theta$ (i.e.~with $P_\text{branch}$ marginalized over the variables not shown) at the first step in the jet evolution $t=1$, as well as at a later step $t=10$. (See \Fig{Coordinates} for definitions of $z$ and $\theta$ and \Eq{TransformedCoordinates} for their ranges in the data.) This shows the dependency of the branching function on the evolving jet representation $h^{(t)}$, which we will discuss in detail in \Sec{DiscussionGlobal}. We see that for these direct predictions, \JUNIPR fits the validation data almost perfectly. Note that in \Fig{validate_splitting_fns} (top) soft wide-angle emissions are the norm at the earliest $t$ steps, as expected with the C/A clustering algorithm. In \Fig{validate_splitting_fns} (bottom) one can see that later in the clustering trees, harder more-collinear branchings are commonplace. It bears repeating that these trends are highly dependent on the chosen clustering algorithm and have no precise connection to the underlying physical processes generating the data.

\subsection{Increasing the Branching Function Resolution}
\label{sec:TrainingResolution}

In this section, we discuss increasing the resolution of the branching function 
\begin{equation}
    P(x) \equiv P_\text{branch}\big(k_{d_1}^{(t+1)}, k_{d_2}^{(t+1)} \big| k_m^{(t)}, h^{(t)}\big)
\end{equation}
including the case where $P(x)$ is an energy model over continuous $x=(z,\theta,\phi,\delta)$. (The $x$ coordinates were defined in \Fig{Coordinates}.) This technical section can easily be skipped without loss of the logical flow of the paper.

We begin by briefly discussing increasing the resolution of the branching function over discrete $x$, the case described in \Sec{NNModel}. The first thing to note is that with a softmax over 4-dimensional $x$, the size of the matrix multiplication required in a dense network is quartic in the number of bins used for each dimension. We generically use 10 bins for each of $z,\theta,\phi,\delta$ resulting in an output size of $10^4$. 
(In fact we use 10 linearly spaced bins in the transformed coordinates of \Eq{TransformedCoordinates}, and this can be seen on the logarithmic axes of \Fig{validate_splitting_fns}, but this detail is not conceptually important.)
Given this quartic scaling, simply increasing the number of discrete $x$ cells quickly becomes prohibitively computationally expensive. Potential solutions to this problem include: (i) using a hierarchical softmax \cite{HierarchicalSoftmax1,HierarchicalSoftmax2}, and (ii) simply interpolating between the discrete bins of the model.

In a hierarchical softmax, a low-resolution probability is predicted first, say with $5^4$ cells, then another $5^4$-celled distribution is predicted inside the chosen low-resolution cell. In principle, this gives $25^4$ resolution at only twice the computational time required for $5^4$ resolution. We briefly implemented the hierarchical softmax, and preliminary tests found it to work efficiently, but perhaps with a decrease in training stability. We chose not to pursue the hierarchical softmax further in this work, primarily because we have not seen the need for resolution much higher than $10^4$ discrete $x$ cells. 

Due to its ease of use, we do employ linear interpolation between the discrete bins in our baseline model with resolution $10^4$. This comes at no extra training cost, and removes most of the effects of discretization on the observable distributions generated by sampling from \JUNIPR; see \Sec{ResultsSampling}.

We now turn to the continuous version of \JUNIPR in which the branching function $P(x)$ is given by an undirected energy model:
\eqn{
\label{eq:ZDef}
P(x) = \frac{e^{E(x)}}{Z}, \quad \text{where} \quad Z = \int dx \; e^{E(x)}.
}
To model $E(x)$, we again use a fully-connected network with hidden layer of size 100, as used everywhere else, except here the output layer is left to be linear. We perform the integral over $Z$ using importance sampling:
\eqn{
\label{eq:ZApprox}
Z = \int dx \; q(x) \; \frac{e^{E(x)}}{q(x)}
= \bigg\langle \frac{e^{E(x)}}{q(x)} \bigg\rangle_q
\approx \frac{1}{|S|} \sum_{x_s \in S} \frac{e^{E(x_s)}}{q(x_s)}
= \widehat Z(S)
}
where $S$ is the set of $x_s$'s sampled from the importance distribution $q$.

Unlike the discrete-$x$ version of \JUNIPR, where training is relatively straightforward, the continuous-$x$ version requires a non-standard technique in training the branching function $P(x)$. This is because, although \Eq{ZApprox} provides an unbiased approximation to $Z$,
\eqn{
\langle \widehat Z \rangle_{S\sim q} = Z,
}
this leads to a biased estimate of the log likelihood, since
\eqn{
\label{eq:bias}
\big\langle \log \widehat Z \,\big\rangle_{S\sim q} \,<\,  \log \big\langle \widehat Z \,\big\rangle_{S\sim q} = \log Z
}
by Jensen's inequality. Thus, every gradient step taken is systematically different from the true gradient, and this bias derails training, especially near convergence when the true gradient becomes small.

To overcome this problem, we start by computing the sample variance on our estimate $\widehat Z(S)$, which is
\eqn{
\sigma(\widehat Z)^2 = \frac{1}{|S|-1} \sum_{x_s \in S} \bigg( \frac{e^{E(x_s)}}{q(x_s)} - \widehat Z(S) \bigg)^2.
}
Then the percent-error $\Delta$ in our biased estimate of the gradient is approximately
\eqn{
\Delta = \frac{1}{\sqrt{|S|}} \, \frac{\sigma(\widehat Z)}{\widehat Z}.
}
This error propagates into the log likelihood, causing the bias in \Eq{bias}.
To mitigate this, we adopt a policy of monitoring $\Delta$ during training, and whenever $\Delta$ increases above some value $\Delta_\text{threshold}$ (a hyperparameter that we set to 2\%) we double the sample size $|S|$ used to compute $\widehat Z(S)$. This slows down training considerably, but it effectively reduces the bias in our gradient estimates. Note that while generic importance sampling typically fails in higher dimensions, our branching function lives in only 4 dimensions, so this approach is robust using any reasonable importance distribution $q$. Indeed, we found that a uniform distribution over the transformed coordinates of \Eq{TransformedCoordinates} is a fine choice for $q$. 

\begin{figure}[t]
\centering
\includegraphics[width=0.5\linewidth]{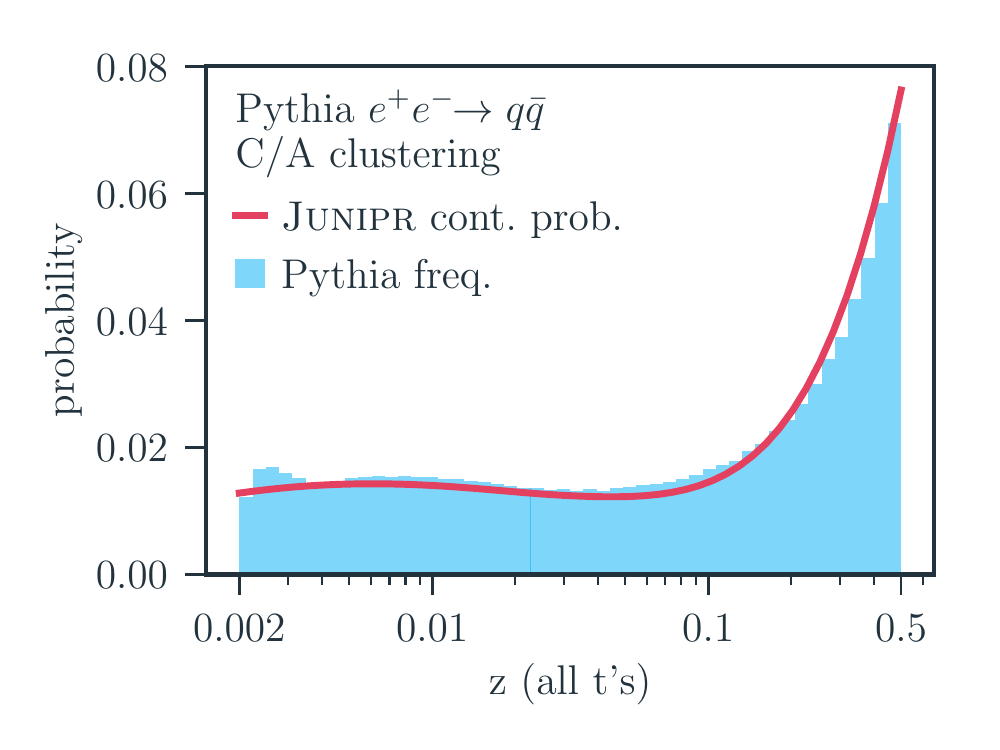}
\caption{Branching function modelled by a deep undirected energy model over continuous variables $z,\theta,\phi,\delta$ that parameterize the branching. Shown is the marginalized distribution over $z$, averaged over all $t$ steps. Comparison is made between actual outcomes in the validation set of \Pythia jets and \JUNIPR's probabilistic predictions for these jets. }
\label{fig:continuous_splitting_fn}
\end{figure}

In \Fig{continuous_splitting_fn} we show results for \JUNIPR trained with the continuous branching function as described above. In this case, we can use arbitrarily high-resolution binning, as \JUNIPR has learned a fully continuous probability density. \Fig{continuous_splitting_fn} can be roughly compared to \Fig{validate_splitting_fns}, where we were required to use 10 bins for each dimension of $x$.

To close this section, we note that in most cases, we expect the discretized branching function with 10 bins per dimension of $x$ to be sufficient, especially if one performs a linear interpolation on the output cells. This simple case is certainly faster to train and does not require the technique described here to avoid biased gradient estimates.


\section{Applications and Results}
\label{sec:Results}

\begin{figure}[t]
\centering
\begin{tikzpicture}
\node at (0,0) {\includegraphics[width=0.74\linewidth]{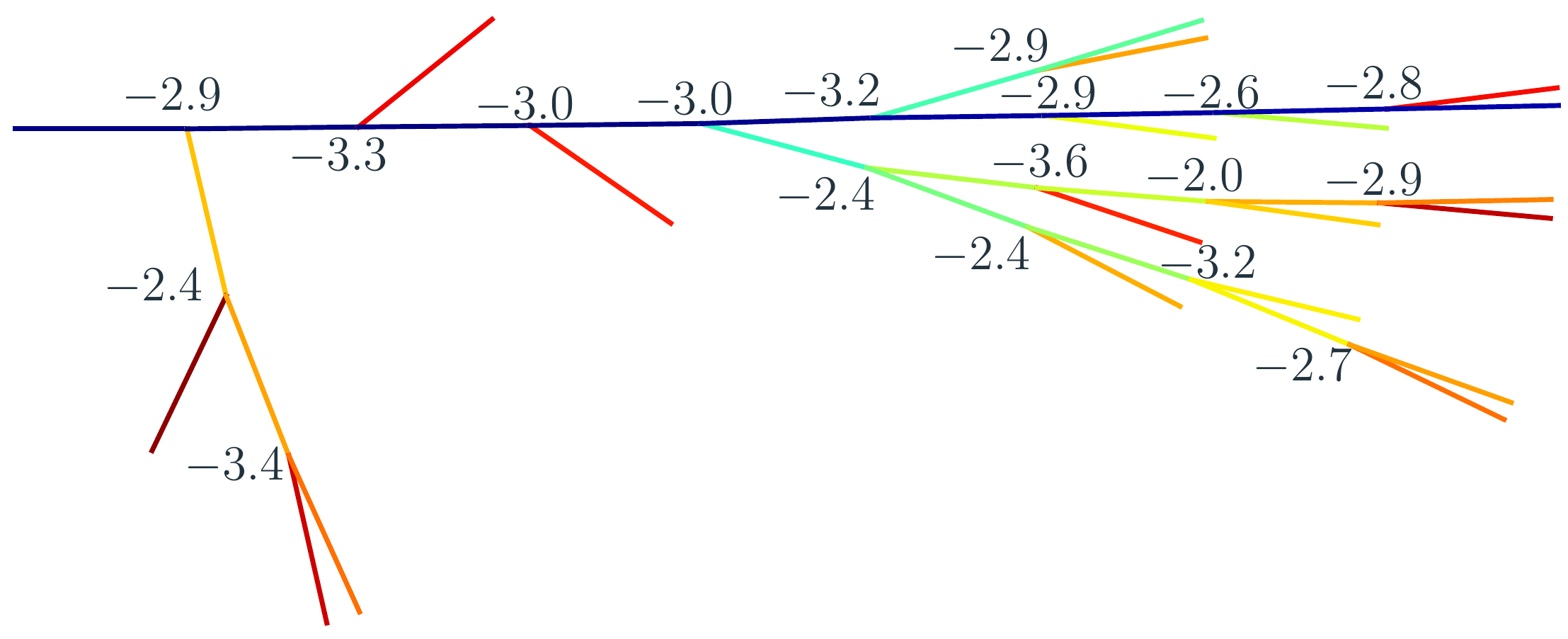}};
\node[above] at (1.6,3.4) {$P_{\text{end}}\cdot P_{\text{mother}}\cdot P_{\text{branch}}$};
\draw[decoration={brace,raise=6pt},decorate,thick] (-0.25,3.1) -- (3.55,3.1);
\node at (0.92,3) {$P_{t=18}=(10^{-0.7})(10^{-0.1})(10^{-2.0})$};
\draw[thick,->,rounded corners=10pt] (3.7,3) -| (4.45,2);
\node at (2,-1.7) {\includegraphics[width=0.5\linewidth]{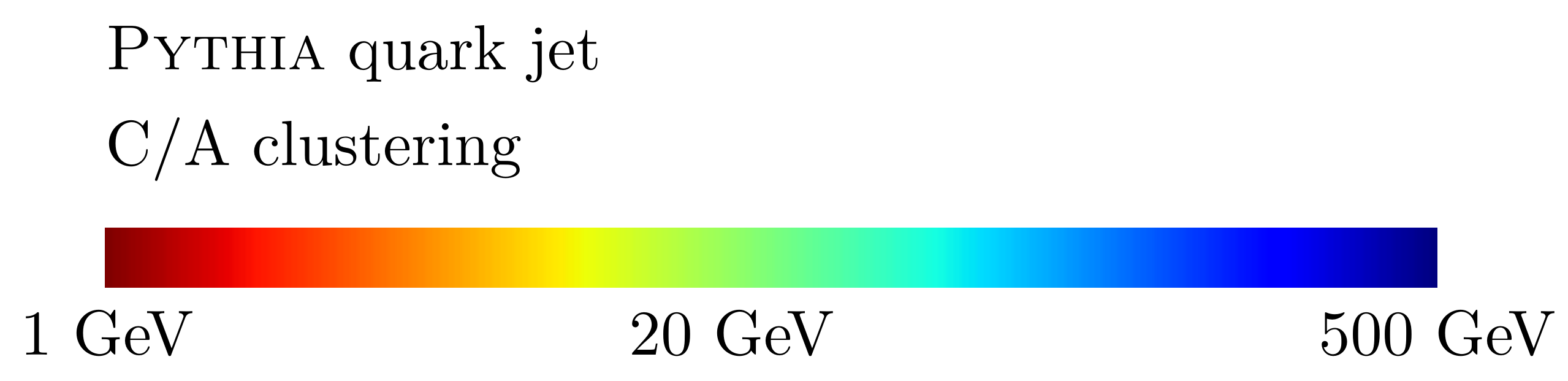}};
\node at (7.8,1.58) {$\implies ~ P_{\text{jet}} = 10^{-51.6}$};
\end{tikzpicture}
\caption{\JUNIPR-computed probability asigned to example \Pythia jet and sequentially decomposed along its C/A clustering tree.
Nodes are labeled with $\log_{10}P_t$, where $P_t = P_\text{end} \cdot P_\text{mother} \cdot P_\text{branch}$ includes the product of all three components of the probability at step $t$, as shown in \Eq{IndividTimeStepModel}. Color corresponds to energy and opening angle corresponds to 3-dimensional branching angle.
Probabilities are small and discrete due to the discretized branching function used in \JUNIPR's implementation.}
\label{fig:probability_tree}
\end{figure}
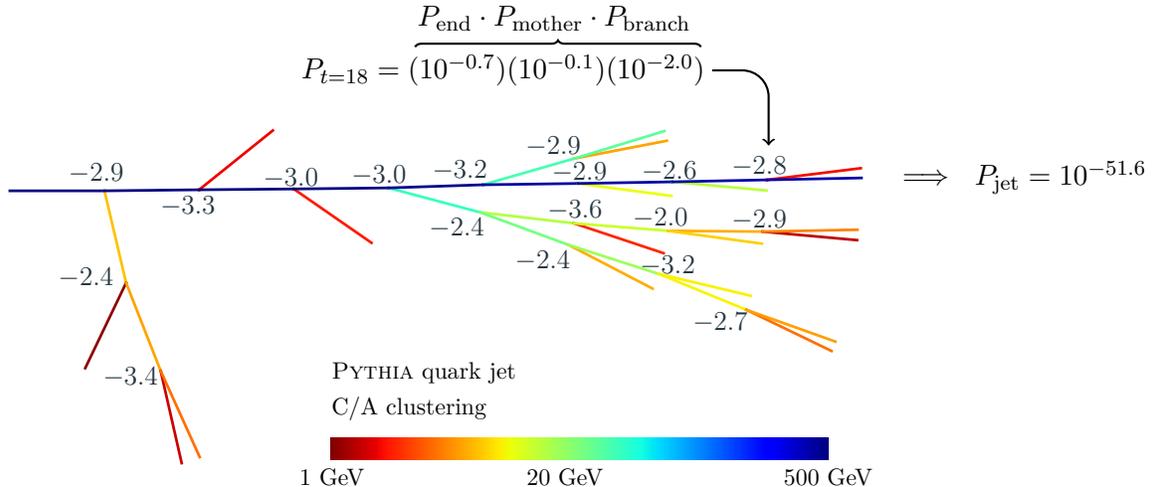

With \JUNIPR trained and validated, we turn to some of the most interesting results it enables. 
Given a jet, \JUNIPR can compute the probability density associated with the momenta inside the jet, conditioned on the criteria used to select the training data. 
To visualize this, we show a C/A-clustered \Pythia jet in \Fig{probability_tree} with the \JUNIPR-computed probability associated with each branching written near that node in the tree. 
Note that these are small discretized probabilities due to the discretized implementation of \JUNIPR's branching function described in \Sec{Model}.
This is shown primarily to conceptualize the model, which is constructed to be quite interpretable as it is broken down to compute the probability of each step in the clustering history of a jet.

A direct and powerful application of the \JUNIPR framework, 
enabled by having access to separate probabilistic models of different data sources,
is in discrimination based on likelihood ratios. 
We discuss discrimination in \Sec{ResultsLikelihood}, along with a highly intuitive way of visualizing it. In contrast, an instinctive but indirect use of \JUNIPR as a probabilistic model is in sampling new jets from it. We discuss the observable distributions generated through sampling in \Sec{ResultsSampling}. However, sampling from a probabilistic model is often inefficient (e.g.~slower than \Pythia) compared to evaluating probabilities of jets directly. In \Sec{ResultsReweighting} we discuss reweighting samples from one simulator to match those of another distribution. In principle, this could be used to tweak \Pythia samples to match observed collider data simply by reweighting.

\subsection{Likelihood Ratio Discrimination}
\label{sec:ResultsLikelihood}

We expect that one of the most exciting applications of \JUNIPR will be in discriminating the underlying physics that could have created a jet.\footnote{We thank Kyle Cranmer for an early discussion on this topic.} For example, suppose we had two sets of jets, one set corresponding to decays of a boosted $Z$ boson, the other set simply high-energy quarks. We could then train one copy of \JUNIPR on just the boosted $Z$ sample, giving the probability distribution $P_Z$, and another copy of \JUNIPR on just the quark jets, giving $P_q$. Finally, for any new jet we could determine whether the jet was initiated by a boosted $Z$ or by a high-energy quark by looking at the likelihood ratio:
\eqn{
\frac{P_Z(\text{jet})}{P_q(\text{jet})} > \text{threshold} \quad\implies\quad \text{jet is boosted }Z
}
where the threshold is set according to the location on the ROC (receiver operating characteristic) curve desired for the discrimination task at hand. In contrast to approaches that try to compute likelihood ratios like this using QCD~\cite{Soper:2011cr,Soper:2014rya}, the \JUNIPR approach can learn the separate probability distributions directly from samples of training data. 

Discrimination based on the likelihood ratio theoretically provides the most statistically powerful discriminant between two hypotheses \cite{Neyman289}. 
Moreover, our setup takes into account all the momenta that define a specific type of jet. 
Note also that for the task of pairwise discrimination between $N$ jet types, this unsupervised approach requires training $N$ probabilistic models, whereas a supervised learning approach would require training $N(N-1)/2$ classifiers. 
Thus, we expect likelihood-ratio discrimination using \JUNIPR to provide a powerful tool. 

We note further that we do not even require pure samples of the two underlying processes between which we would like to discriminate \cite{Metodiev:2017vrx}. Thus, it would be feasible to discriminate based solely on real collider data. In our $Z$/quark example above, we would simply train one copy of \JUNIPR on a sample of \emph{predominantly} boosted-$Z$ jets, and train another copy on \emph{predominantly} quark jets, and the likelihood ratio of those two models would still be theoretically optimal for $Z$/quark discrimination. 

In order to get a first look at the potential of likelihood-ratio discrimination using \JUNIPR, we continue with the $Z$/quark example discussed above. We use \Pythia to simulate $e^+e^- \to q\bar{q}$ and $e^+e^- \to ZZ$ events at a center-of-mass energy of 1 TeV. We impose a very tight mass window, 90.7 -- 91.7 GeV, on the jets in each data set, so that no discrimination power can be gleaned from the jet mass. More details on the generation of the data sets were given in \Sec{TrainingData}. We admit that a more compelling example of discrimination power would be for quark and gluon jets at hadron colliders, but we leave a proper treatment of that important case to future work. 
The toy scenario studied here serves both to prove that the probabilities output by \JUNIPR are meaningful, and that likelihood ratio discrimination using unsupervised probabilistic models is a promising application of the \JUNIPR framework.

\begin{figure}[t]
\centering
\includegraphics[width=0.5\linewidth]{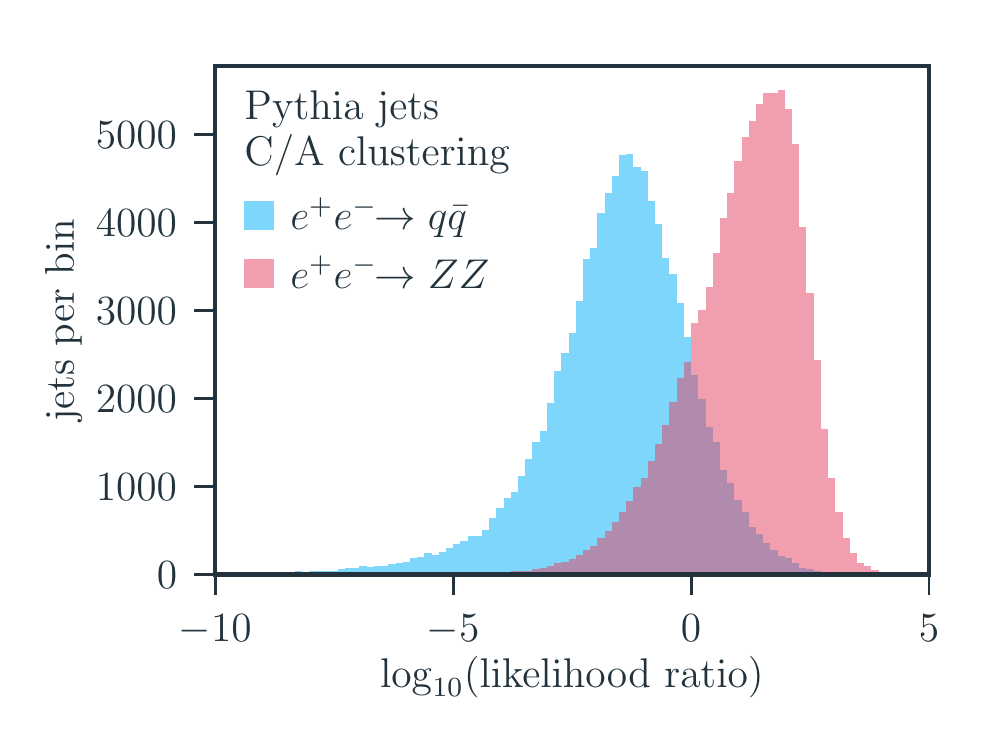}~
\includegraphics[width=0.5\linewidth]{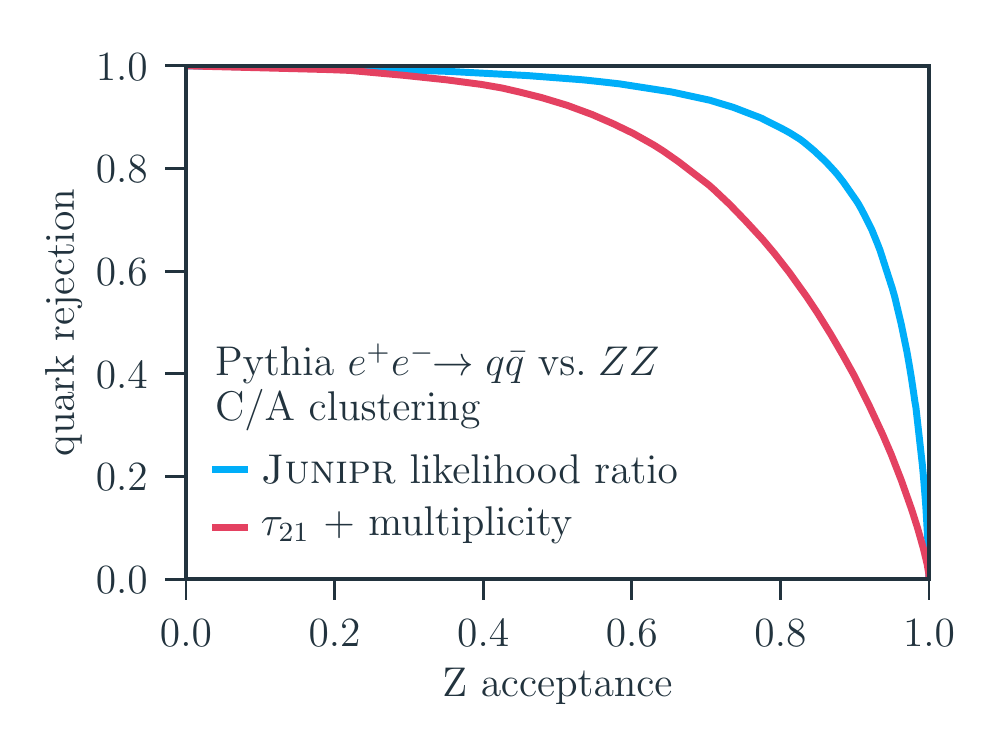}
\caption{(Left) Likelihood ratio $P_Z(\text{jet})/P_q(\text{jet})$ evaluated on \Pythia jets in the validation set. (Right) ROC curve for discrimination based on \JUNIPR's likelihood ratio, in comparison to the empirical 2D distribution using 2-subjettiness and constituent multiplicity. All jets used in this study have masses between 90.7 and 91.7 GeV.}
\label{fig:likelihood_ratio}
\end{figure}

In \Fig{likelihood_ratio} we show the $Z$/quark separation power achieved by \JUNIPR, both in terms of full likelihood ratio distributions for validation sets of $Z$ and quark jets, as well as the resulting ROC curve. For comparison, in \Fig{likelihood_ratio} we also show the ROC curve achieved using a 2D likelihood ratio discriminant based on 2-subjettiness \cite{Thaler:2010tr} and multiplicity. \JUNIPR's likelihood-ratio discrimination is clearly superior to that based on combining the most natural observables: 2-subjettiness, multiplicity, (and keep in mind the tight mass cut). Of course, these observables do not provide state-of-the-art discrimination power even in this toy scenario, but we include the comparison in this proof-of-concept to provide a sense of scale on the plot.

\begin{figure}[t]
\begin{tikzpicture}
\node at (0,0) {\includegraphics[width=0.74\linewidth]{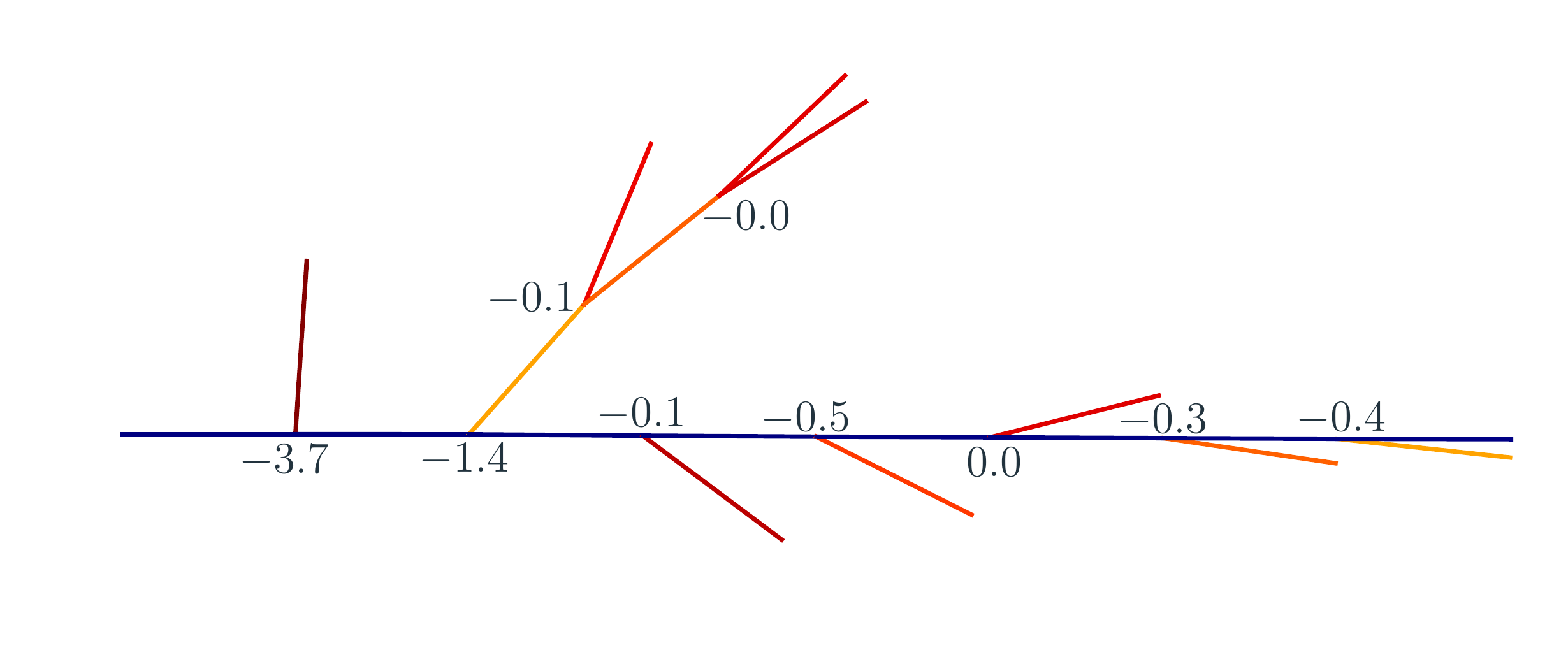}};
\node at (0,-4) {\includegraphics[width=0.74\linewidth]{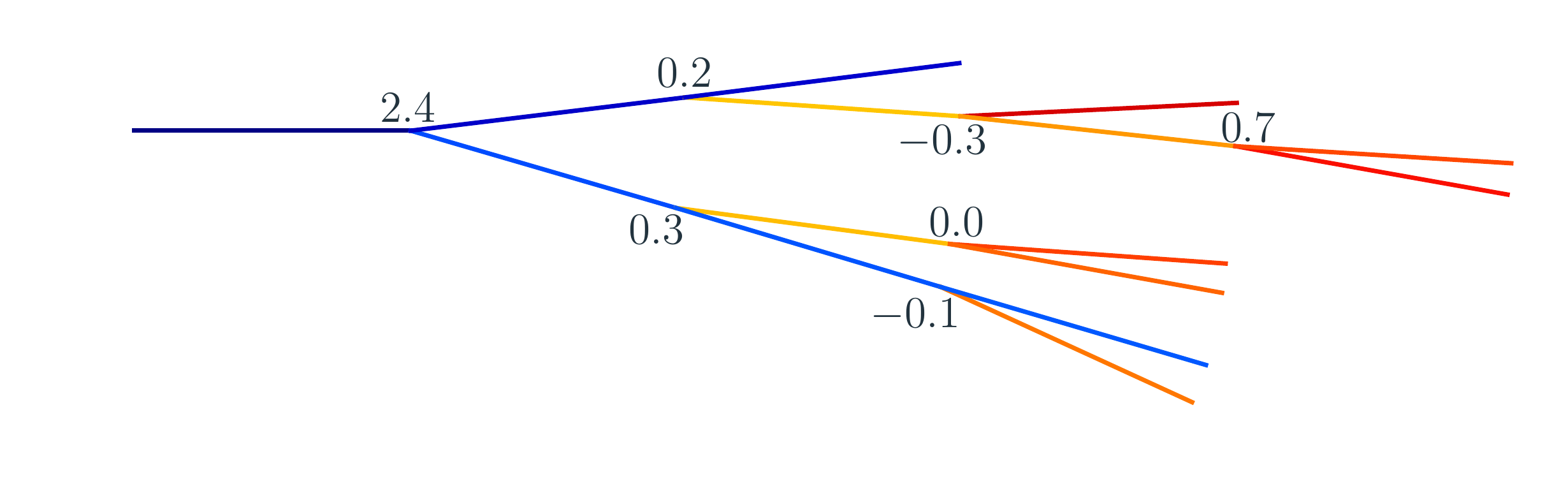}};
\node at (7.5,-1.5) {\includegraphics[width=0.12\linewidth]{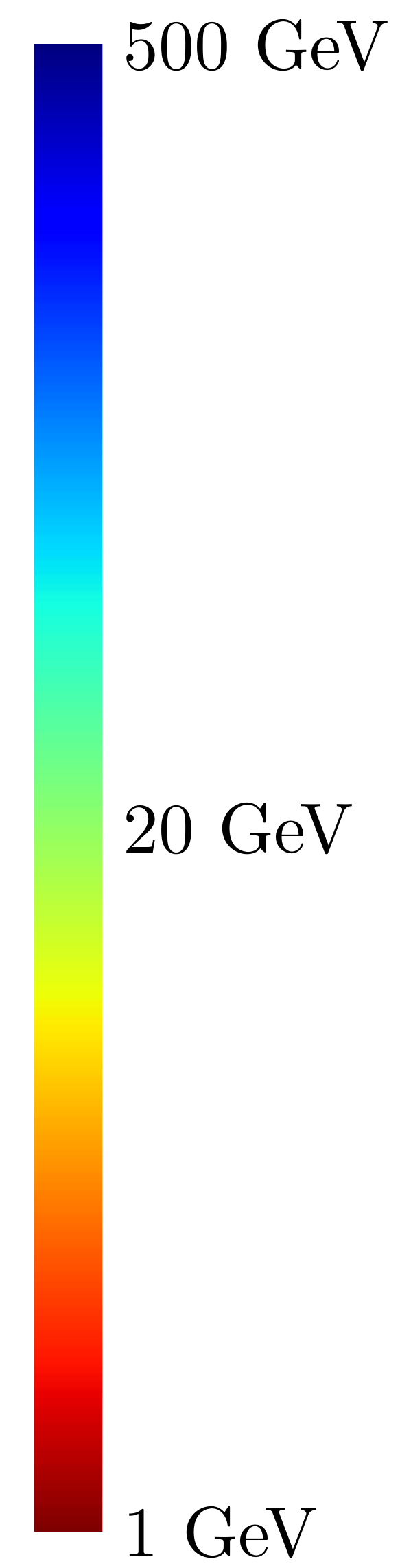}};
\node at (5,1) {\includegraphics[width=0.2\linewidth]{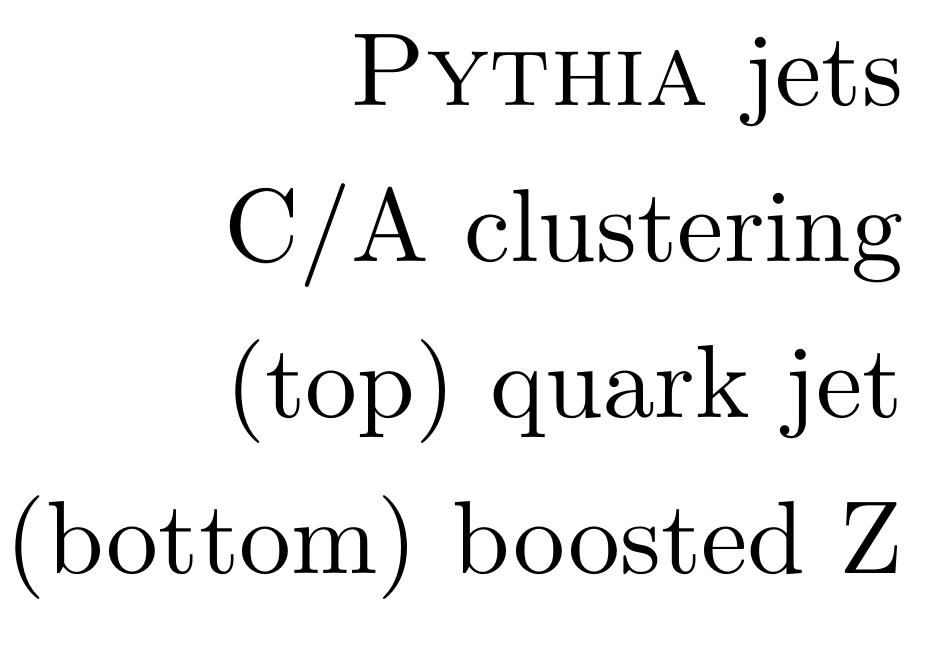}};
\end{tikzpicture}
\caption{\JUNIPR trees for visualization of discrimination power at individual nodes in the clustering history. Each node is labeled with the component of $\log_{10} {P_Z(\text{jet}) \big/ P_q(\text{jet})}$ associated with that $t$ step. Colors represent energies, and opening angles represent physical 3-dimensional branching angles. The top figure is a quark jet generated using \Pythia, with mass between between 90.7 and 91.7 GeV; the bottom figure is a boosted-$Z$ jet. The role that the energy distribution, opening angles, multiplicity, and branching pattern play in high-performance discrimination can be understood from such pictures.}
\label{fig:discrimination_trees}
\end{figure}

By design, \JUNIPR naturally processes the information in jets via a recurrent mechanism that tracks the evolution of their clustering trees, and this allows users to peer inside at this structure and access the probabilities at each branching. In particular, we can consider the likelihood ratio at each step in the clustering trees to understand which branchings give rise to the greatest discrimination power. We show this in \Fig{discrimination_trees}, where it is clear that \JUNIPR can extract useful discriminatory information at most branchings.

Indeed, visualizing jets as in \Fig{discrimination_trees} can provide a number of insights. Unsurprisingly, we see for the quark jet (on the top) that the likelihood ratio of the first branching is rather extreme, at $10^{-3.7}$, since it is unlike the energy-balanced first branching associated with boosted-$Z$ jets. However, we also see that almost all subsequent branchings are also unlike those expected in boosted-$Z$ jets, and they combine to provide comparable discrimination power to the first branching alone. Many effects probably contribute to this separation power at later branchings, including that quark jets often gain their mass throughout their evolution instead of solely at the first branching, and that the quark jet is color-connected to other objects in the global event. Such effects have proven to be useful for discrimination in other contexts \cite{Chien:2017xrb}. 

Similarly, considering the boosted-$Z$ jet on the bottom of \Fig{discrimination_trees} shows that significant discrimination power comes not only from the first branching, but also from subsequent splittings, as the boosted-$Z$ jet evolves as a color-singlet $q \bar q$ pair. Note the presence of the predictive secondary emissions sent from one quark-subjet toward the other. This is reminiscent of the pull observable, which has proven useful for discrimination in other contexts \cite{Gallicchio:2010sw}. 
More generally, the importance of the energy distribution, opening angles, multiplicity, and branching pattern in high-performance discrimination can be understood from such pictures.

We are very excited by the prospect of visualizing \JUNIPR's discrimination power on jets, based on the likelihood ratio it assigns at each branching in their clustering trees, as in \Fig{discrimination_trees}. Such visualizations could provide intuition that leads to the development of new, human-interpretable, perhaps calculable observables for discrimination in important contexts.

We would like to make one side note about discrimination, before moving on to the next application of \JUNIPR. The statement that likelihood-ratio discrimination is optimal of course only applies in the limit of perfect models. Since this limit is never fully realized, one may worry that discrimination with \JUNIPR may in fact be suboptimal. Since the two probabilistic models we use for discrimination are each trained individually to replicate a certain type of jet, they are not conditioned to focus on the differences between the two jet types, which may be very subtle in the case of a difficult discrimination task. In the realistic case of slightly imperfect models, it may be advantageous for discrimination purposes to instead train the two models to focus on the differences. To be specific, one could train the two models on the two data sets simultaneously, with the goal being to maximize the likelihood ratio on one data set and minimize it on the other. Following this method in the particular example of $Z$/quark discrimination used above, one would train the $P_Z$ and $P_q$ models on data sets $D_Z$ and $D_q$ to maximize the following quantity:
\begin{equation}
\label{eq:modified_discrimination}
    \sum_{\text{jet} \in D_Z} \log{P_Z(\text{jet}) \over P_q(\text{jet})}
    ~- \sum_{\text{jet} \in D_q} \log{P_Z(\text{jet}) \over P_q(\text{jet})}.
\end{equation}
Compare this to the approach we have taken above, namley training $P_Z$ and $P_q$ to separately maximize the log likelihood of \Eq{LogLike} on their corresponding sets of training data. This alternative training method would correspond to optimizing \JUNIPR for the application of discrimination, leaving intact our ability to visualize discrimination power in clustering trees, but sacrificing the probabilistic interpretation of the model's output. 
We have not tested training with \Eq{modified_discrimination}, and thus cannot attest to its practicality, but we suspect an approach along these lines may be useful in certain contexts.

\subsection{Generation from \JUNIPRtitle}
\label{sec:ResultsSampling}

We now turn to a more familiar approach to jet physics, but a somewhat less appropriate usage of \JUNIPR models: sampling new jets from the learned probability distribution to generate traditional observable distributions. We include this application here, not only to demonstrate this capability, but also to further validate the distribution learned by \JUNIPR during unsupervised training.

\begin{figure}[t]
\centering
\includegraphics[width=0.8\linewidth]{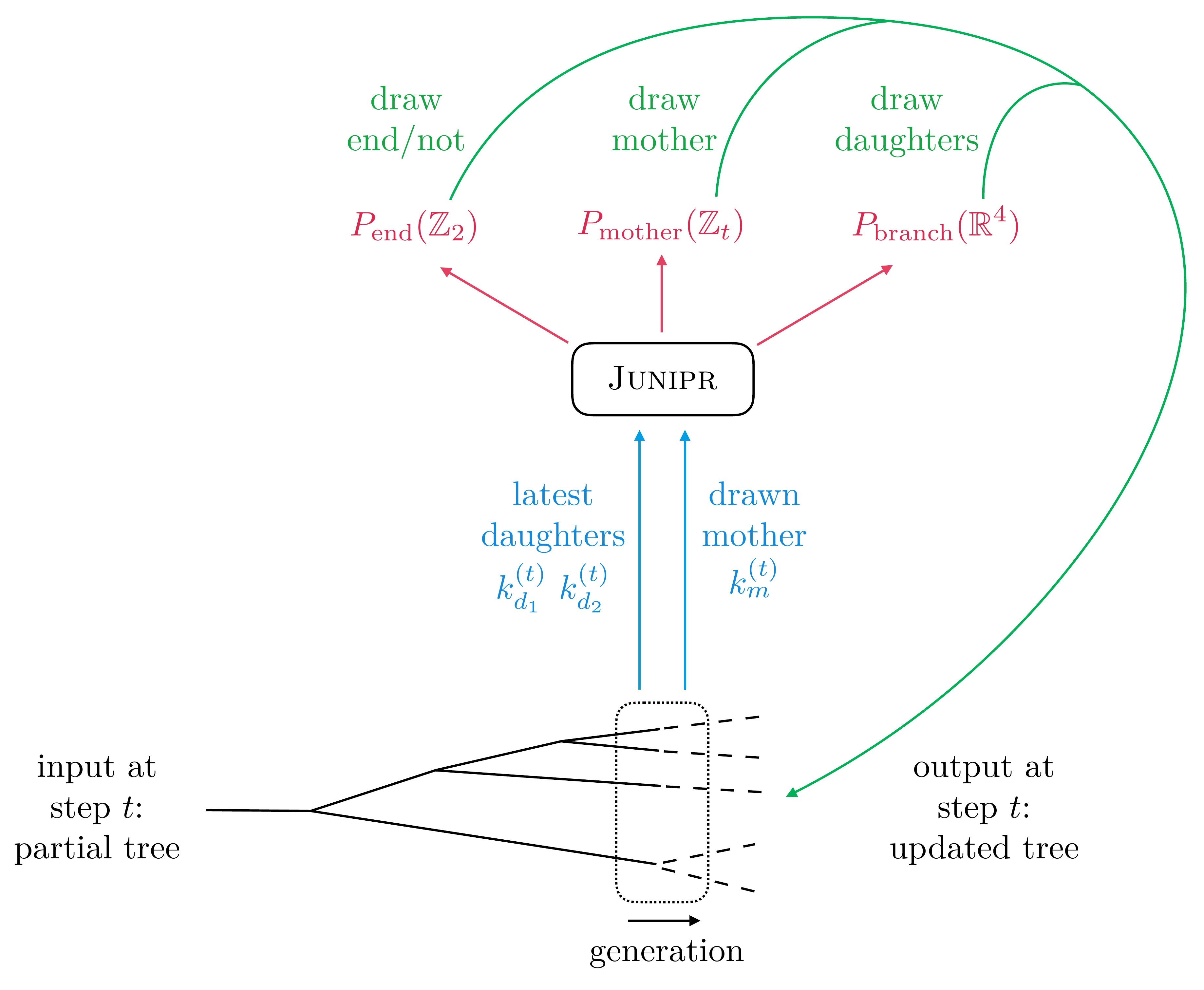}
\caption{Sampling from \JUNIPR to generate jets. Draws from low-dimensional distributions at each step $t$ are fed forward to subsequent steps to ultimately generate a full jet.}
\label{fig:generation_mode}
\end{figure}

Sampling from \JUNIPR is relatively efficient; one simply samples from the low dimensional distributions at each step $t$ and feeds those samples forward as input to subsequent steps. In this way, one generates a full jet in many steps, as detailed in \Fig{generation_mode}.

\begin{figure}[t]
\centering
\includegraphics[width=0.5\linewidth]{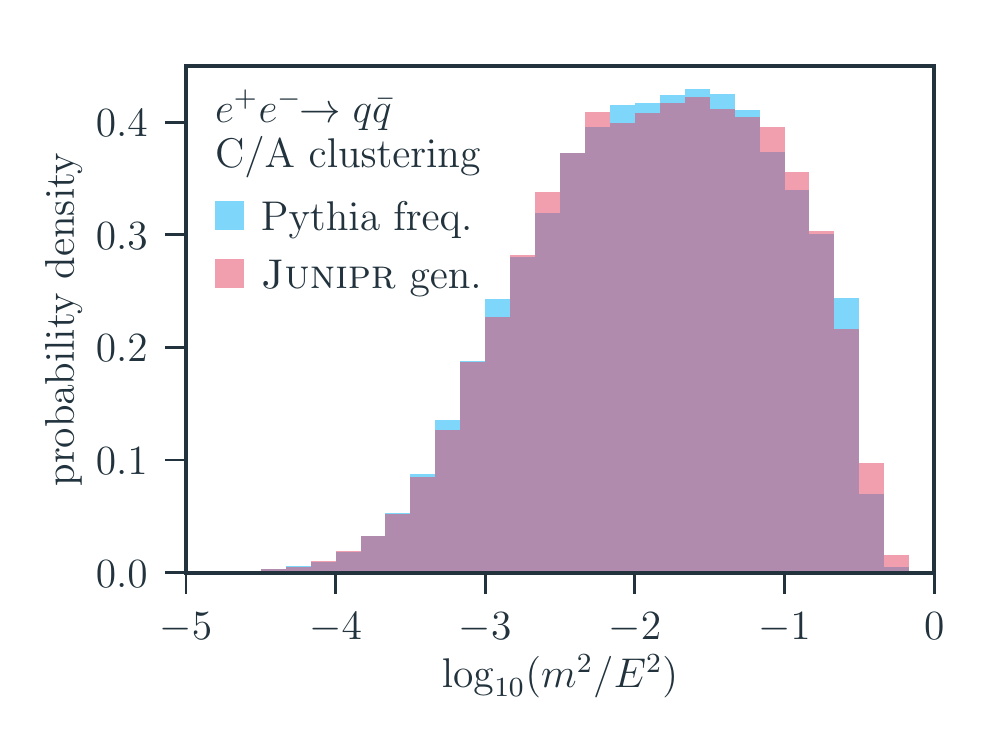}~
\includegraphics[width=0.5\linewidth]{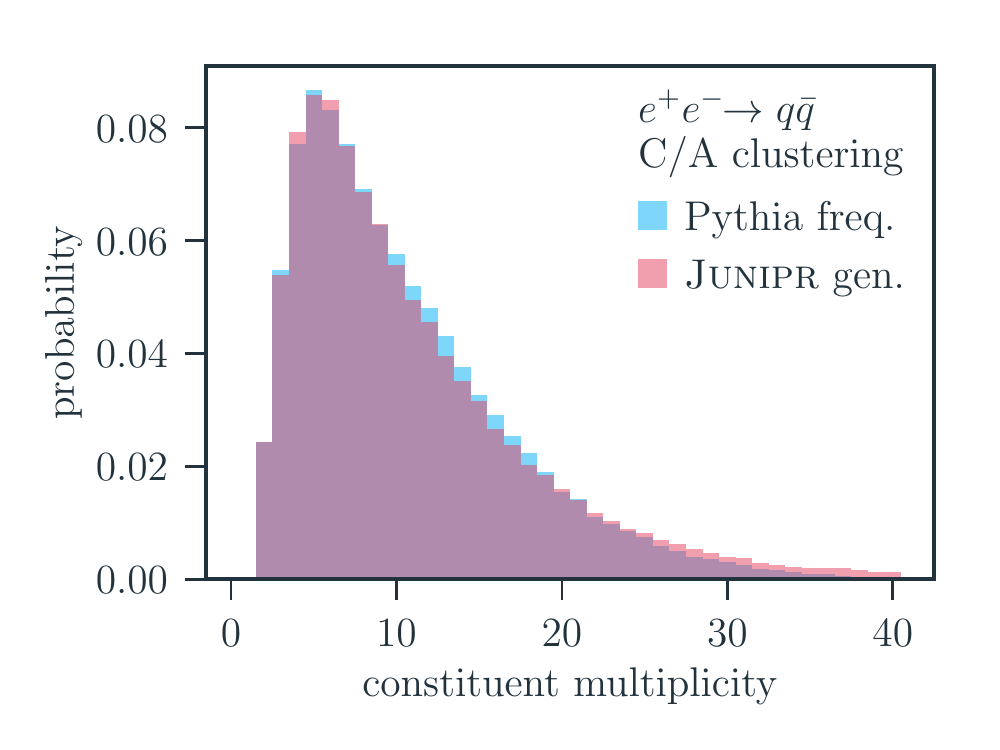}
\caption{Jet mass (left) and constituent multiplicity (right) distributions computed on jets sampled from \JUNIPR and compared against \Pythia jets in the validation set.}
\label{fig:mass_mult}
\end{figure}

\begin{figure}[t]
\centering
\includegraphics[width=0.5\linewidth]{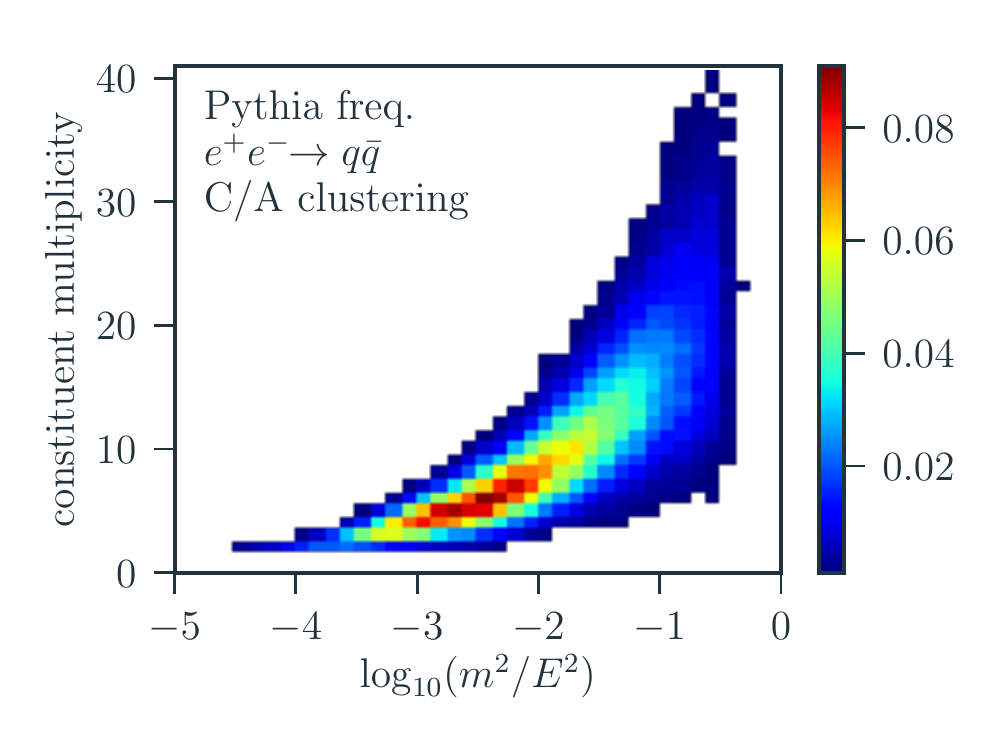}~
\includegraphics[width=0.5\linewidth]{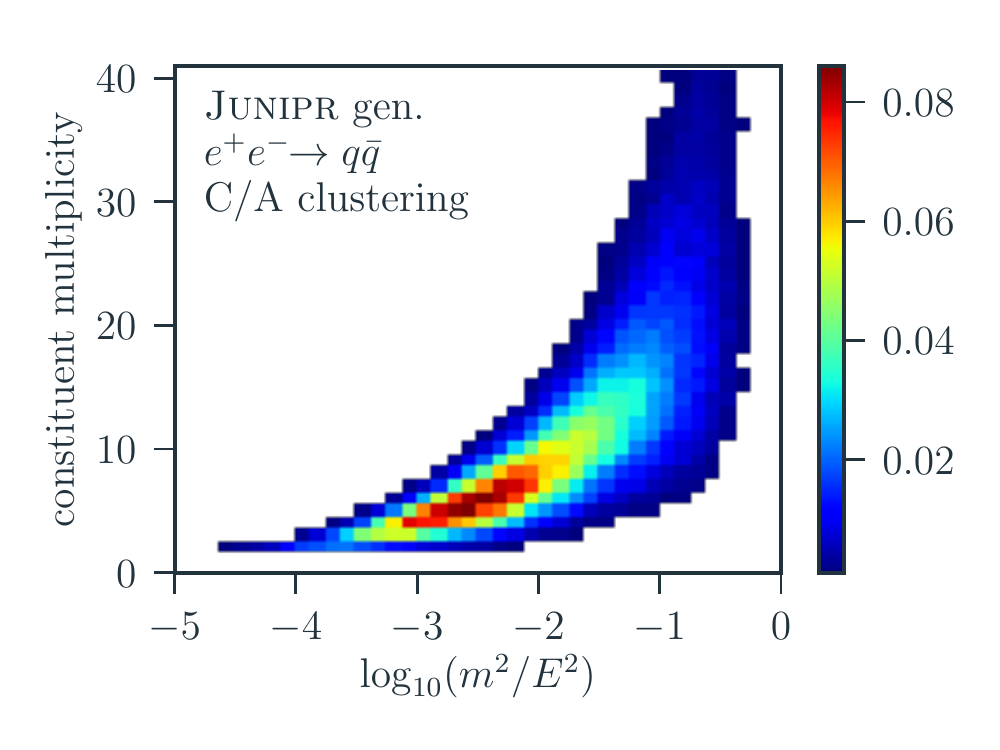}
\caption{2-dimensional probability distributions with respect to jet mass and constituent multiplicity. (Left) Distribution computed using validation set of \Pythia jets. (Right) Distribution computed using jets sampled from \JUNIPR.}
\label{fig:mass_mult_2D}
\end{figure}

We used the baseline implementation of \JUNIPR trained on quark jets, as described in \Sec{Training}, to generate 100k jets in this way. The resulting jet mass and constituent multiplicity distributions are plotted in \Fig{mass_mult} where both distributions sampled from \JUNIPR match those created from our validation set of 100k \Pythia jets withheld from training. Reasonable agreement can also be seen in the 2D distributions of \Fig{mass_mult_2D}.

However, there are two reasons why we do not consider \JUNIPR to be built for generation. 
(These drawbacks could be avoided with a generative model; see \cite{deOliveira:2017pjk,Paganini:2017hrr,Paganini:2017dwg}.)
The first is simply that sampling from probability distributions is generally difficult. As we just showed, it turns out that \JUNIPR is relatively easy to sample from, due to its sequential structure and the fact that distributions are low-dimensional at each $t$ step. Despite this, sampling jets from \JUNIPR is still much slower than generation with, for example, \Pythia.

The second reason is more fundamental. With a sequential model structured as \JUNIPR is, probability distributions at late $t$ steps in generation are highly sensitive to the draws made at earlier $t$ steps. Very small defects in the probability distributions at early steps cause feedback in the model that amplifies those errors. Furthermore, as a partially generated jet becomes more misrepresentative of the training data, the resulting probability distributions used at later steps are less trained, which can result in a run-away effect. All of this is to say that, for the purpose of generating jets, \JUNIPR's accuracy at early $t$ steps is disproportionately important. This is in tension with the training method undertaken in \Sec{TrainingApproach}, namely the maximization of the log-likelihood, which prioritizes all branchings equally. Thus, we should expect that some observable distributions generated by sampling jets from \JUNIPR might agree worse with the validation set of \Pythia data than otherwise expected.
We mention in passing that this second drawback could be mitigated by reweighting jets after generation, as detailed in \Sec{ResultsReweighting} below.

\begin{figure}[t]
\centering
\includegraphics[width=0.5\linewidth]{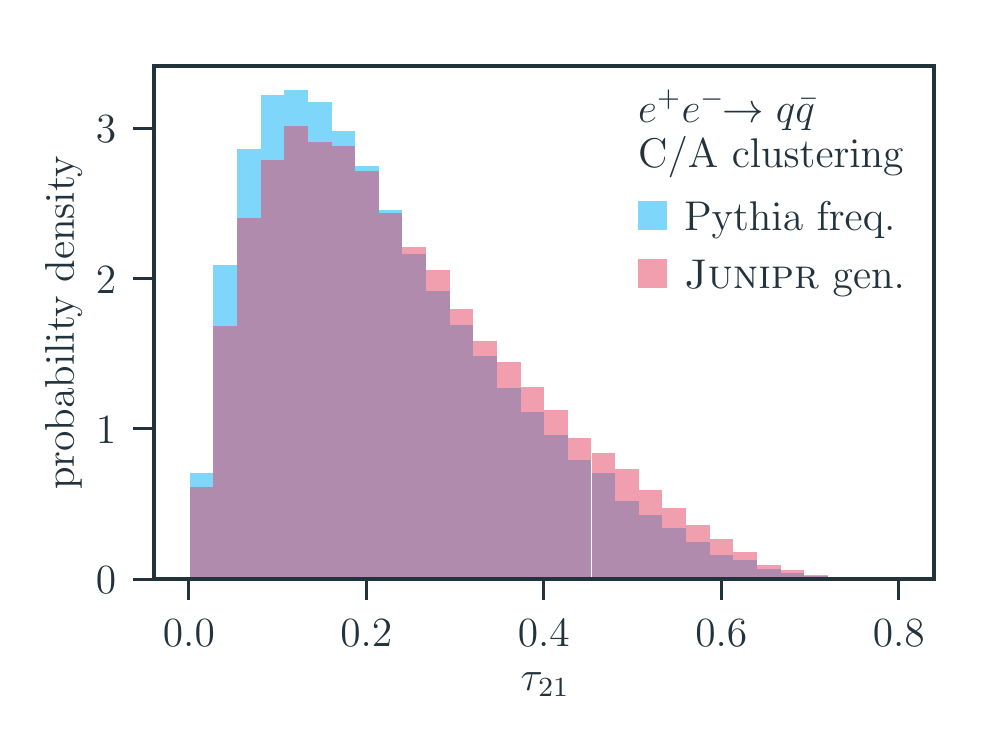}
\caption{2-subjettiness ratio observable computed on jets sampled from \JUNIPR. Disagreement with the distribution on \Pythia jets, due to the feedback involved in sampling from \JUNIPR, is visible. This disagreement is amended in \Fig{Reweighting}.}
\label{fig:tau21_badmatch}
\end{figure}

In fact, we have found empirically that the N-subjettiness ratio observables computed by sampling from \JUNIPR do not match the held-out \Pythia data perfectly. This can be seen in \Fig{tau21_badmatch} with the 2-subjettiness distribution, where the difference between the two distributions is more significant. 

We consider this disagreement to be both expected and non-diminishing of \JUNIPR's potential. Indeed, in the next section we will show how to overcome this issue, by generating samples consistent with \JUNIPR's learned probabilistic model, without ever sampling from it.
In particular, the disagreement in \Fig{tau21_badmatch} will be rectified in \Fig{Reweighting}.

\subsection{Reweighting Monte Carlo Events}
\label{sec:ResultsReweighting}
Another application of the \JUNIPR framework is to reweight events.
For example, suppose we trained \JUNIPR on data from the Large Hadron Collider (LHC) to yield a probabilistic model $P_\text{LHC}$. Then one could generate a sample of new events using a relatively accurate Monte Carlo simulator, train another instance of \JUNIPR on that sample to yield $P_\text{sim}$, and finally reweight the simulated events by $P_\text{LHC} / P_\text{sim}$ evaluated
on an event-by-event basis. This process yields a sample of events that is theoretically equivalent to the LHC data used in training $P_\text{LHC}$.
The advantage of such an approach is that \JUNIPR can correct the simulated events on different levels, for example using the data reclustered in $R_\text{sub}=0.1$ subjets as we have done in this paper. However, the full simulated event
has the complete hadron distributions and can thereby be interfaced with a detector simulation. This is in many ways a simpler approach than trying
to improve the simulation directly through the dark art of Monte-Carlo tuning.

This reweighting is identical to importance sampling from a proposal distribution given by the simulated data distribution $P_\text{sim}$. For example, suppose one wanted to measure the distribution of an observable $\mathcal{O}(\text{jet})$ at the LHC, which is given by
\eqna{
P(\mathcal{O}) &= \int d[\text{jet}] \, P_\text{LHC}(\text{jet}) \, \delta(\mathcal{O} - \mathcal{O}(\text{jet})) \\[5pt]
&\approx \frac{1}{N} \sum_{\text{jet} \sim P_\text{LHC}}
\delta(\mathcal{O} - \mathcal{O}(\text{jet}))
}
where the last approximation is associated with collecting a finite amount $N$ of LHC data in order to measure the distribution. (The reader can substitute discretized delta functions appropriate for histogramming if averse to the singular notation used in these equations.) Instead of using real data, if say a public version of $P_\text{LHC}$ were available, then anyone could calculate this observable distribution using only simulated data sampled from $P_\text{sim}$ as follows:
\eqna{
P(\mathcal{O}) &= \int d[\text{jet}] \, P_\text{sim}(\text{jet}) \, \delta(\mathcal{O} - \mathcal{O}(\text{jet})) \,  \frac{P_\text{LHC}(\text{jet})}{P_\text{sim}(\text{jet})} 
\\[5pt]
&\approx \frac{1}{N} \sum_{\text{jet} \sim P_\text{sim}}
\delta(\mathcal{O} - \mathcal{O}(\text{jet})) \, \frac{P_\text{LHC}(\text{jet})}{P_\text{sim}(\text{jet})}\,.
}

In this way, one could efficiently obtain samples of arbitrary size from $P_\text{LHC}$ by reweighting samples generated by an efficient simulator. The only limitation to this process is that the simulated data must be similar to the actual target data, so that they have overlapping regions of support (formal requirement) and the weights are not too far from unity (efficiency requirement).

As with the likelihood-ratio discrimination in \Sec{ResultsLikelihood}, here we will show results in a toy scenario as a proof-of-principle. Ideally a model trained on LHC data, with all related complications, would be used to reweight Monte Carlo jets to make the simulated data indiscernible from LHC data; we leave a proper study of this to future work.

Instead, here we use two samples of jets generated using two different versions of \Pythia. We reweight jets from one of the samples and demonstrate their agreement with the other sample. In particular, we use our baseline \JUNIPR model trained on \Pythia-generated quark jets as our ``true distribution''. For the moment, we will refer to this model as $P_{\alpha_s=0.1365}$, since its training data was generated using \Pythia's default value of $\alpha_s(m_Z) = 0.1365$ in the final state shower. As our ``simulated distribution'' we will use $P_{\alpha_s=0.11}$, which was trained on quark jets generated with coupling parameter changed to $\alpha_s(m_Z) = 0.11$ in \Pythia's final-state shower. (See \Sec{TrainingData} for a more in-depth description of the training data used.) Our goal is to show that reweighting jets from the ``simulated distribution'' according to the likelihood ratio $P_{\alpha_s = 0.1365} / P_{\alpha_s = 0.11}$ leads to observables in agreement with the ``true distribution''. 

\begin{figure}[t]
\centering
\includegraphics[width=0.5\linewidth]{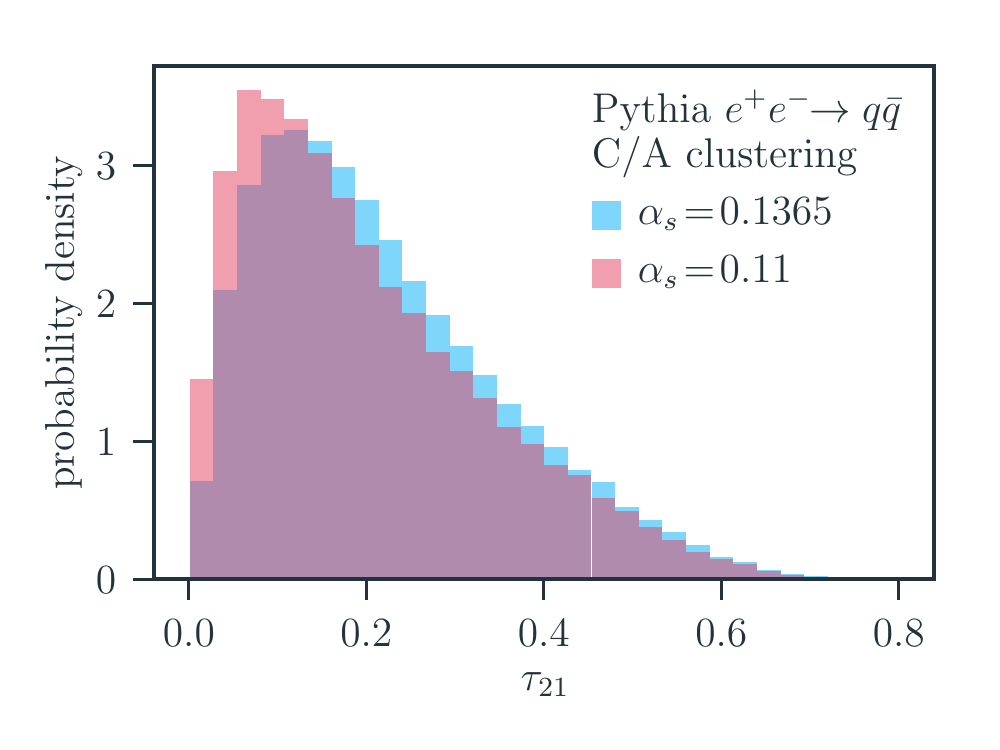}~
\includegraphics[width=0.5\linewidth]{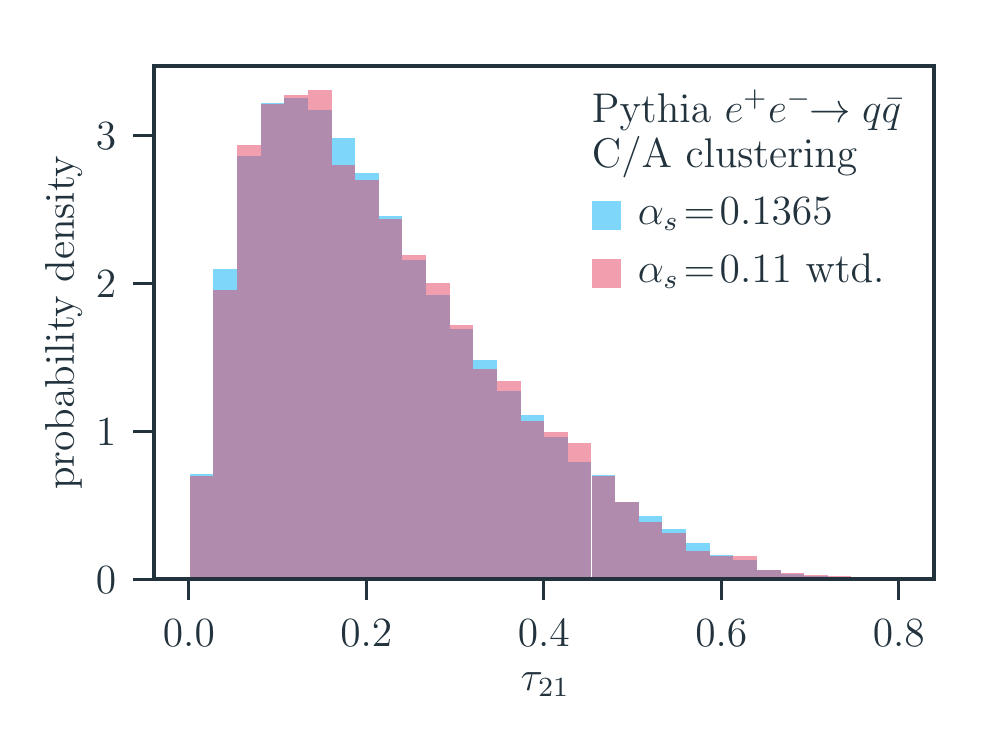}
\\
\includegraphics[width=0.5\linewidth]{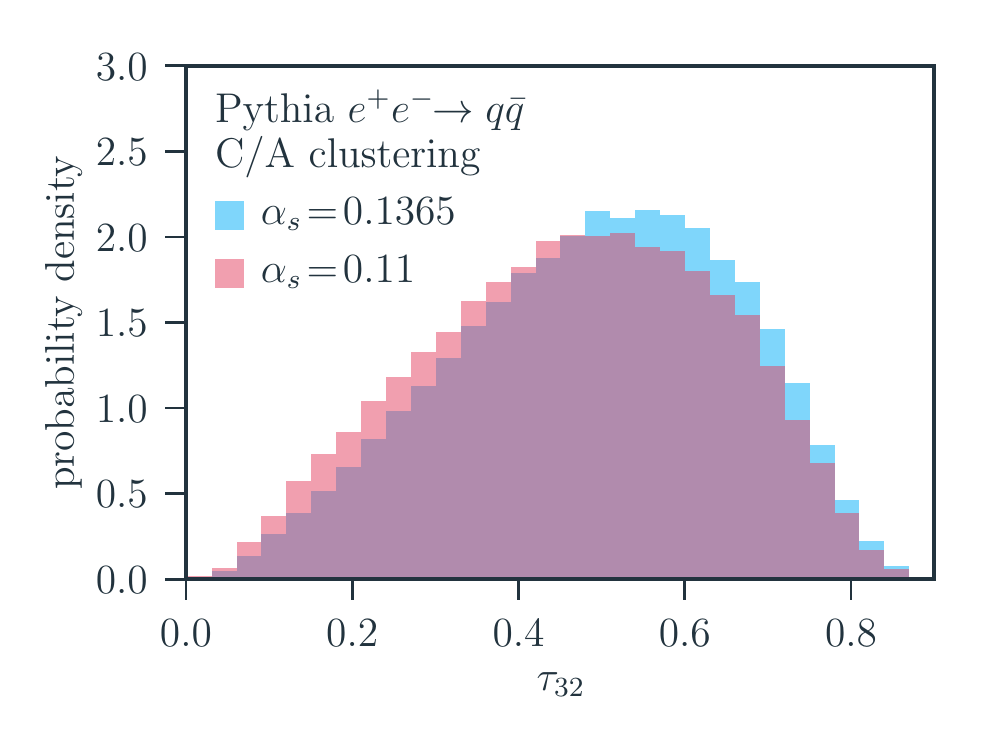}~
\includegraphics[width=0.5\linewidth]{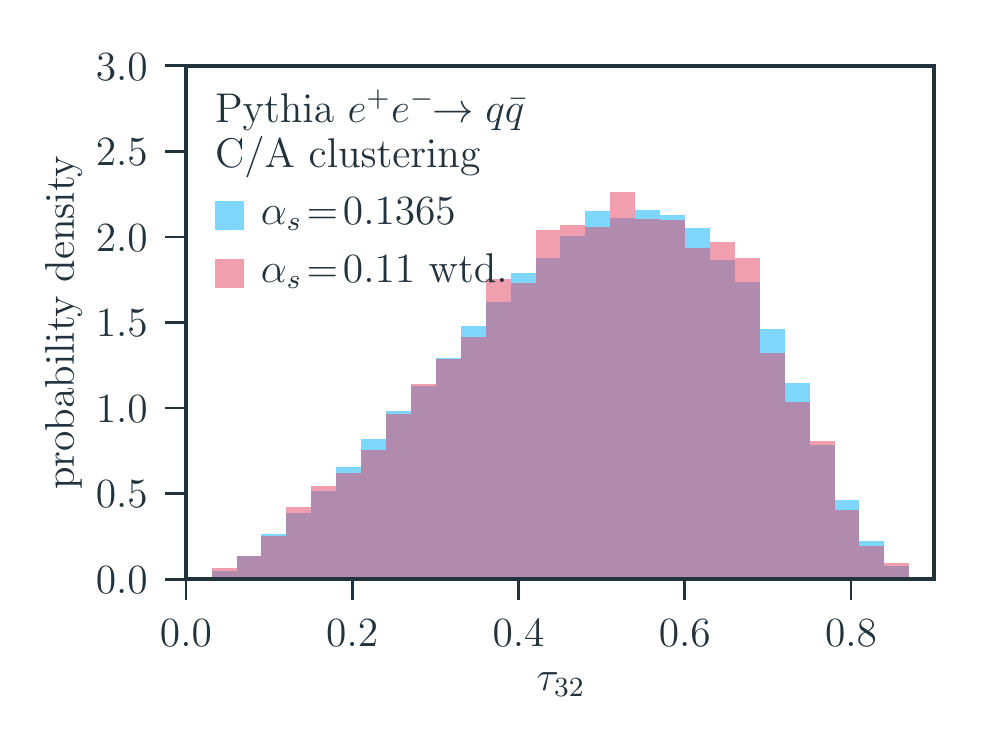}
\\
\includegraphics[width=0.5\linewidth]{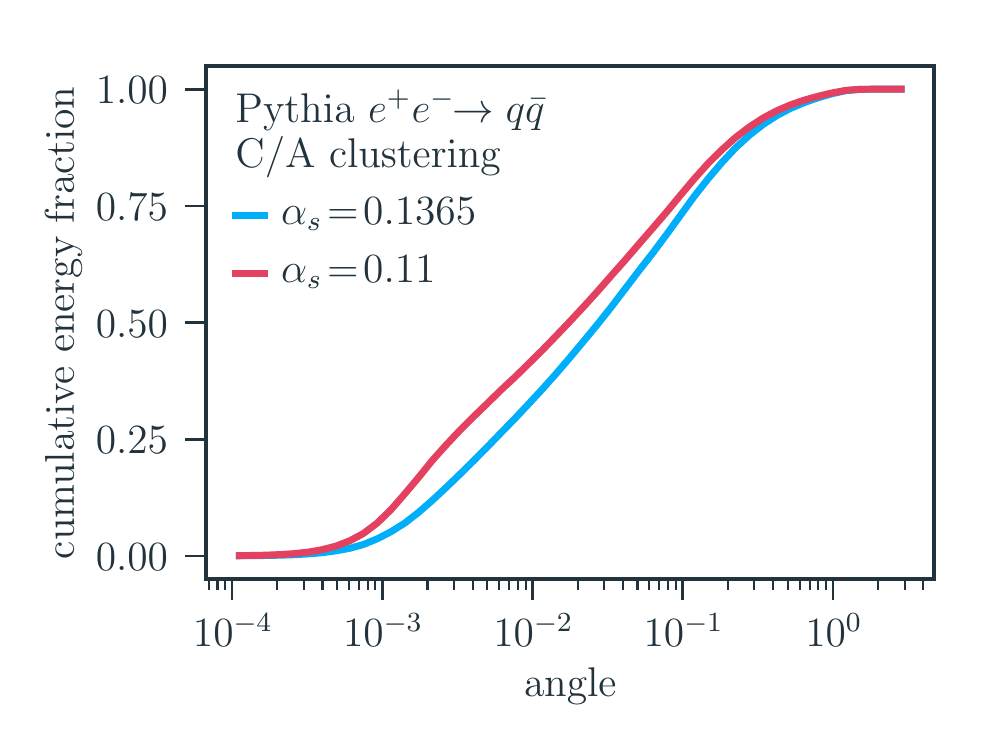}~
\includegraphics[width=0.5\linewidth]{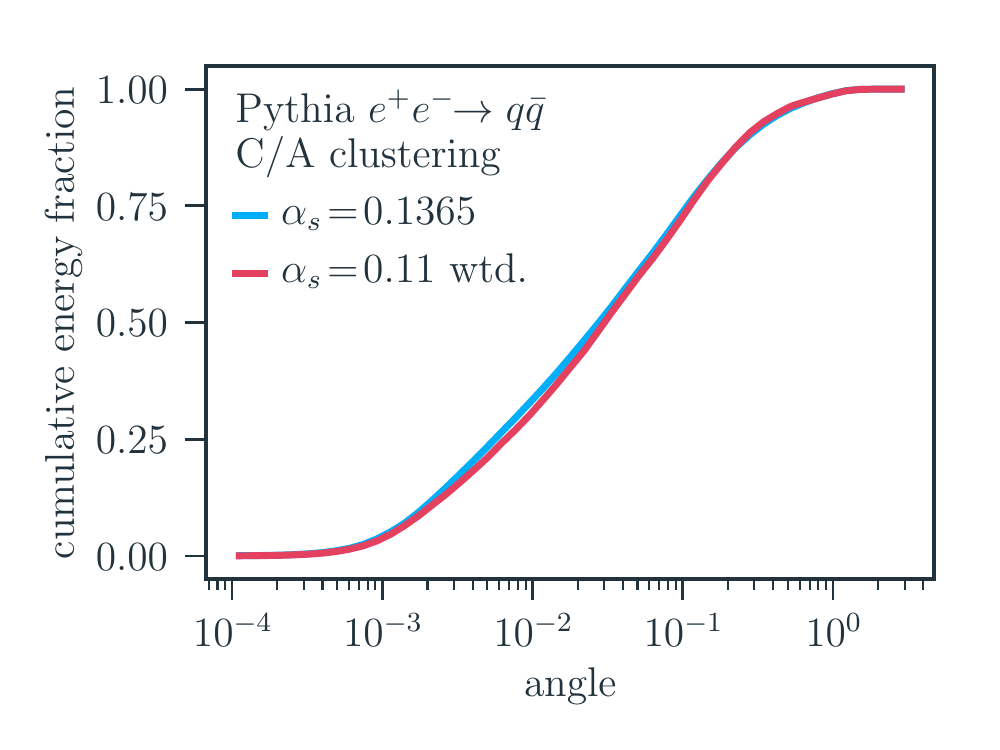}
\caption{(Left) Disagreement in observable distributions for two \Pythia tunes of $\alpha_s$. Observables are the 2-subjettiness and 3-subjettiness ratio observables and the jet shape, from top to bottom. (Right) Upon reweighting the $\alpha_s = 0.11$ jets by the ratio $P_{\alpha_s = 0.1365} / P_{\alpha_s = 0.11}$ of learned underlying probability distributions, observable distributions exhibit good agreement.}
\label{fig:Reweighting}
\end{figure}

In \Fig{Reweighting} we demonstrate that this is indeed the case. We check this for both the 2-subjettiness and 3-subjettiness ratio observables, as well as the jet shape observable. On the left side of \Fig{Reweighting}, one can see that in all cases, the $\alpha_s=0.11$ distribution is clearly different from the $\alpha_s=0.1365$ distribution. On the right side of \Fig{Reweighting}, one finds that the two distributions come into relatively good agreement once the $\alpha_s=0.11$ jets are reweighted by $P_{\alpha_s = 0.1365} / P_{\alpha_s = 0.11}$.
This also provides further confirmation that \JUNIPR learns subtle correlations between constituent momenta inside jets.

Note that it was the 2-subjettiness ratio observable that \JUNIPR struggled to predict well through direct sampling (see \Fig{tau21_badmatch}), whereas when reweighting another set of samples, \JUNIPR matches the data well on this observable (see top-right of \Fig{Reweighting}). This corroborates the discussion in \Sec{ResultsSampling} concerning the difficulties in sampling directly from \JUNIPR. 

Before closing this section, let us reiterate one point mentioned above. For the procedure of reweighting events to be practical, the weights used should not be radically different from unity, meaning that the two distributions generating the two samples should not be too different. If this condition is not satisfied, then away from the limit of infinite statistics, a few events with very large weights could vastly overpower the rest of the events, leading to a choppy reweighted distribution with large statistical uncertainties. To avoid this problem in the toy scenario explored in this section, we found it necessary to discard roughly $0.1\%$ of the jets in the $\alpha_s = 0.11$ sample which were outliers with $P_{\alpha_s = 0.1365} / P_{\alpha_s = 0.11} > 100$. These outliers were uncorrelated with the observables shown, and we believe they resulted from imperfections in the trained model. 
It is clear that much more needs to be understood about the application of reweighting, but this would perhaps be more effectively done in the context of a specific task of interest involving LHC data.


\section{Factorization and \JUNIPRtitle}
\label{sec:Discussion}

In the previous section, we showed some preliminary but very exciting results for likelihood-ratio discrimination and for the generation of observables by reweighting simulated jets. Both of these applications require access to an unsupervised probabilistic model. Next we discuss some of the more subtle internal workings of \JUNIPR, which are intimately related to the underlying physics of factorization.

In particular, we show that the hidden representation $h^{(t)}$ indeed stores important global information about intermediate states of jets in \Sec{DiscussionGlobal}. We then discuss the clustering-algorithm independence of \JUNIPR by considering two distinct clustering algorithms: a ``printer'' algorithm in \Sec{DiscussionPrinterJets}, where momenta are processed left-to-right and top-to-bottom as if by an inkjet printer; and the anti-$k_t$ algorithm in \Sec{DiscussionAntikt}, which allows us to present another counterintuitive result, the anti-$k_t$ shower generator.

\subsection{The Encoding of Global Information}
\label{sec:DiscussionGlobal}

We have constructed \JUNIPR so that all global information about the jet is contained in the RNN's hidden state $h^{(t)}$. Only the branching function $P_\text{branch}$ receives the local $1\to2$ branching information in addition to $h^{(t)}$. This forces $h^{(t)}$ to contain all the information needed to predict when the shower should end, $P_\text{end}$, to predict which momentum should branch next, $P_\text{mother}$, and to inform the branching function $P_\text{branch}$ of the relevant global structure. As the primary feature vector for all three of these distinct tasks, $h^{(t)}$ must learn an effective representation of the jet at evolution step $t$. 

\begin{figure}
\centering
\includegraphics[width=0.5\linewidth]{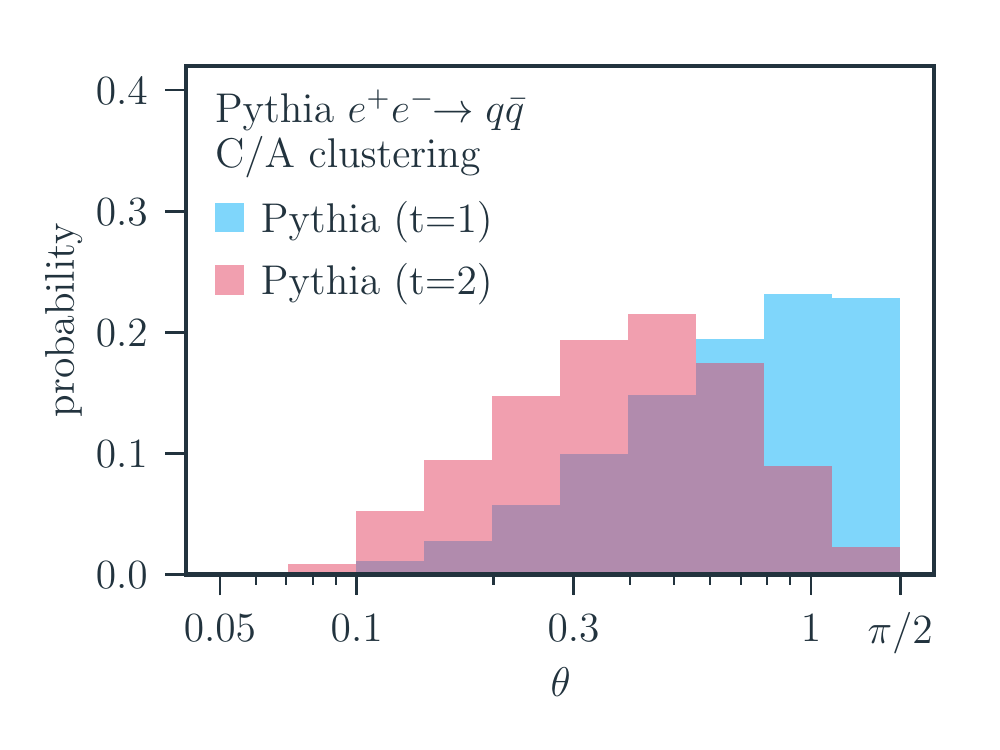}~
\includegraphics[width=0.5\linewidth]{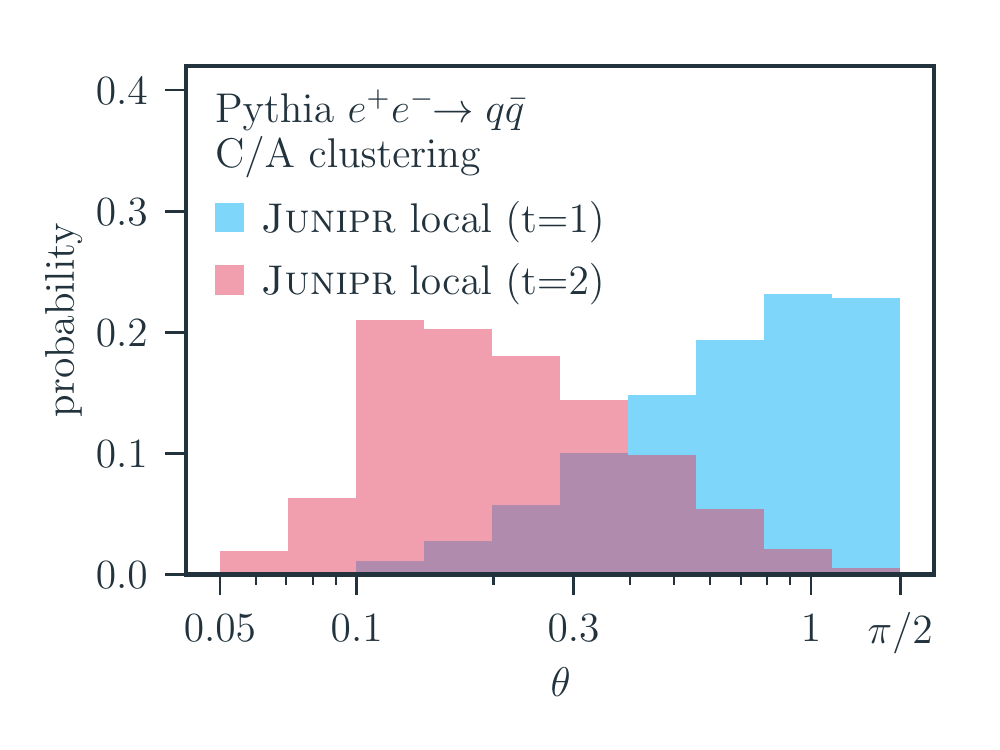}
\caption{(Left) Evolution of the $\theta$ distribution from $t=1$ to $t=2$ in the validation set of \Pythia jets. (Right) Corresponding evolution of the branching function as predicted by a ``local'' branching model without access to the hidden representation $h^{(t)}$. Disagreement between \Pythia and this local model is clear. Not shown is the result using our baseline (global) model, which agrees perfectly with \Pythia, as expected from \Fig{validate_splitting_fns}.}
\label{fig:local_splitting_model}
\end{figure}

To explicitly show that $h^{(t)}$ stores important global information about the intermediate jet state at step $t$, we train a new model on our baseline quark jet data (see \Sec{TrainingData}) with the difference that we remove $h^{(t)}$ as an input to the branching function $P_\text{branch}$. We expect that such a ``local'' branching model will not evolve correctly as the global jet structure evolves, since all global information is being withheld. This is indeed what we find, as can be seen in \Fig{local_splitting_model}. On the left side of that figure, the evolution of the $\theta$ distribution (defined in \Fig{Coordinates}) from $t=1$ to $t=2$ is shown using 100k \Pythia jets from our held-out set of validation data. There we see the gradual decrease in angle as expected for C/A trees. On the right side of \Fig{local_splitting_model}, the evolution of the branching function is shown for the ``local'' branching model, and the disagreement between this damaged model and \Pythia is clear. 
Note that this prediction of incorrect distributions at intermediate branchings in the C/A tree will inevitably lead to an incorrect probability distribution $P_\text{jet}(\{p_1, \ldots, p_n\})$ over final-state momenta.

While we do not show the corresponding results from our baseline (global) model in \Fig{local_splitting_model} to avoid clutter, the agreement with \Pythia is essentially perfect, as one would expect from the similar check performed in \Fig{validate_splitting_fns}. This confirms the success of the jet representation $h^{(t)}$ in supplying the branching function $P_\text{branch}$ with important information about the global structure.

\subsection{Clustering Algorithm Independence}
\label{sec:DiscussionPrinterJets}

Another subtle aspect of \JUNIPR is its theoretical clustering algorithm independence. In principle, the model as described in \Sec{ProbModel} is indeed independent of the chosen algorithm, which is fixed simply to avoid a sum over all possible trees consistent with the final-state momenta. That is, for each clustering procedure chosen by the user, a different model is learned, but one that describes the same probability distribution over final-state momenta, at least formally.

However, it is not guaranteed that a given neural-network implementation of \JUNIPR will work well for every clustering algorithm. We have chosen an architecture that stores the global jet physics in the RNN's hidden state $h^{(t)}$ and the local $1 \to 2$ branching physics in the branching function $P_\text{branch}$. This architecture is motivated by the factorizing structure of QCD, and thus \JUNIPR will most easily learn jet trees that are most similar to QCD --- our primary reason for predominantly using the C/A algorithm. Consequently, though the model described in \Sec{ProbModel} is formally independent of clustering algorithm, the particular implementation adopted in \Sec{NNModel} may weakly depend on the chosen algorithm by virtue of the ease with which it can learn the data.

\begin{figure}[t]
\centering
\begin{tikzpicture}
\node at (7,0.5) {\includegraphics[width=0.35\linewidth]{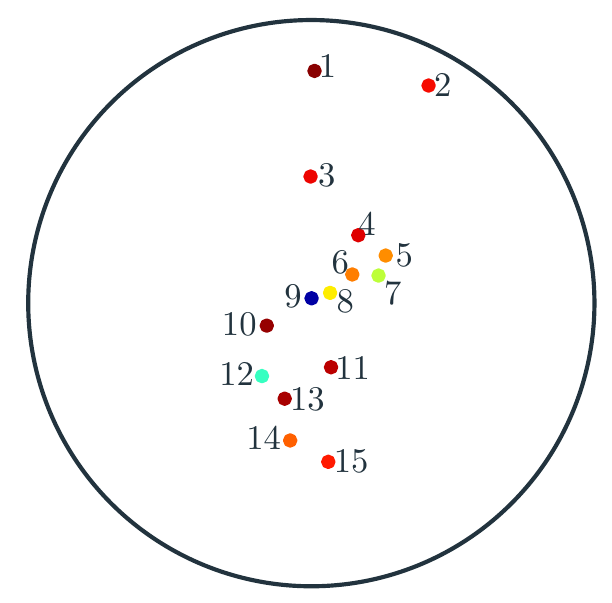}};
\node at (0,1) {\includegraphics[width=0.5\linewidth]{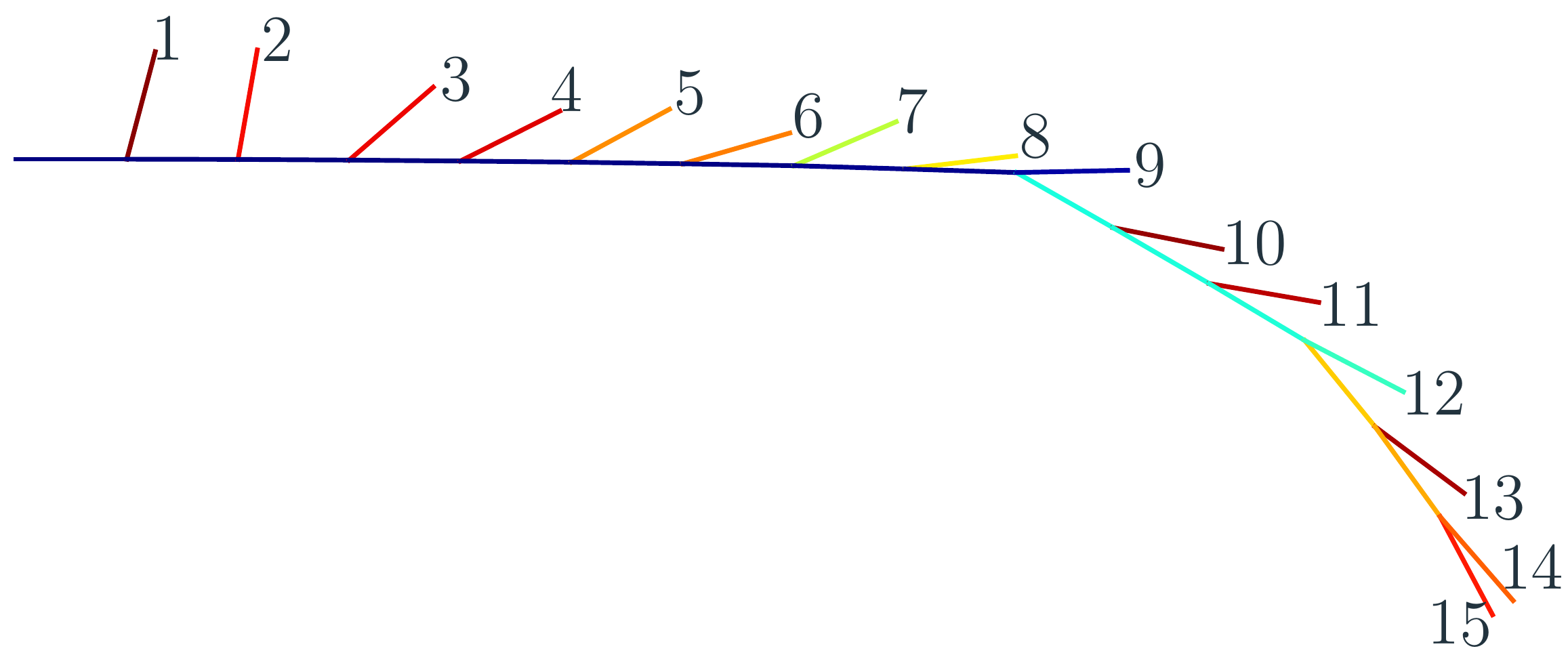}};
\node at (0,-1.5) {\includegraphics[width=0.5\linewidth]{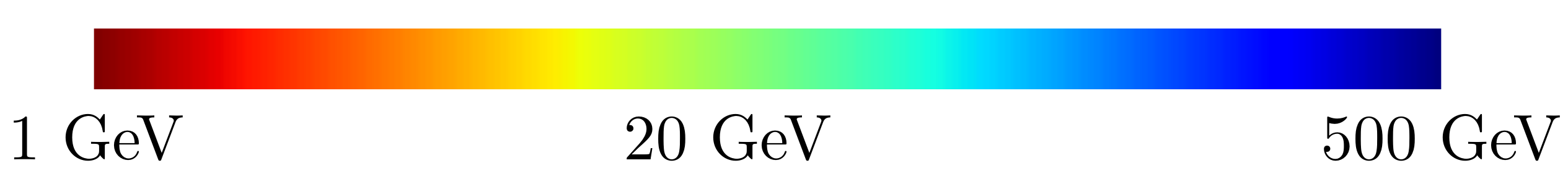}};
\node at (-2,-0.5) {\includegraphics[width=0.2\linewidth]{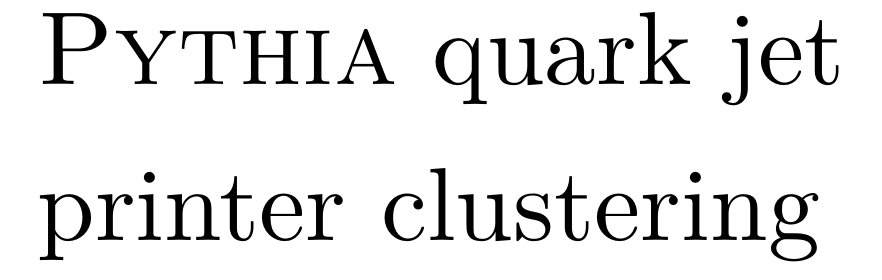}};
\end{tikzpicture}
\caption{A single \Pythia jet clustered using the printer algorithm. Shown are its clustering tree (left) and jet image (right) in which colors correspond to energies and polar coordinates correspond to the $\theta$ and $\phi$ values of the momenta. Each momentum is labelled by its corresponding step $t$ in the clustering tree.}
\label{fig:PythiaPrinterJet}
\end{figure}

To put this to the test, we have introduced a jet clustering algorithm that is nothing like QCD, but more like a 2D printer.\footnote{We thank Eric Metodiev for this suggestion.} The ``printer'' clustering algorithm scans the 2D jet image (i.e.~the cross sectional image perpendicular to the jet axis) from right-to-left and bottom-to-top, clustering particles as it encounters them. Run in reverse (i.e.~as a shower) particles are emitted from the jet core from left-to-right and top-to-bottom; this is how a jet image would be printed by an inkjet printer with a single printing tip. In \Fig{PythiaPrinterJet} we show a single \Pythia jet clustered using the printer algorithm. As can be seen in the jet image on the right side of \Fig{PythiaPrinterJet}, momenta are indeed emitted top-to-bottom. On the left side of \Fig{PythiaPrinterJet}, we see that any collinear branching structure is completely absent from the clustering tree; instead, particles are steadily emitted up-and-to-the-left.

\begin{figure}[t]
\centering 
\begin{tikzpicture}
\node at (6.5,0) {\includegraphics[width=0.35\linewidth]{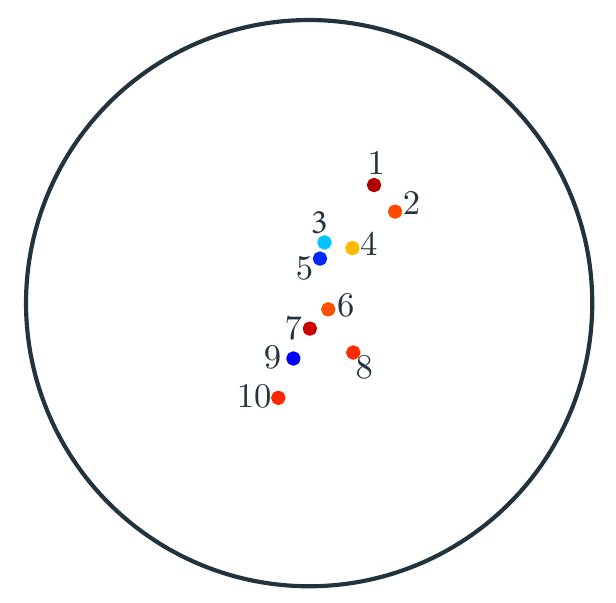}};
\node at (0,1) {\includegraphics[width=0.45\linewidth]{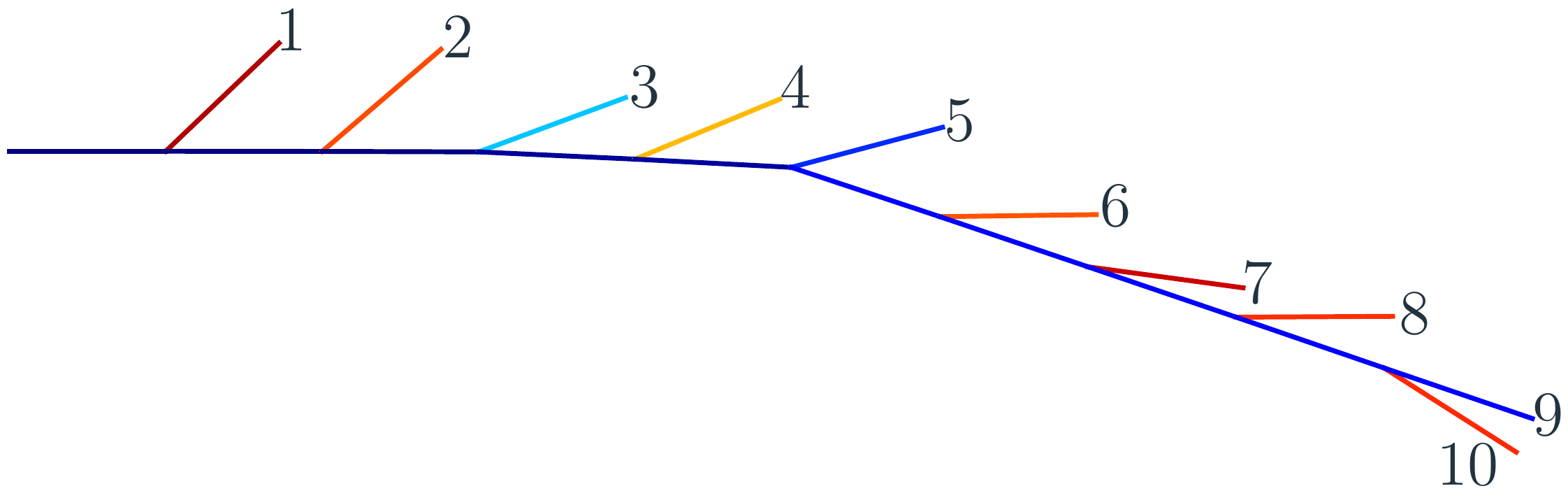}};
\node at (0,-2) {\includegraphics[width=0.5\linewidth]{figs/JUNIPR_printer_jet_legend.pdf}};
\node at (-2,-1) {\includegraphics[width=0.19\linewidth]{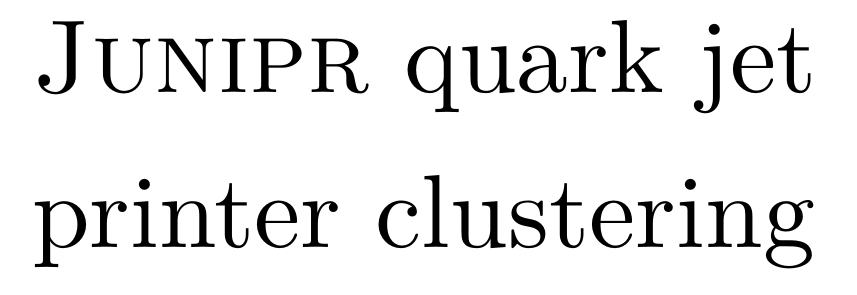}};
\end{tikzpicture}
\caption{A single jet sampled from \JUNIPR, which was trained on \Pythia-generated quark jets that were clustered using the printer algorithm. The sampled jet emits with the correct printer structure, as can be seen by its emission tree (left) and jet image (right). Each momentum is labelled by the step $t$ at which it was emitted during generation from \JUNIPR.}
\label{fig:RNNPrinterJetImage}
\end{figure}

Though \JUNIPR's neural network architecture is not optimized for the informational structure of the printer algorithm, it is still able to learn the structure, by relying much more heavily on the the jet representation $h^{(t)}$. We demonstrate this by training \JUNIPR on our data set of \Pythia-generated quark jets (see \Sec{TrainingData}) clustered with the printer algorithm, thus yielding the probabilistic model $P_\text{printer}$. Indeed, in \Fig{RNNPrinterJetImage} one can see a jet sampled from $P_\text{printer}$, which correctly follows the printer structure. 

\begin{figure}[t]
\centering
\includegraphics[width=0.5\linewidth]{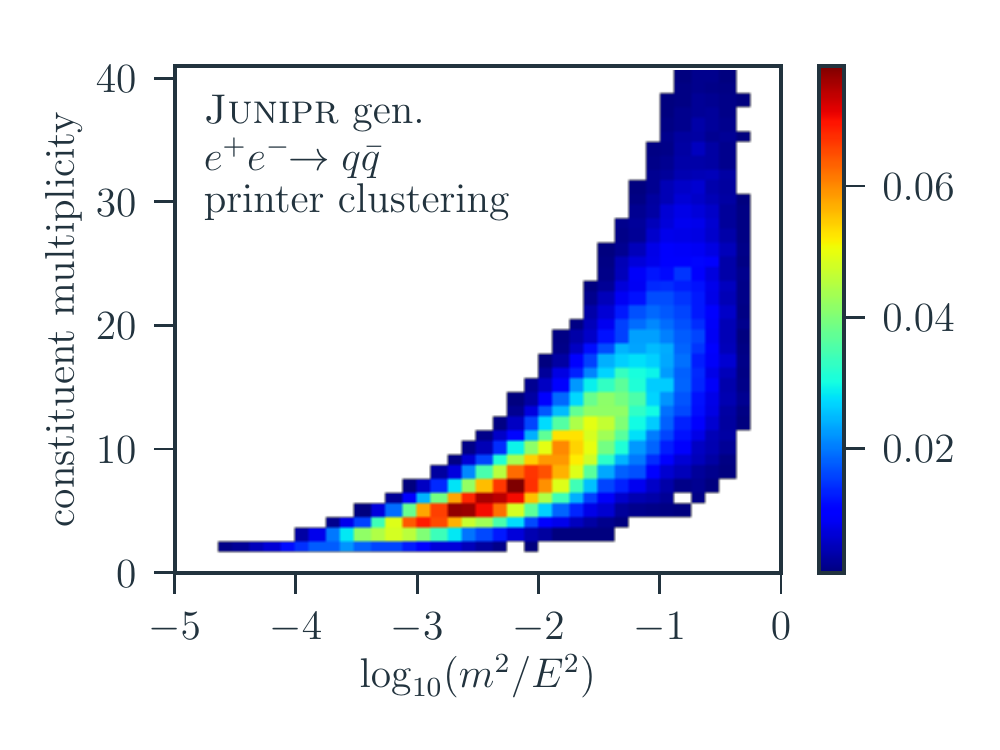}~
\includegraphics[width=0.5\linewidth]{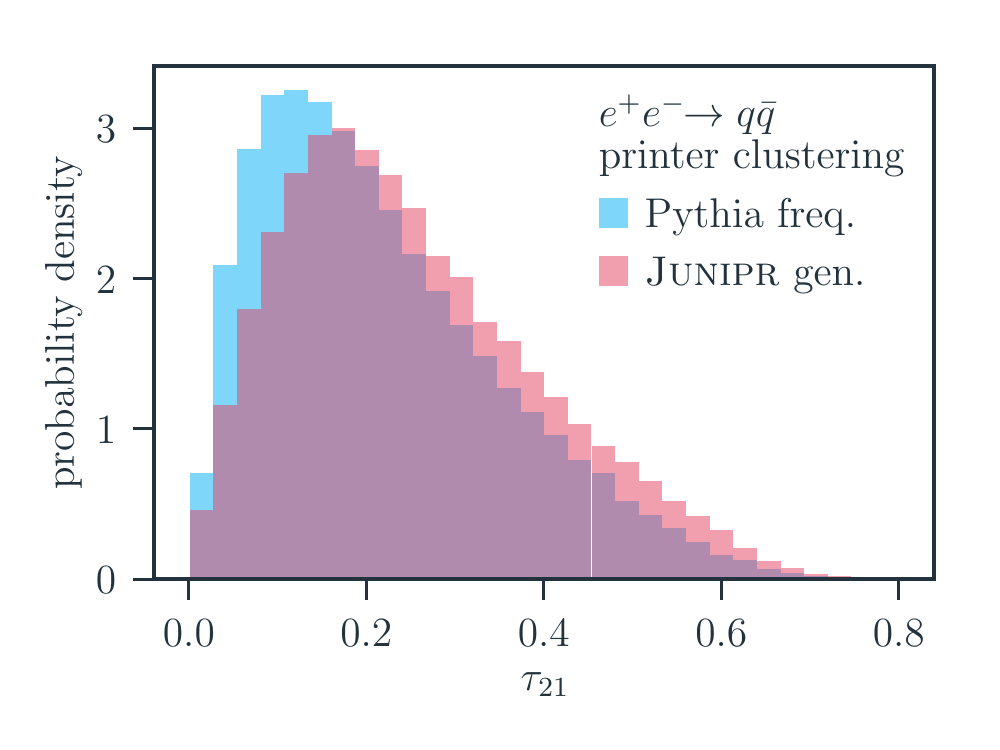}
\caption{(Left) 2-dimensional distribution with respect to jet mass and constituent multiplicity, calculated by sampling jets directly from $P_\text{printer}$, an instance of \JUNIPR trained on jets clustered with the printer algorithm. (Right) 2-subjettiness ratio observable distribution generated using $P_\text{printer}$ and compared to the corresponding distribution on \Pythia jets in the validation set.}
\label{fig:PrinterJetDistributions}
\end{figure}

As expected, however, the distributions sampled from $P_\text{printer}$ are not quite as good as our C/A results. On the left side of \Fig{PrinterJetDistributions} we show the 2-dimensional distribution over jet mass and constituent multiplicity generated using 100k jets sampled directly from $P_\text{printer}$. Comparing to the distribution generated by \Pythia (see the left side of \Fig{mass_mult_2D}) this distribution matches well. However, for the 2-subjettiness ratio observable on the right side of \Fig{PrinterJetDistributions} we get a somewhat worse match to the \Pythia validation data; compare this to the results of the C/A model in \Fig{tau21_badmatch}. Of course, we discussed in \Sec{ResultsSampling} why we do not expect direct sampling from \JUNIPR to be perfectly reliable (and we discussed a way around this in \Sec{ResultsReweighting}), but it is still clear that such distributions are comparably worse when using the printer clustering algorithm, instead of the more natural C/A algorithm.

\subsection{Anti-$k_t$ Shower Generator}
\label{sec:DiscussionAntikt}

Reassured by the results of the previous section, we next consider \JUNIPR trained on \Pythia jets reclustered with anti-$k_t$ \cite{Cacciari:2008gp}. Like the printer algorithm, anti-$k_t$ does not approximate the natural collinear structure of QCD. Unlike the printer algorithm, however, anti-$k_t$ is a very commonly used tool. For the latter reason we explore anti-$k_t$ jets here. 

Perhaps the most interesting result associated with an anti-$k_t$ version of \JUNIPR is that it provides access to an anti-$k_t$ shower generator. Generating an anti-$k_t$ shower is counterintuitive, because the anti-$k_t$ algorithm generally clusters soft emissions one-by-one with the hard jet core. Thus, a generator must remember where previous emissions landed in order to send subsequent emissions nearby. This is required to reproduce the correct collinear structure in the distribution of final-state of momenta. Said in another way, since the collinear factorization of QCD is not built into the anti-$k_t$ clustering algorithm, a local (or factorized) anti-$k_t$ generator could not produce emissions with the correct collinear distribution. Thus, we should expect that, in an anti-$k_t$ version of \JUNIPR, higher demands will be placed on the jet representation $h^{(t)}$ to monitor all the radiation in the jet. This is certainly possible, but not the task for which our neural network architecture is optimized.

To see to what extent an anti-$k_t$ implementation of \JUNIPR relies on the global information stored in $h^{(t)}$, we trained two models on \Pythia-generated quark jets clustered with anti-$k_t$ (see \Sec{TrainingData} for more details on the training data used). One model, $P_\text{anti}$, has the baseline architecture outlined in \Sec{Model}. The other, $P_\text{anti-local}$, is a local branching model like the one used in \Sec{DiscussionGlobal}, in which the global representation $h^{(t)}$ is withheld as input to the branching function. 

\begin{figure}[t]
\centering
\begin{tikzpicture}
\node at (0,-2) {\includegraphics[width=0.8\linewidth]{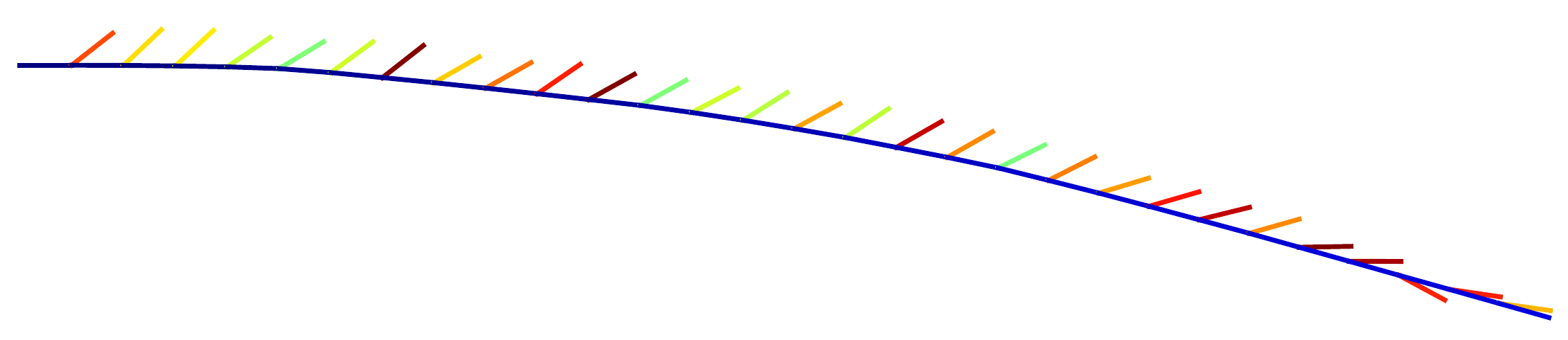}};
\node at (0,0) {\includegraphics[width=0.8\linewidth]{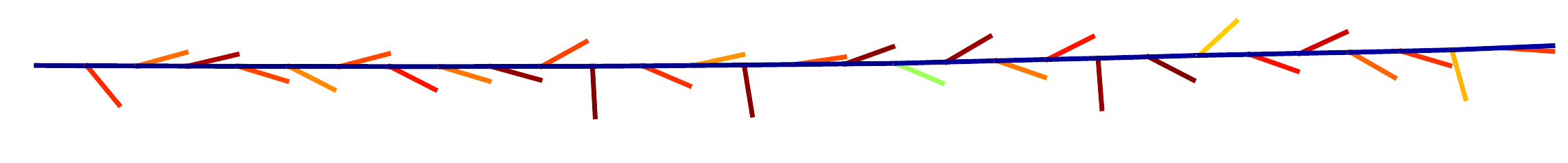}};
\node at (0,-4.5) {\includegraphics[width=0.5\linewidth]{figs/JUNIPR_printer_jet_legend.pdf}};
\node at (-2,-3.0) {\includegraphics[width=0.2\linewidth]{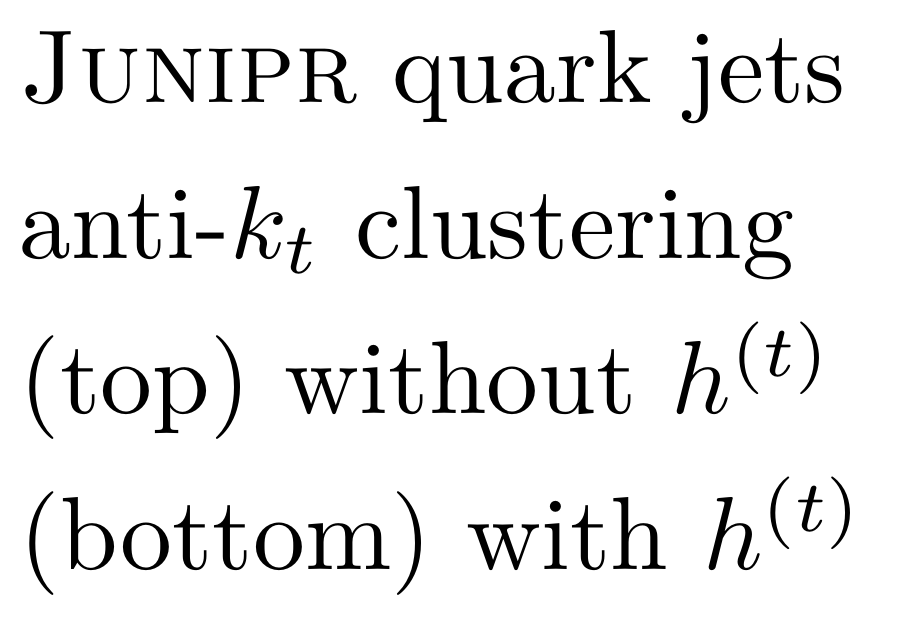}};
\end{tikzpicture}
\caption{(Top) Shower sampled from an anti-$k_t$ version of \JUNIPR, but one in which the global representation $h^{(t)}$ is withheld from the branching function. 
Correlation between emission directions is absent in this case.
(Bottom) Shower sampled from an anti-$k_t$ version of \JUNIPR, using the standard architecture complete with $h^{(t)}$. 
Strong coherence in emission directions is clearly evident.}
\label{fig:Anti-ktShowers}
\end{figure}

In \Fig{Anti-ktShowers} (bottom) we show a jet sampled from $P_\text{anti}$. In this case, though the tree itself does not properly guide the collinear structure of emissions, one can see that the emission directions are highly correlated with one another, demonstrating the success of the jet representation $h^{(t)}$ in tracking the global branching pattern. 
In \Fig{Anti-ktShowers} (top) we show for comparison a jet sampled from $P_\text{anti-local}$, in which the branching function does not receive $h^{(t)}$. In the latter case, all correlation between the emission directions is lost. This shows that the global representation $h^{(t)}$ is crucial for a successful anti-$k_t$ branching model.

\begin{figure}[t]
\centering
\includegraphics[width=0.5\linewidth]{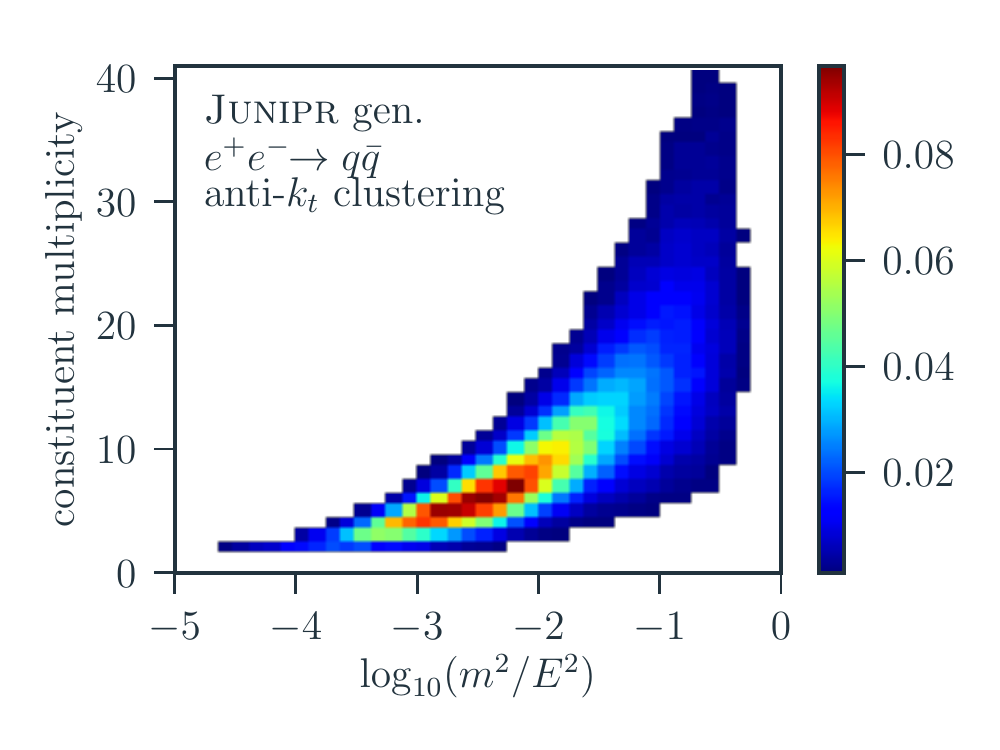}~
\includegraphics[width=0.5\linewidth]{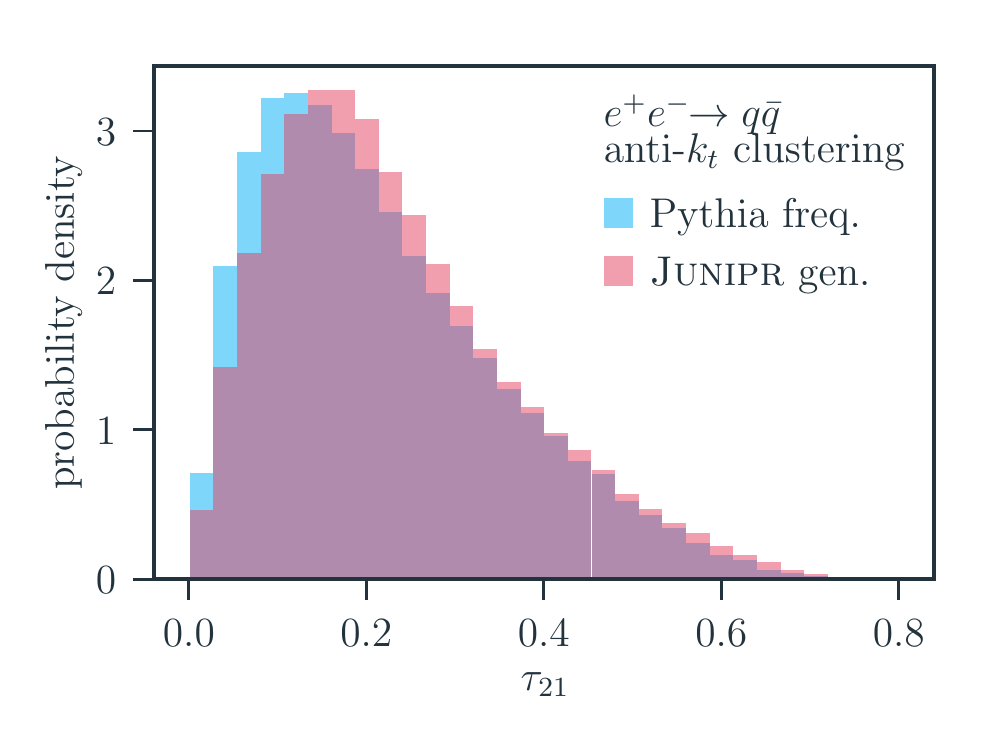}
\caption{(Left) 2-dimensional distribution over jet mass and constituent multiplicity, calculated by sampling jets directly from an anti-$k_t$ implementation of \JUNIPR. (Right) 2-subjettiness ratio observable distribution sampled from this model and compared to the distribution on \Pythia jets in the validation set.}
\label{fig:Anti-ktDistributions}
\end{figure}

In \Fig{Anti-ktDistributions} we show the 2-dimensional distribution over jet mass and constituent multiplicity, as well as the 2-subjettiness distribution, generated with $P_\text{anti}$. One can see that the former distribution is consistent with the distribution generated by \Pythia in \Fig{mass_mult_2D}. Mild disagreement between $P_\text{anti}$'s 2-subjettiness distribution and \Pythia's can be seen on the right side of \Fig{Anti-ktDistributions}. This is on par with the agreement obtained by sampling from the C/A model in \Fig{tau21_badmatch}.

In \Sec{DiscussionGlobal} we saw that the RNN's hidden state $h^{(t)}$ manages the global information in \JUNIPR's neural network architecture. This is an efficient and natural way to characterize QCD-like jets, and therefore also C/A clustering trees. Though \JUNIPR is formally independent of jet algorithm (i.e.~in the infinite-capacity and perfect-training limit), we might expect \JUNIPR's performance to degrade somewhat when paired with clustering algorithms that require significantly more information to be stored in $h^{(t)}$. This was explored in \Secs{DiscussionPrinterJets}{DiscussionAntikt} using two separate non-QCD-like clustering algorithms, namely the ``printer'' and  anti-$k_t$ algorithms. Despite these clustering algorithms being unnatural choices for \JUNIPR, we were able to demonstrate conceptually interesting and novel results, such as the anti-$k_t$ shower generator. This further demonstrates that \JUNIPR can continue to function well, even when the clustering algorithm chosen for implementation bears little resemblance to the underlying physical processes that generate the data.


\section{Conclusions and Outlook}
\label{sec:Conclusions}

In this paper, we have introduced \JUNIPR as a framework for unsupervised machine learning in particle physics. 
The framework calls for a neural network architecture designed to efficiently describe the leading-order physics of $1\to2$ splittings, 
alongside a representation of the global jet physics. 
This requires the momenta in a jet to be clustered into a binary tree.
The choice of clustering algorithm is not essential to \JUNIPR's performance, but choosing an algorithm that has some correspondence with an underlying physical model, 
such as the angular-ordered parton shower in quantum chromodynamics, gives improved performance and allows for intrerpretability of the network. 
At \JUNIPR's core is a recurrent neural network with three interconnected components. 
It moves along the jet's clustering tree, evaluating the likelihood of each branching. 
More generally, \JUNIPR is a function that acts on a set of 4-momenta in an event to compute their relative differential cross section, 
i.e.~the probability density for this event to occur, given the event selection criteria used to select the training sample.
One of the appealing features of \JUNIPR is its interpretability: 
it provides a desconstruction of the probability density into contributions from each point in the clustering history.

There are many promising applications of \JUNIPR, and we have only been able to touch on a few proof-of-concept tests in this introductory work. 
One exciting use case is discrimination. 
In contrast to supervised models which directly learn to discriminate between two samples, \JUNIPR learns the features of the samples separately. 
It then discriminates by comparing the likelihood of a given event with respect to alternative models of the underlying physics.
The resulting likelihood ratio provides theoretically optimal statistical power. 
As an example, we showed that \JUNIPR can discriminate between boosted $Z$ bosons and quark jets (in a very tight mass window around $m_Z$) in $e^+e^-$ events 
when trained on the two samples separately. 
With \JUNIPR, it is not only possible to perform powerful discrimination using unsupervised learning, 
but the discrimination power can be visualized over the entire clustering tree of each jet, as in \Fig{discrimination_trees}.
This opens new avenues for physicists to gain intuition about the physics underlying high-performance discrimination. 
Such studies might even inspire the construction of new calculable observables. 

Another exciting potential application of \JUNIPR is the reweighting of Monte Carlo events, in order to improve agreement with real collider data. 
A proof-of-concept of this idea was given in \Fig{Reweighting}, where jets generated with one \Pythia tune were reweighted to match jets generated with another. 
The reason this application is important is that current Monte Carlo event generators do an excellent job of simulating events on average, 
but are limited by the models and parameters within them. 
It may be easier to correct for systematic bias in event generation by a small reweighting factor appropriate for a particular data sample, 
rather than by trying to isolate and improve faulty components of the model. 
In this context, \JUNIPR can be thought of as providing small but highly granular tweaks to simulations in order to improve agreement with data.  

The \JUNIPR framework was used here to compute the likelihood that a given set of particle momenta will arise inside a jet. 
One can also imagine more general models that act on all the momenta in an entire event, including particle identification tags, or even low-level detector data. 
A particularly interesting direction would be to consider applying \JUNIPR to heavy ion collisions, 
in which the medium where the jets are produced and evolve is not yet well understood. 
In this case, comparing the probabilities in data to those of simulation could give insights into how to improve the simulations, 
or more optimistically, to improve our understanding of the underlying physics.

\acknowledgments

We benefited from interesting discussions with D.~Barber, E.~Bernton, A.~Botev, Y.-T.~Chien, K.~Cranmer, R.~Elayavalli, M.~Freytsis, B.~Gaujac, R.~Habib, P.~Komiske, E.~Metodiev, B.~Nachman, and J.~Thaler.
AA, CF, and MDS are supported in part by the Department of Energy under contract DE-SC0013607. 
Support for AA and CF was provided in part by the Harvard Data Science Initiative.


\bibliography{unsupervised_jets}

\providecommand{\href}[2]{#2}\begingroup\raggedright\begin{thebibliography}{10}

\bibitem{krizhevsky2012imagenet}
A.~Krizhevsky, I.~Sutskever and G.~E. Hinton, \emph{Imagenet classification
  with deep convolutional neural networks},  in \emph{Advances in neural
  information processing systems}, pp.~1097--1105, 2012.

\bibitem{ResNet}
K.~He, X.~Zhang, S.~Ren and J.~Sun, \emph{Deep residual learning for image
  recognition},  pp.~770--778, 2016.
\newblock \href{http://arxiv.org/abs/1512.03385}{{\tt 1512.03385}}.

\bibitem{DenseNet}
G.~Huang, Z.~Liu and K.~Q. Weinberger, \emph{Densely connected convolutional
  networks},  2017.
\newblock \href{http://arxiv.org/abs/1608.06993}{{\tt 1608.06993}}.

\bibitem{NMT}
D.~Bahdanau, K.~Cho and Y.~Bengio, \emph{Neural machine translation by jointly
  learning to align and translate},  2014.
\newblock \href{http://arxiv.org/abs/1409.0473}{{\tt 1409.0473}}.

\bibitem{GoogleNMT}
Y.~Wu, M.~Schuster, Z.~Chen, Q.~V. Le, M.~Norouzi, W.~Macherey et~al.,
  \emph{Google's neural machine translation system: Bridging the gap between
  human and machine translation},  \href{http://arxiv.org/abs/1609.08144}{{\tt
  1609.08144}}.

\bibitem{Transcription}
A.~Graves and N.~Jaitly, \emph{Towards end-to-end speech recognition with
  recurrent neural networks},  2014.

\bibitem{Wavenet}
A.~Van Den~Oord, S.~Dieleman, H.~Zen, K.~Simonyan, O.~Vinyals, A.~Graves
  et~al., \emph{Wavenet: A generative model for raw audio},  2016.
\newblock \href{http://arxiv.org/abs/1609.03499}{{\tt 1609.03499}}.

\bibitem{BackProp}
D.~E. Rumelhart, G.~E. Hinton and R.~J. Williams, \emph{Learning
  representations by back-propagating errors}, {\emph{nature} {\bf 323} (1986)
  533}.

\bibitem{mikolov2010recurrent}
T.~Mikolov, M.~Karafi{\'a}t, L.~Burget, J.~{\v{C}}ernock{\`y} and S.~Khudanpur,
  \emph{Recurrent neural network based language model},  in \emph{Eleventh
  Annual Conference of the International Speech Communication Association},
  2010.

\bibitem{Aad:2014yva}
{\scshape ATLAS} collaboration, G.~Aad et~al., \emph{{A neural network
  clustering algorithm for the ATLAS silicon pixel detector}},
  \href{http://dx.doi.org/10.1088/1748-0221/9/09/P09009}{\emph{JINST} {\bf 9}
  (2014) P09009}, [\href{http://arxiv.org/abs/1406.7690}{{\tt 1406.7690}}].

\bibitem{Aad:2015ydr}
{\scshape ATLAS} collaboration, G.~Aad et~al., \emph{{Performance of $b$-Jet
  Identification in the ATLAS Experiment}},
  \href{http://dx.doi.org/10.1088/1748-0221/11/04/P04008}{\emph{JINST} {\bf 11}
  (2016) P04008}, [\href{http://arxiv.org/abs/1512.01094}{{\tt 1512.01094}}].

\bibitem{Chatrchyan:2012zz}
{\scshape CMS} collaboration, S.~Chatrchyan et~al., \emph{{Performance of
  tau-lepton reconstruction and identification in CMS}},
  \href{http://dx.doi.org/10.1088/1748-0221/7/01/P01001}{\emph{JINST} {\bf 7}
  (2012) P01001}, [\href{http://arxiv.org/abs/1109.6034}{{\tt 1109.6034}}].

\bibitem{Datta:2017rhs}
K.~Datta and A.~Larkoski, \emph{{How Much Information is in a Jet?}},
  \href{http://dx.doi.org/10.1007/JHEP06(2017)073}{\emph{JHEP} {\bf 06} (2017)
  073}, [\href{http://arxiv.org/abs/1704.08249}{{\tt 1704.08249}}].

\bibitem{Datta:2017lxt}
K.~Datta and A.~J. Larkoski, \emph{{Novel Jet Observables from Machine
  Learning}}, \href{http://dx.doi.org/10.1007/JHEP03(2018)086}{\emph{JHEP} {\bf
  03} (2018) 086}, [\href{http://arxiv.org/abs/1710.01305}{{\tt 1710.01305}}].

\bibitem{Luo:2017ncs}
H.~Luo, M.-x. Luo, K.~Wang, T.~Xu and G.~Zhu, \emph{{Quark jet versus gluon
  jet: deep neural networks with high-level features}},
  \href{http://arxiv.org/abs/1712.03634}{{\tt 1712.03634}}.

\bibitem{Komiske:2017aww}
P.~T. Komiske, E.~M. Metodiev and J.~Thaler, \emph{{Energy flow polynomials: A
  complete linear basis for jet substructure}},
  \href{http://arxiv.org/abs/1712.07124}{{\tt 1712.07124}}.

\bibitem{Gallicchio:2010dq}
J.~Gallicchio, J.~Huth, M.~Kagan, M.~D. Schwartz, K.~Black and B.~Tweedie,
  \emph{{Multivariate discrimination and the Higgs + W/Z search}},
  \href{http://dx.doi.org/10.1007/JHEP04(2011)069}{\emph{JHEP} {\bf 04} (2011)
  069}, [\href{http://arxiv.org/abs/1010.3698}{{\tt 1010.3698}}].

\bibitem{ATL-PHYS-PUB-2017-004}
{\scshape ATLAS Collaboration} collaboration, \emph{{Identification of
  Hadronically-Decaying W Bosons and Top Quarks Using High-Level Features as
  Input to Boosted Decision Trees and Deep Neural Networks in ATLAS at
  $\sqrt{s}$ = 13 TeV}},  Tech. Rep. ATL-PHYS-PUB-2017-004, CERN, Geneva, Apr,
  2017.

\bibitem{Cogan:2014oua}
J.~Cogan, M.~Kagan, E.~Strauss and A.~Schwarztman, \emph{{Jet-Images: Computer
  Vision Inspired Techniques for Jet Tagging}},
  \href{http://dx.doi.org/10.1007/JHEP02(2015)118}{\emph{JHEP} {\bf 02} (2015)
  118}, [\href{http://arxiv.org/abs/1407.5675}{{\tt 1407.5675}}].

\bibitem{deOliveira:2015xxd}
L.~de~Oliveira, M.~Kagan, L.~Mackey, B.~Nachman and A.~Schwartzman,
  \emph{{Jet-images — deep learning edition}},
  \href{http://dx.doi.org/10.1007/JHEP07(2016)069}{\emph{JHEP} {\bf 07} (2016)
  069}, [\href{http://arxiv.org/abs/1511.05190}{{\tt 1511.05190}}].

\bibitem{Komiske:2016rsd}
P.~T. Komiske, E.~M. Metodiev and M.~D. Schwartz, \emph{{Deep learning in
  color: towards automated quark/gluon jet discrimination}},
  \href{http://dx.doi.org/10.1007/JHEP01(2017)110}{\emph{JHEP} {\bf 01} (2017)
  110}, [\href{http://arxiv.org/abs/1612.01551}{{\tt 1612.01551}}].

\bibitem{Komiske:2017ubm}
P.~T. Komiske, E.~M. Metodiev, B.~Nachman and M.~D. Schwartz, \emph{{Pileup
  Mitigation with Machine Learning (PUMML)}},
  \href{http://dx.doi.org/10.1007/JHEP12(2017)051}{\emph{JHEP} {\bf 12} (2017)
  051}, [\href{http://arxiv.org/abs/1707.08600}{{\tt 1707.08600}}].

\bibitem{Kasieczka:2017nvn}
G.~Kasieczka, T.~Plehn, M.~Russell and T.~Schell, \emph{{Deep-learning Top
  Taggers or The End of QCD?}},
  \href{http://dx.doi.org/10.1007/JHEP05(2017)006}{\emph{JHEP} {\bf 05} (2017)
  006}, [\href{http://arxiv.org/abs/1701.08784}{{\tt 1701.08784}}].

\bibitem{Bhimji:2017qvb}
W.~Bhimji, S.~A. Farrell, T.~Kurth, M.~Paganini, Prabhat and E.~Racah,
  \emph{{Deep Neural Networks for Physics Analysis on low-level whole-detector
  data at the LHC}},  in \emph{{18th International Workshop on Advanced
  Computing and Analysis Techniques in Physics Research (ACAT 2017) Seattle,
  WA, USA, August 21-25, 2017}}, 2017.
\newblock \href{http://arxiv.org/abs/1711.03573}{{\tt 1711.03573}}.

\bibitem{ATL-PHYS-PUB-2017-017}
{\scshape ATLAS Collaboration} collaboration, \emph{{Quark versus Gluon Jet
  Tagging Using Jet Images with the ATLAS Detector}},  Tech. Rep.
  ATL-PHYS-PUB-2017-017, CERN, Geneva, Jul, 2017.

\bibitem{Macaluso:2018tck}
S.~Macaluso and D.~Shih, \emph{{Pulling Out All the Tops with Computer Vision
  and Deep Learning}},  \href{http://arxiv.org/abs/1803.00107}{{\tt
  1803.00107}}.

\bibitem{Chien:2018dfn}
Y.-T. Chien and R.~Kunnawalkam~Elayavalli, \emph{{Probing heavy ion collisions
  using quark and gluon jet substructure}},
  \href{http://arxiv.org/abs/1803.03589}{{\tt 1803.03589}}.

\bibitem{Pearkes:2017hku}
J.~Pearkes, W.~Fedorko, A.~Lister and C.~Gay, \emph{{Jet Constituents for Deep
  Neural Network Based Top Quark Tagging}},
  \href{http://arxiv.org/abs/1704.02124}{{\tt 1704.02124}}.

\bibitem{Louppe:2017ipp}
G.~Louppe, K.~Cho, C.~Becot and K.~Cranmer, \emph{{QCD-Aware Recursive Neural
  Networks for Jet Physics}},  \href{http://arxiv.org/abs/1702.00748}{{\tt
  1702.00748}}.

\bibitem{Cheng:2017rdo}
T.~Cheng, \emph{{Recursive Neural Networks in Quark/Gluon Tagging}},
  \href{http://arxiv.org/abs/1711.02633}{{\tt 1711.02633}}.

\bibitem{Egan:2017ojy}
S.~Egan, W.~Fedorko, A.~Lister, J.~Pearkes and C.~Gay, \emph{{Long Short-Term
  Memory (LSTM) networks with jet constituents for boosted top tagging at the
  LHC}},  \href{http://arxiv.org/abs/1711.09059}{{\tt 1711.09059}}.

\bibitem{Fraser:2018ieu}
K.~Fraser and M.~D. Schwartz, \emph{{Jet Charge and Machine Learning}},
  \href{http://arxiv.org/abs/1803.08066}{{\tt 1803.08066}}.

\bibitem{Guest:2016iqz}
D.~Guest, J.~Collado, P.~Baldi, S.-C. Hsu, G.~Urban and D.~Whiteson, \emph{{Jet
  Flavor Classification in High-Energy Physics with Deep Neural Networks}},
  \href{http://dx.doi.org/10.1103/PhysRevD.94.112002}{\emph{Phys. Rev.} {\bf
  D94} (2016) 112002}, [\href{http://arxiv.org/abs/1607.08633}{{\tt
  1607.08633}}].

\bibitem{ATL-PHYS-PUB-2017-003}
{\scshape ATLAS Collaboration} collaboration, \emph{{Identification of Jets
  Containing $b$-Hadrons with Recurrent Neural Networks at the ATLAS
  Experiment}},  Tech. Rep. ATL-PHYS-PUB-2017-003, CERN, Geneva, Mar, 2017.

\bibitem{Metodiev:2017vrx}
E.~M. Metodiev, B.~Nachman and J.~Thaler, \emph{{Classification without labels:
  Learning from mixed samples in high energy physics}},
  \href{http://dx.doi.org/10.1007/JHEP10(2017)174}{\emph{JHEP} {\bf 10} (2017)
  174}, [\href{http://arxiv.org/abs/1708.02949}{{\tt 1708.02949}}].

\bibitem{Cohen:2017exh}
T.~Cohen, M.~Freytsis and B.~Ostdiek, \emph{{(Machine) Learning to Do More with
  Less}}, \href{http://dx.doi.org/10.1007/JHEP02(2018)034}{\emph{JHEP} {\bf 02}
  (2018) 034}, [\href{http://arxiv.org/abs/1706.09451}{{\tt 1706.09451}}].

\bibitem{Komiske:2018oaa}
P.~T. Komiske, E.~M. Metodiev, B.~Nachman and M.~D. Schwartz, \emph{{Learning
  to Classify from Impure Samples}},
  \href{http://arxiv.org/abs/1801.10158}{{\tt 1801.10158}}.

\bibitem{Metodiev:2018ftz}
E.~M. Metodiev and J.~Thaler, \emph{{On the Topic of Jets}},
  \href{http://arxiv.org/abs/1802.00008}{{\tt 1802.00008}}.

\bibitem{deOliveira:2017pjk}
L.~de~Oliveira, M.~Paganini and B.~Nachman, \emph{{Learning Particle Physics by
  Example: Location-Aware Generative Adversarial Networks for Physics
  Synthesis}}, \href{http://dx.doi.org/10.1007/s41781-017-0004-6}{\emph{Comput.
  Softw. Big Sci.} {\bf 1} (2017) 4},
  [\href{http://arxiv.org/abs/1701.05927}{{\tt 1701.05927}}].

\bibitem{Paganini:2017hrr}
M.~Paganini, L.~de~Oliveira and B.~Nachman, \emph{{Accelerating Science with
  Generative Adversarial Networks: An Application to 3D Particle Showers in
  Multilayer Calorimeters}},
  \href{http://dx.doi.org/10.1103/PhysRevLett.120.042003}{\emph{Phys. Rev.
  Lett.} {\bf 120} (2018) 042003}, [\href{http://arxiv.org/abs/1705.02355}{{\tt
  1705.02355}}].

\bibitem{Paganini:2017dwg}
M.~Paganini, L.~de~Oliveira and B.~Nachman, \emph{{CaloGAN : Simulating 3D high
  energy particle showers in multilayer electromagnetic calorimeters with
  generative adversarial networks}},
  \href{http://dx.doi.org/10.1103/PhysRevD.97.014021}{\emph{Phys. Rev.} {\bf
  D97} (2018) 014021}, [\href{http://arxiv.org/abs/1712.10321}{{\tt
  1712.10321}}].

\bibitem{Neyman289}
J.~Neyman and E.~S. Pearson, \emph{Ix. on the problem of the most efficient
  tests of statistical hypotheses},
  \href{http://dx.doi.org/10.1098/rsta.1933.0009}{\emph{Philosophical
  Transactions of the Royal Society of London A: Mathematical, Physical and
  Engineering Sciences} {\bf 231} (1933) 289--337},
  [\href{http://arxiv.org/abs/http://rsta.royalsocietypublishing.org/content/231/694-706/289.full.pdf}{{\tt
  http://rsta.royalsocietypublishing.org/content/231/694-706/289.full.pdf}}].

\bibitem{Butter:2017cot}
A.~Butter, G.~Kasieczka, T.~Plehn and M.~Russell, \emph{{Deep-learned Top
  Tagging with a Lorentz Layer}},  \href{http://arxiv.org/abs/1707.08966}{{\tt
  1707.08966}}.

\bibitem{Coleman:1965xm}
S.~Coleman and R.~Norton, \emph{{Singularities in the physical region}},
  \href{http://dx.doi.org/10.1007/BF02750472}{\emph{Nuovo Cim.} {\bf 38} (1965)
  438--442}.

\bibitem{Collins:1985ue}
J.~C. Collins, D.~E. Soper and G.~F. Sterman, \emph{{Factorization for Short
  Distance Hadron - Hadron Scattering}},
  \href{http://dx.doi.org/10.1016/0550-3213(85)90565-6}{\emph{Nucl.Phys.} {\bf
  B261} (1985) 104}.

\bibitem{Collins:1988ig}
J.~C. Collins, D.~E. Soper and G.~F. Sterman, \emph{{Soft Gluons and
  Factorization}},
  \href{http://dx.doi.org/10.1016/0550-3213(88)90130-7}{\emph{Nucl.Phys.} {\bf
  B308} (1988) 833}.

\bibitem{feige2014hard}
I.~Feige and M.~D. Schwartz, \emph{Hard-soft-collinear factorization to all
  orders}, {\emph{Physical Review D} {\bf 90} (2014) 105020}.

\bibitem{feige2013shell}
I.~Feige and M.~D. Schwartz, \emph{An on-shell approach to factorization},
  {\emph{Physical Review D} {\bf 88} (2013) 065021}.

\bibitem{Catani:1993hr}
S.~Catani, Y.~L. Dokshitzer, M.~H. Seymour and B.~R. Webber,
  \emph{{Longitudinally invariant $K_t$ clustering algorithms for hadron hadron
  collisions}},
  \href{http://dx.doi.org/10.1016/0550-3213(93)90166-M}{\emph{Nucl. Phys.} {\bf
  B406} (1993) 187--224}.

\bibitem{Ellis:1993tq}
S.~D. Ellis and D.~E. Soper, \emph{{Successive combination jet algorithm for
  hadron collisions}},
  \href{http://dx.doi.org/10.1103/PhysRevD.48.3160}{\emph{Phys. Rev.} {\bf D48}
  (1993) 3160--3166}, [\href{http://arxiv.org/abs/hep-ph/9305266}{{\tt
  hep-ph/9305266}}].

\bibitem{Dokshitzer:1997in}
Y.~L. Dokshitzer, G.~D. Leder, S.~Moretti and B.~R. Webber, \emph{{Better jet
  clustering algorithms}},
  \href{http://dx.doi.org/10.1088/1126-6708/1997/08/001}{\emph{JHEP} {\bf 08}
  (1997) 001}, [\href{http://arxiv.org/abs/hep-ph/9707323}{{\tt
  hep-ph/9707323}}].

\bibitem{Wobisch:1998wt}
M.~Wobisch and T.~Wengler, \emph{{Hadronization corrections to jet
  cross-sections in deep inelastic scattering}},  in \emph{{Monte Carlo
  generators for HERA physics. Proceedings, Workshop, Hamburg, Germany,
  1998-1999}}, pp.~270--279, 1998.
\newblock \href{http://arxiv.org/abs/hep-ph/9907280}{{\tt hep-ph/9907280}}.

\bibitem{Ellis:2012sn}
S.~D. Ellis, A.~Hornig, T.~S. Roy, D.~Krohn and M.~D. Schwartz, \emph{{Qjets: A
  Non-Deterministic Approach to Tree-Based Jet Substructure}},
  \href{http://dx.doi.org/10.1103/PhysRevLett.108.182003}{\emph{Phys. Rev.
  Lett.} {\bf 108} (2012) 182003}, [\href{http://arxiv.org/abs/1201.1914}{{\tt
  1201.1914}}].

\bibitem{Kahawala:2013sba}
D.~Kahawala, D.~Krohn and M.~D. Schwartz, \emph{{Jet Sampling: Improving Event
  Reconstruction through Multiple Interpretations}},
  \href{http://dx.doi.org/10.1007/JHEP06(2013)006}{\emph{JHEP} {\bf 06} (2013)
  006}, [\href{http://arxiv.org/abs/1304.2394}{{\tt 1304.2394}}].

\bibitem{Mackey:2015hwa}
L.~Mackey, B.~Nachman, A.~Schwartzman and C.~Stansbury, \emph{{Fuzzy Jets}},
  \href{http://dx.doi.org/10.1007/JHEP06(2016)010}{\emph{JHEP} {\bf 06} (2016)
  010}, [\href{http://arxiv.org/abs/1509.02216}{{\tt 1509.02216}}].

\bibitem{Soper:2011cr}
D.~E. Soper and M.~Spannowsky, \emph{{Finding physics signals with shower
  deconstruction}},
  \href{http://dx.doi.org/10.1103/PhysRevD.84.074002}{\emph{Phys. Rev.} {\bf
  D84} (2011) 074002}, [\href{http://arxiv.org/abs/1102.3480}{{\tt
  1102.3480}}].

\bibitem{Soper:2014rya}
D.~E. Soper and M.~Spannowsky, \emph{{Finding physics signals with event
  deconstruction}},
  \href{http://dx.doi.org/10.1103/PhysRevD.89.094005}{\emph{Phys. Rev.} {\bf
  D89} (2014) 094005}, [\href{http://arxiv.org/abs/1402.1189}{{\tt
  1402.1189}}].

\bibitem{Cacciari:2008gp}
M.~Cacciari, G.~P. Salam and G.~Soyez, \emph{{The Anti-k(t) jet clustering
  algorithm}},
  \href{http://dx.doi.org/10.1088/1126-6708/2008/04/063}{\emph{JHEP} {\bf 04}
  (2008) 063}, [\href{http://arxiv.org/abs/0802.1189}{{\tt 0802.1189}}].

\bibitem{Cacciari:2011ma}
M.~Cacciari, G.~P. Salam and G.~Soyez, \emph{{FastJet User Manual}},
  \href{http://dx.doi.org/10.1140/epjc/s10052-012-1896-2}{\emph{Eur. Phys. J.}
  {\bf C72} (2012) 1896}, [\href{http://arxiv.org/abs/1111.6097}{{\tt
  1111.6097}}].

\bibitem{DeepLearningBook}
I.~Goodfellow, Y.~Bengio and A.~Courville, \emph{Deep Learning}.
\newblock MIT Press, 2016.

\bibitem{GRUs}
K.~Cho, B.~van Merrienboer, C.~Gulcehre, D.~Bahdanau, F.~Bougares, H.~Schwenk
  et~al., \emph{Learning phrase representations using rnn encoder--decoder for
  statistical machine translation},  in \emph{Proceedings of the 2014
  Conference on Empirical Methods in Natural Language Processing (EMNLP)},
  pp.~1724--1734, Association for Computational Linguistics, 2014.
\newblock \href{http://dx.doi.org/10.3115/v1/D14-1179}{DOI}.

\bibitem{LSTMs}
S.~Hochreiter and J.~Schmidhuber, \emph{Long short-term memory},
  \href{http://dx.doi.org/10.1162/neco.1997.9.8.1735}{\emph{Neural Comput.}
  {\bf 9} (Nov., 1997) 1735--1780}.

\bibitem{Sjostrand:2006za}
T.~Sjostrand, S.~Mrenna and P.~Z. Skands, \emph{{PYTHIA 6.4 Physics and
  Manual}}, \href{http://dx.doi.org/10.1088/1126-6708/2006/05/026}{\emph{JHEP}
  {\bf 05} (2006) 026}, [\href{http://arxiv.org/abs/hep-ph/0603175}{{\tt
  hep-ph/0603175}}].

\bibitem{Sjostrand:2014zea}
T.~Sjöstrand, S.~Ask, J.~R. Christiansen, R.~Corke, N.~Desai, P.~Ilten et~al.,
  \emph{{An Introduction to PYTHIA 8.2}},
  \href{http://dx.doi.org/10.1016/j.cpc.2015.01.024}{\emph{Comput. Phys.
  Commun.} {\bf 191} (2015) 159--177},
  [\href{http://arxiv.org/abs/1410.3012}{{\tt 1410.3012}}].

\bibitem{Theano}
{\scshape Theano Development Team} collaboration, R.~Al-Rfou, G.~Alain,
  A.~Almahairi, C.~Angermueller, D.~Bahdanau, N.~Ballas et~al., \emph{{Theano:
  A {Python} framework for fast computation of mathematical expressions}},
  {\emph{arXiv e-prints} {\bf abs/1605.02688} (May, 2016) }.

\bibitem{HierarchicalSoftmax1}
F.~Morin and Y.~Bengio, \emph{Hierarchical probabilistic neural network
  language model},  in \emph{AISTATS’05}, pp.~246--252, 2005.

\bibitem{HierarchicalSoftmax2}
A.~Mnih and G.~E. Hinton, \emph{A scalable hierarchical distributed language
  model},  in \emph{Advances in Neural Information Processing Systems 21}
  (D.~Koller, D.~Schuurmans, Y.~Bengio and L.~Bottou, eds.), pp.~1081--1088.
\newblock Curran Associates, Inc., 2009.

\bibitem{Thaler:2010tr}
J.~Thaler and K.~Van~Tilburg, \emph{{Identifying Boosted Objects with
  N-subjettiness}},
  \href{http://dx.doi.org/10.1007/JHEP03(2011)015}{\emph{JHEP} {\bf 03} (2011)
  015}, [\href{http://arxiv.org/abs/1011.2268}{{\tt 1011.2268}}].

\bibitem{Chien:2017xrb}
Y.-T. Chien, A.~Emerman, S.-C. Hsu, S.~Meehan and Z.~Montague,
  \emph{{Telescoping jet substructure}},
  \href{http://arxiv.org/abs/1711.11041}{{\tt 1711.11041}}.

\bibitem{Gallicchio:2010sw}
J.~Gallicchio and M.~D. Schwartz, \emph{{Seeing in Color: Jet Superstructure}},
  \href{http://dx.doi.org/10.1103/PhysRevLett.105.022001}{\emph{Phys. Rev.
  Lett.} {\bf 105} (2010) 022001}, [\href{http://arxiv.org/abs/1001.5027}{{\tt
  1001.5027}}].

\end{thebibliography}\endgroup
\bibliographystyle{JHEP}

\end{document}